\documentclass[11pt,fleqn,a4paper,twoside]{book}
\usepackage{latexsym}
\usepackage{amsmath}
\usepackage{amssymb}
\usepackage{graphicx}
\usepackage{makeidx}
\makeindex \usepackage{fancyhdr} \usepackage{amsfonts}
\usepackage{amsxtra}
\usepackage{pifont} 
\usepackage{a4} \usepackage[dvips]{color}
\usepackage{bbm}
\newcommand{\dif}{\mbox{\rm{d}}}

\newcommand{\gleich}{\overset{!}{=}}
                                   
\newtheorem{lemma}{Lemma}[chapter]
\newtheorem{theorem}{Theorem}[chapter]

\newtheorem{defi}{Definition}[chapter]

\begin{document}
    
\newcounter{exa}
\newenvironment*{exa}{
   \begin{sloppypar} \hspace{2cm} \stepcounter{exa} 
      \begin{minipage}[t]{10.9cm}  \small \bf{Example} \arabic{exa} : \rm \normalsize
\setlength{\mathindent}{1cm}}{\\ \end{minipage}   \end{sloppypar} }

\newenvironment{proof}
{\begin{sloppypar}\small \it{Proof}  :\rm  
\normalsize}{
\hfill {$\Box$}
\end{sloppypar} }

 \begin{center}\thispagestyle{empty}
 \renewcommand{\baselinestretch}{1.0}
 \begin{figure}[t]
 \vspace{-1.0cm}\hspace{3.5cm}{\includegraphics[width=0.3\textwidth]{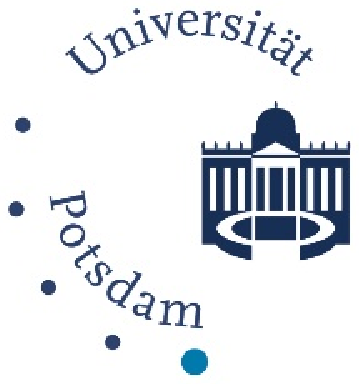}}\\
 \vspace{0.5cm}
 \end{figure}

 \Large Diploma Thesis \\
 \vspace{1.5cm}
 
 \LARGE \scshape Estimating the Degree of Entanglement of Unknown
 Gaussian States \rm\\
 \vspace{1cm}
 \Large Janet Anders \\
 \vspace{0cm}\normalsize
 Potsdam, 30.09.2003\\
 \vspace{6cm}
 
 Advisors: \:  Prof. Dr. Jens Eisert and  Prof. Dr.  Martin Wilkens \\
 \vspace{3.5cm}

 \normalsize
 Thesis submitted in partial fulfillment of the requirements of   the  \\
 University of Potsdam, \\
 Institute of Physics,\\  
 for the degree of diploma.  
 
 \rm
 \end{center}
 
 \renewcommand{\baselinestretch}{1}
 \fancyhead[LO]{\bfseries \rightmark}
 \fancyhead[RO]{\thepage}
 \fancyhead[RE]{\bfseries \leftmark}
 \fancyhead[LE]{\thepage}
 \fancyfoot[CO,CE]{}
 \renewcommand{\headrulewidth}{0.4pt}
 \renewcommand{\footrulewidth}{0.4pt}

 \tableofcontents
 \mainmatter

\chapter{Introduction}

\vspace{-8cm}\begin{figure}[h]
 \hspace{8.8cm}{\includegraphics[width=0.3\textwidth]{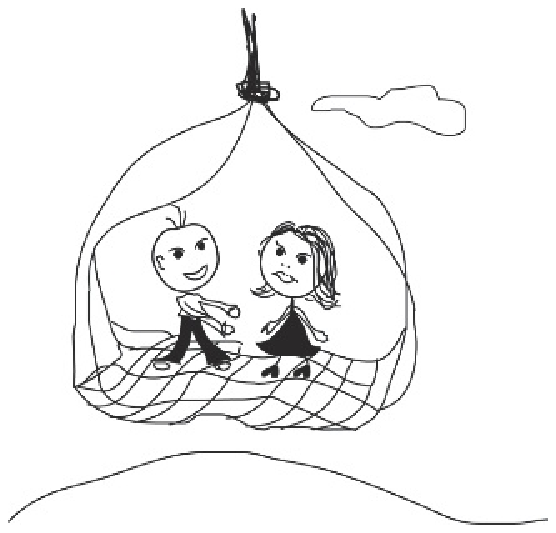}}
 
 \hspace{8.8cm} Are you entangled?\end{figure}
  
 \vspace{1.5cm}
\section{Overview and Objectives}

In the last two decades quantum information theory has emerged
as a branch of quantum physics which links concepts 
of quantum physics to ideas from information theory. 
On the one hand it became
evident that the  language from classical information theory
is appropriate to grasp several important open questions
in quantum mechanics. On the other hand, it was recognised that
quantum mechanical systems would allow for processing of
classical information that could not be realised with 
physical systems belonging to the realm of classical physics.
Most prominently perhaps, factoring of large numbers is a 
problem that can be solved in polynomial time on
a quantum computer. In computer science, 
this problem belongs to the complexity class of NP problems, 
which means that there is no known polynomial time algorithm
for a classical computer. Frankly, 
to factor large numbers is simply not possible in any reasonable time
scale with  a classical computer. 
The insight that there
exist quantum algorithms that so massively outperform the best
known classical algorithms 
resulted in a strong interest in quantum computation.  Soon other
subjects of information theory were reviewed from the 
quantum physics 
perspective, and many other applications were found.

Furthermore it was recognised that entanglement plays a major role for
those applications which can not be realised with classical schemes or
setups.  Today,  entanglement is seen as a resource, which is consumed
when using it for quantum information processing.
The dictionary describes \emph{entangled} as follows:
\begin{quote}
    If something is entangled in something else, such as a rope, wire, or
    net, it is caught in it very firmly.
    
    If you are entangled in something, you are involved in difficulties
    from which it is hard to escape.
    
    If you are entangled with someone, you are involved in a relationship
    with them that causes problems or difficulties.
\end{quote}

In quantum physics, entangled states can be understood as states where
degrees of freedom of parts of the system are tightly correlated.  In
contrast  to classical correlations, entanglement can not be generated by a preparation
using only  local operations on parts of the system and thus offers new 
dimensions for information theory and leads to manifestly 
counterintuitive consequences.
Since entanglement is the necessary ingredient for
so many applications, one naturally wants to know how to quantify it. 
Different entanglement measures were proposed, and the concepts of
 distillable entanglement and entanglement cost were introduced.

Until recently, research in the field of 
the theory of entanglement and its applications
has almost exclusively
dealt with finite dimensional quantum systems such as 
two-level systems, in analogy to the classical
\emph{bits} called \emph{qubits}.
The theoretical concept of
two-level systems can  be realised in nature by the two directions  of
a spin in a magnetic field or by the polarisation degrees of freedom of light.  It
was, for example,
shown that teleportation of an unknown two-level
state between two parties is possible, if both share a pair of
entangled states.  The experimental realisation was done  in 1998, Ref. \cite{tele}.   \\
Discrete systems, like the two-level systems, are
associated with finite-dimens\-ional Hilbert spaces.
In recent years, continuous variable systems (CVS) came into the focus of
research and fruitful attempts were done to employ the concepts and insights from
discrete systems on continuous variable states, living in an infinite
dimensional Hilbert space.  Examples are the quantised
electromagnetic field, known as a set of harmonic oscillators, 
the vibrational degree of freedom of ions in a trap, or
mechanical oscillators in the quantum domain.
The simplest nontrivial yet very important 
class of CVS are the Gaussian
states determined by their first and second moments alone.  They show
most of the interesting effects like entanglement and squeezing and are
mathematically easy to handle.  Furthermore Gaussian states are
experimentally available in optical settings.

We will investigate the entanglement properties of Gaussian states  and
formulate them on covariance matrices, associated to the states. 
Furthermore, we do not only want to quantify the degree of entanglement,
but also to estimate how much a given state is entangled.  This question
is not easy to answer even in the case when the state is fully known, but
additionally one is interested in good strategies of how to estimate or
give a minimum amount of entanglement of an unknown or partly known
state.  The young theory of entanglement witnesses is introduced and
provides a powerful tool to answer this question.  As we will present in
the sixth chapter, it is indeed possible to use entanglement witnesses
in order to estimate the degree of
entanglement of an unknown state with experimentally available
measurements.  To quantify entanglement we will  throughout this thesis 
use the  logarithmic negativity, being  so far the only computable 
entanglement measure for general states.

\section{Outline}
This thesis is divided in seven chapters, and an appendix collecting some 
small definitions and proofs.
The material is organised as follows:

\subsubsection{Basics}
In this chapter several concepts of continuous variable quantum mechanics are
reviewed, namely the commutation relations, the concept of phase space and 
of canonical (symplectic) transformations. We introduce the moments of a state 
and the quantum analogue to the classical  characteristic function. 
Finally, we define entanglement  and state the famous PPT-criterion.

\subsubsection{Gaussian States}
In the third chapter we introduce the set of 
Gaussian states starting with some familiar
examples as the coherent states of a harmonic oscillator.  We define the covariance matrix as
the major quantity determining most of the  important properties of a Gaussian
state.  Furthermore two useful normal forms of covariance matrices
will be introduced connected to invariant quantities, the symplectic
eigenvalues and the Simon invariants, whose properties we will analyse. 
Additionally we see that the Gaussian states maximise the entropy.

\subsubsection{Gaussian Operations}
This chapter mainly focuses on homodyne measurements but also shortly
discusses the allowed operations one could perform on a Gaussian
state without destroying this property.  We show that there exists an
efficient setup to project an arbitrary  state  on a coherent state 
using  homodyne measurements and other Gaussian operations.  
As we see in the subsequent chapter, the derived setup is
necessary for our scheme to determine the degree of entanglement.

\subsubsection{Measuring Entanglement of Two-Mode States}

In this and the subsequent chapter the methods of estimating 
the degree of entanglement are introduced. First, we present a scheme of
how to measure the degree of entanglement of an unknown quantum state.
We propose a setup to give the exact
logarithmic negativity of a Gaussian  two-mode  state with only nine
kinds of measurements instead of estimating all entries of the covariance
matrix with homodyne detections. 

\subsubsection{Entanglement Witnesses}

In this chapter we introduce the theory of entanglement
witnesses and show that  it is a helpful tool to answer our question for
entanglement properties of states.  We notice that the required
measurements are easily available in the laboratory and show that it is
possible not only to decide experimentally whether an unknown given
state is entangled or not, but also to give a lower bound of its
logarithmic negativity.  To do this we introduce the new quantities \emph{
$p-$separability} and \emph{minimal entanglement witness,} which
characterise the entanglement properties of covariance matrices.
Finally, we present a general strategy to estimate the degree of 
entanglement of an unknown Gaussian state, using the concept of 
entanglement witnesses. 
As the  closing words, we  give an example and shortly discuss 
the Duan-criterion as a special case of the general criterion we stated.

\subsubsection{Summary and Outlook}
We finally summarise the main results of this thesis and give some possible 
issues of  further research.

\chapter{Basics}

In this thesis we will discuss aspects of quantum information theory
(QIT)\index{QIT, quantum information theory} with continuous variable states -
especially Gaussian states.  These states can be described in a
different formalism than the one  of the familiar discrete qubit systems. 
We therefore start off with some basic concepts for continuous
variable states (CVS)\index{CVS, continuous variable states}.

\section{Commutation Relations}\label{CR}

In classical mechanics a system is described completely by its degrees of freedom, 
usually the coordinates ($ x_n $) and  momenta ($p_n $) of all $ N $ particles 
(with $ n= 1,..., N $   in one-dimensional space).
These canonical variables  fulfil the canonical relations 
\begin{equation}
\{R_i , R_j \} = \sigma_{ij}
\end{equation}
 where $\{ ~, ~\}$ denotes the \it Poisson bracket \rm defined in
 Def.  \ref{Poisson}, with $ R \in \mathbbm{R}^{2 N}$ being the vector of
 canonical variables $ R =(x_1,p_1,~...~, x_N, p_N)^T $ and $
 \sigma\index{$\sigma =\bigoplus_{i= 1}^N
 \left(\begin{array}{rr}0&1\\-1&0\end{array}\right)$, symplectic
 matrix} $ the \it symplectic matrix \rm defined as:
\begin{equation}\label{3}
    \sigma = \bigoplus_{i= 1}^N
    \left(\begin{array}{rr}0&1\\-1&0\end{array}\right)
    =\left(\begin{array}{rrrrrr}0&1&..&..&0&0\\-1&0&..&..&0&0\\..&..&..&..&..&..\\
    ..&..&..&..&..&..\\0&0&..&..&0&1\\0&0&..&..&-1&0\end{array}\right).
\end{equation}
The phase space $ ( \mathbbm{R}^{2N}, \sigma) $  spanned by the canonical coordinates is 
 then a symplectic vector space (see Def. \ref{det=1}) with a \it
 symplectic scalar product \rm 
\begin{equation*}
( \xi , \eta) : = \sigma(\xi , \eta) = \eta ^T \sigma \xi = - \sigma(\eta , \xi), 
\quad \mbox{ for } \xi , \eta \in \mathbbm{R}^{2N},
\end{equation*} 
 which is preserved under
 symplectic basis transformations, defined in the next section.  All
 $\sigma$ appearing in this thesis denote symplectic matrices of
 appropriate size. \\

In quantum mechanics we describe the canonical degrees of freedom by the set of operators 
$\{\hat x_n, \hat p_n\}_n$ or equivalently by the set of ladder  operators 
$\{\hat a_n, \hat a^{\dag}_n\}_n$ with $ n= 1, \ldots, N $.  
The generalised position operators $ \hat x_n $ act in the position representation on a vector $ \psi \in
\mathcal{H} $ as multiplications
\begin{equation*}\hat x_n ~ \psi (x) = x_n ~ \psi (x)\end{equation*} 
while the generalised momentum operators $\hat p_n$ act as derivatives
\begin{equation*} \hat p_n ~ \psi (x) = -i ~ {\partial \psi (x) \over \partial x_n}. \end{equation*}
These are unbounded linear, selfadjoint operators on the Hilbert space
$ \mathcal {H} =\mathcal{L}^{2}(\mathbbm{R}^{N},\mathbbm{C}) $.  
When quantising a system the Poisson brackets translate to commutators 
and the classical canonical relations become the canonical commutation
relations (CCRs)\index{CCR, canonical commutation relations}:
\begin{equation}\label{hat R}
    [\hat R_j,\hat R_k] = i \sigma_{jk} \, \hat{\mathbbm{1}}
\end{equation} 
 with $ \hat R\index{$\hat R =(\hat R_1, ~...~,\hat R_{2N})^T =(\hat x_1,\hat
 p_1,~...~,\hat x_N,\hat p_N)^T $} =(\hat R_1, ~...~,\hat R_{2N})^T =(\hat x_1,\hat
 p_1,~...~,\hat x_N,\hat p_N)^T $ and $\sigma$ being the symplectic
 matrix. Sometimes we will also use the ladder operators, connected to 
the position and momentum operators via  
\begin{eqnarray}\label{ladder}
\hat a_n : = \frac{\hat x_n + i \hat p_n}{\sqrt 2} &\mbox{ and }& 
\hat a^{\dag}_n : = \frac{\hat x_n - i \hat p_n}{\sqrt 2} ,\\
\hat x_n = \frac{\hat a_n +  \hat a^{\dag}_n }{\sqrt 2} &\mbox{ and }&
\hat p_n = \frac{\hat a_n -  \hat a^{\dag}_n }{i \sqrt 2},
\end{eqnarray}
and the translated CCRs  read
$[\hat a_j,\hat a^{\dag}_k] = \delta_{jk}\, \hat{\mathbbm{1}}$, with
all other commutators equal to zero.

\section{Symplectic Transformations}\label{12}

When changing our coordinate system or doing  any transformation on a quantum 
mechanical state  we use canonical transformations, 
$ T : \hat R \mapsto \hat R'$,  which leave the basic kinematic relations unchanged,
e.g., they leave the CCRs unchanged: $ [\hat R'_j,\hat  R'_k ] = i \, \sigma _{jk} \hat{\mathbbm{1}}$. 
Apart from shifts of the coordinate system's origin the simplest canonical 
transformations are linear homogeneous ones.
These are called \it symplectic transformations \rm (see
Ref. \cite{ADMS,ABKLR,HZ}) which form the real symplectic group, being one
of the three major semisimple Lie groups besides the real orthogonal
groups and the complex unitary groups.
\begin{defi} The symplectic group will be denoted by
\begin{equation}
Sp(2N, \mathbbm{R})\index{$Sp(2N, \mathbbm{R})$, symplectic
transformations, usually denoted by $S$} = \{ S |\mbox{ real } 2N
\times 2N \mbox{ matrix } , S \sigma S^T = \sigma \},
\end{equation}
\end{defi}

where the last condition comes from the CCRs, 
which shall still be fulfilled when expressing the new coordinates
as linear functions of the old ones via
\begin{equation} \label{1b}
    S : \hat R \mapsto \hat R'= S \hat R,
\end{equation} 
with $ S $ being a symplectic transformation.\\

We find the following properties for the matrices 
$ S \in Sp(2 N, \mathbbm{R}) :$
\begin{eqnarray}
    1.&& \dim ( Sp ( 2N, \mathbbm{R})) = N (2N + 1), \nonumber\\
    2.&& \sigma \in Sp( 2N, \mathbbm{R}), \\
    3.&& S \in  Sp( 2N, \mathbbm{R}) 
    \Rightarrow - S, S^{-1}, S^T \in Sp( 2N, \mathbbm{R}),\nonumber \\ 
    4.&& \det S = + 1. \nonumber
\end{eqnarray}
The first property comes from the restrictions on the entries of $ S $  
given by $ S \sigma S^T = \sigma $,  $ 2 $ and $ 3 $ are easy to check while the last 
condition is rather subtle and will be proved in the appendix (see
Def. \ref{det=1} and the following discussion).  Note, that all $2 \times
2$ matrices with determinant one are symplectic matrices, especially
all matrices from $SO(2,\mathbbm{R})$.\\

\subsection{Subsets of the Symplectic Group}

Sometimes it is useful to divide a
symplectic transformation in transformations on subsystems.  With $S =
\left( \begin{array}{cc} A & B \\
               C & D 
            \end{array} \right)$ and $ \sigma $ 
in a basis where $ \sigma $ looks like 
 $ \sigma ' = \small \left( \begin{array}{cc} 
                                     0 & \mathbbm{1}_{N \times N}\\
                                     - \mathbbm{1}_{N \times N} & 0 
                                  \end{array}\right)  \normalsize, $
the properties of the  symplectic  $ S$  may  be translated to:
\begin{eqnarray}\label{submatrices}
A B^T , C D^T , A^T C, B^T D&& \mbox{ symmetric,} \nonumber\\
A D^T - B C^T = ~ \mathbbm{1}_{ N \times N},  &&\ \\
A^T D - C^T B = ~ \mathbbm{1}_{N \times N}.  &&\nonumber 
\end{eqnarray}

Obviously, the symplectic transformations do in general not preserve 
angles between vectors. 
We therefore introduce the important compact, maximally complete
subgroup $ K(N) = Sp(2 N , \mathbbm{R}) \cap SO(2 N) $ of $ Sp(2 N ,
\mathbbm{R})$ and find that $ K(N)$ is isomorphic to the unitary group
$ U (N) $.\\

That is because every unitary $ N \times N $ matrix can be expanded 
$ U = X + i Y $ where $ X $ and $ Y $ are real $ N \times N$ matrices with 
\begin{eqnarray*}
 X^T X + Y^T Y & = \mathbbm{1}_{N \times N} = & X X^T + Y Y^T,\\
& X^T Y, X Y^T & \mbox{ symmetric.}
\end{eqnarray*}
The submatrices of any  $S \in K(N)$ 
fulfil exactly the above conditions   
required for any element of $ U(N)$. Hence $K(N)$ is isomorphic to $U(N)$.
We write the set $K (N)$ as :
\setlength{\mathindent}{0cm}
\begin{eqnarray} \label{passive}
\small{K(N) = \left\{ K |  K = 
                      \left( \begin{array}{cc}
                                 X & -Y \\
                                Y & X 
                              \end{array} \right) 
\mbox{ with } X + i Y \in U(N) ; \,  X , Y \mbox{ real} \right\}}.\normalsize
\end{eqnarray}
\setlength{\mathindent}{2cm}

These linear homogeneous transformations  will be called \it passive transformations \rm
because they conserve the energy of the system. \\

\begin{exa}  The two-mode (50:50) beam splitter is an element of $K(N)$.
\begin{eqnarray*}
S^{BS} = {1 \over \sqrt{2}} 
  \left(
     \begin{array}{cc}
        \mathbbm{1}_{2 \times 2} & - \mathbbm{1}_{2 \times 2}\\
        \mathbbm{1}_{2 \times 2} & \mathbbm{1}_{2 \times 2}. 
     \end{array}
  \right) 
\end{eqnarray*}
\end{exa} 

Another subset of $ Sp(2 N, \mathbbm{R}) $, although not a group,  are  the transformations 
of the form:
\begin{eqnarray}
\Pi(N)\index{$\Pi(N)$, active symplectic transformations} = \left\{ P | P
\in Sp(2 N, \mathbbm{R} ) , P = P^T, P \ge 0 \right\}.
\end{eqnarray}
These transformations will be called  \it active transformations \rm 
since they change the energy of the system.  Generally these
transformations are difficult to implement experimentally, e.g.  for
 states of light  strong nonlinear optical media are required to realise 
an active squeezing transformation.  \rm\\

\begin{exa} The one-mode-squeezer changes the variances of the quadratures of a 
state and is an element of $ \Pi(N)$.
\begin{eqnarray*}
S^{SQ} = 
  \left(
     \begin{array}{cc}
        d & 0\\
        0 & {1 \over d} 
     \end{array}
  \right)  \quad  d >0 \, \mbox{ is the  squeezing parameter}
\end{eqnarray*}
Remark: We will investigate squeezing in the subsequent chapter.
\end{exa}

The only matrix which have both, active and passive transformations,
in common is the identity matrix. \vspace{3ex}

\subsection{Normal Forms of Symplectic Transformations}\vspace{1ex}

We introduce some useful decompositions for symplectic matrices.  For a detailed discussion see
Ref. \cite{ADMS} and references therein.

\paragraph{Polar Decomposition}\hspace{10cm}\vspace{1ex}

From matrix analysis we know that every $ N \times N $ matrix $A $  
can be decomposed in two matrices $ P = \sqrt{A \, A^{\dag} } $ 
and $ U $ unitary,  where for $ A$ nonsingular $ U = P^{-1} A $ is uniquely determined.
If $ A $ is a symplectic matrix,  then $ P $ and hence $ U $ are real symplectic as well and 
 $ P \in \Pi(N) $ and $ U \in K(N) $ by definition. 
Now for all $  S \in Sp(2 N, \mathbbm{R}) $ exists a unique $ P \in \Pi(N)$ and a unique $  U \in K(N)$
so that $ S $ can be written as a product of both:
\begin{equation}
S = P \, U. 
\end{equation}
The decomposition shows  that every symplectic transformation can be realised by a
passive transformation, like beam splitters or phase shifters, and an
additional active squeezing operation.

\paragraph{Euler Decomposition (Singular Value Decomposition)}\hspace*{10cm}\vspace{1ex}

Since $ P $ is real symmetric we may transform it orthogonally to a basis where it is 
diagonal: $ D= O^{T} P O$ where $ D $ and $ O $ are again symplectic.
For $ D $ then follows from $ D \sigma D = \sigma$ that either 
$ \sigma_{kl} = 0 $ or $\lambda_k \lambda_l =1 $. 
The diagonal matrix is thus of the form 
 $ D = \mbox{diag}(d_1,{1 \over d_1}, ..., d_N, {1 \over d_N}) \quad \mbox{with } d_i  > 0 
\,\, \forall i = 1,...N $. \\
These diagonal matrices again form  a subgroup $ \tilde{\Pi}(N)$ 
of the symplectic matrices which we name \it squeezing transformations 
\rm or simply \it squeezers\rm.  We will introduce squeezed states in
Section \ref{csts} and discuss squeezing as a property of covariance
matrices in Section \ref{PoCN}.  \setlength{\mathindent}{0cm}
\begin{eqnarray*}\tilde{\Pi}(N) = \{ \tilde{P} | \tilde{P} =
\mbox{diag}(d_1,{1 \over d_1}, ..., d_N, {1 \over d_N}) \quad
\mbox{with } d_i > 0 \} \subset \Pi(N) \subset Sp(2N, \mathbbm{R}).
\end{eqnarray*} 
\setlength{\mathindent}{2cm}

We may now  write every $  S \in Sp(2 N, \mathbbm{R}) $ in the non-unique decomposition 
\begin{equation}
S = U_1 \, \tilde{P} \, U_2 
\end{equation}
with two passive transformations $ U_1, U_2 \in K(N) $ and a squeezing transformation 
$ \tilde{P} \in \tilde{\Pi}(N) $.
The discrete non-uniqueness results from the possibility to arrange the entries of $ \tilde P $ 
in different order. 

\paragraph{Iwasawa Decomposition}\hspace*{10cm}\vspace{1ex}

The polar decomposition enables us to decompose any symplectic matrix  with unique factors, 
but unfortunately one factor is not taken from a subgroup.  The Euler
decomposition allows a decomposition in subgroup factors but is not
unique.  We now introduce a third decomposition which unites both
virtues.\\

The Iwasawa decomposition states that any $ S \in Sp(2N, \mathbbm{R}) $ 
can uniquely be  expressed as a product of three factors 
\begin{equation}
S = N \, \tilde{ P } \, U
\end{equation}
with $N \in \mathcal{N}, \   \tilde{ P } \in \tilde{\Pi}(N)$ and $ U \in K(N) $ where $\mathcal{N} $ 
is a subgroup of $Sp(2N, \mathbbm{R})$ defined as:
\setlength{\mathindent}{0cm}
\begin{equation*}
\mathcal{N} = \left\{ \left(
   \begin{array}{cc} 
{A }& {0 }\\ 
{C}  & {(A^{-1})^T } 
   \end{array} \right) \bigg|
{A} \, =\left( 
   \begin{array}{ccc} 
\scriptstyle{ 1} &&  \scriptstyle{0} \\
&\scriptstyle{ \ddots} &\\  \scriptstyle{ X}  & & \scriptstyle{1} \end{array}
\right), \, \, \scriptstyle{ C A^{-1}} \mbox{ symmetric }\right\}
\subset Sp(2N, \mathbbm{R}).
\end{equation*}\setlength{\mathindent}{2cm}

\section{Moments}\label{13}

With the postulates of quantum mechanics every system can be  
described by an operator called  density matrix or state, 
usually denoted by $\rho$. The density matrices provide a 
more convenient language for quantum mechanical systems whose state is not completely known.  
Additionally every such system can  be completely characterised by the 
moments of its probability distribution,  as we know it from  classical statistics. 
In the quantum world this is the Wigner function, which is in contrast to classical statistics 
in general not positive. Alternatively one can use the characteristic function $\chi$, being  
the Fourier transformed of the Wigner function and allowing to derive the 
moments of the probability distribution  by taking its derivatives. \\

\begin{defi}[Density Matrices]\label{states}\hspace{10cm}
    
The set of \emph{density matrices (states)}\index{states or density
matrices $\rho$ in $S(\mathcal{H})$} on a Hilbert space
$\mathcal{H}$\index{$\mathcal{H}$, Hilbert space, in our case the
Hilbert space is the $\mathcal{L}^{2}(\mathbbm{R}^{N}, \mathbbm{C})$,
the space of square integrable functions over $\mathbbm{R}^{N}$}
fulfilling
\begin{eqnarray*}
    \rho = \rho^{\dag} && \mbox{selfadjoint}\\
        \rho \ge 0 &\quad& \mbox{positive}\\
    tr[\rho] =1 && \mbox{normalised}
\end{eqnarray*} will be denoted by
$S(\mathcal{H})$\index{$S(\mathcal{H})$, closed convex set of states 
on a Hilbert space $\mathcal{H}$, usually denoted by $\rho$}. 
$S(\mathcal{H})$ is a closed convex subset of the bounded linear
operators $ \mathcal{B}(\mathcal{H})$ \index{$\mathcal{B}(\mathcal{H})$,
bounded linear operators on $\mathcal{H}$ } on $\mathcal{H}$. 
Note, that all states  are by definition trace-class operators.
\end{defi}

From now on all $\rho$ appearing in this thesis are elements of
$S(\mathcal{H})$.
The first moments of the canonical observables
of a state $\rho$ are collected in the \it displacement vector \rm
$\vec d$\index{$\vec d$, displacement vector, element of
$\mathbbm{R}^{2N}$, also $\vec D = \sigma \vec d $} and can be
calculated via
\begin{equation} \label{2}
    \vec d_j \quad :=\quad tr[\rho \, \hat R_j ]= \langle \hat R_j\rangle, \,
    \quad j= 1, ...,2N, 
\end{equation}
while the second moments of $\rho$  are collected in the \it covariance matrix
(CM) $\gamma$\index{$\gamma$, covariance matrix or CM, element of
$\Gamma(\mathbbm{R}^{2N})$, also $\Gamma = \sigma \gamma \sigma^{T}$} \rm with
the relation
\begin{eqnarray}\label{1}
    \gamma_{ij}&:=& tr[\rho \, \{\hat R_{i}- \langle \hat R_i \rangle , \hat R_j -
    \langle \hat R_j \rangle\}_+ ] \\
    &=&2 \, tr[\rho \, (\hat R_{i} -\langle \hat R_{i} \rangle)( \hat R_{j} -
\langle \hat R_{j} \rangle )] -i \sigma_{ij}, \, \quad i, j =1, ...,2N.\nonumber
\end{eqnarray} 
The covariance matrix $\gamma$ of the
state is a real, symmetric, $2N \times 2N$ matrix.

But not every matrix fulfilling these conditions is a proper covariance matrix, 
from a simple calculation we see that  $\gamma$ and $\gamma + i \sigma $ 
(and $\gamma - i \sigma $ with complex conjugation) are necessarily positive.\\

Let $\gamma$ be the covariance matrix of system of $ N $ modes.
\begin{eqnarray} \label{15}
\hspace{-2cm}\forall v \in \mathbbm{C}^{2N}: \quad ( v |(\gamma + i \sigma) | v ) 
&=& \sum_{i,j} v^*_{i}( \gamma_{i j} + i \sigma_{i j}) v_{j} \nonumber \\ &=&
2 \, tr[\rho\sum_{i} v^*_{i} (\hat R_{i} -\langle \hat R_{i}\rangle)
\sum_{j} v_{j} ( \hat R_{j} - \langle \hat R_{j} \rangle )
]\nonumber\\ &=& 2 \, tr[\rho \, M^{\dag} M ] \ge 0, 
\end{eqnarray}
with $M =\sum_{j} \, v_{j} \, (\hat R_{j} - \langle \hat R_{j} \rangle ) $. 
The right hand side is a  
positive number for all $ v$ since $\rho$ and  operators of the form $M^{\dag} M$  
are always positive. Therefore we find that $ \gamma + i \sigma $ is positive. 
Similarly one shows that $ \gamma\ge 0 $ and $\gamma - i \sigma\ge 0$. \\

It turns out that the restriction $ \gamma + i \sigma \ge 0$ is
exactly the uncertainty relation for the canonical operators formulated in a
basis independent matrix inequality (e.g., for all quadratures).  Since
there are no other requirements than the CCRs and the uncertainty
relation, all mentioned properties of $ \gamma $ are sufficient to
guarantee that there exists a state $ \rho \in S (\mathcal{H})$ with  covariance matrix of $ \gamma$ .  \\

\begin{exa}
 The covariance matrix of a one-mode state with vanishing displacement
 is:
\begin{eqnarray*}
   \gamma =\left(\begin{array}{cc}
                    2\langle \hat x^2\rangle & 2\langle \hat x \hat p\rangle -i\\
                    2\langle \hat p \hat x \rangle + i & 2\langle \hat p^2\rangle
                 \end{array}\right).
\end{eqnarray*}
Assuming $\langle \hat  x  \hat p\rangle +\langle \hat  p  \hat x\rangle 
= 2 \langle \hat  x  \hat p\rangle -  i =  0$, the uncertainty relation on $\gamma $ reads 
\begin{equation*} 
\gamma + i \sigma = \left(\begin{array}{cc}
                    2\langle \hat x^2\rangle & + i \\
                    - i & 2\langle \hat p^2\rangle 
                           \end{array}\right) \ge 0.
\end{equation*}
Since the trace is positive  anyway the only condition to be fulfilled is a positive determinant 
\begin{equation*} 
4 \langle \hat x^2\rangle \langle \hat p^2\rangle - 1  \ge 0 \quad \mbox{or }  
\quad \langle \hat x^2\rangle \langle \hat p^2\rangle \ge { 1 \over 4}. 
\end{equation*}
But this is exactly what we know as Heisenberg uncertainty relation for the canonical coordinates.
\end{exa}

\begin{defi}
We define the \emph{ states of minimal uncertainty } to  be those states 
where half of the  eigenvalues of 
$ \gamma + i \sigma$ are zero, that is, the uncertainty relations hit equality 
in every mode.  
\end{defi}

\begin{exa} The condition for the  states of minimal uncertainty of only  one mode can easily be derived :\\
$\gamma + i \sigma  =\left(\begin{array}{cc}
                  a & c + i\\ c - i & b
                 \end{array}\right) $,  with eigenvalues $ \lambda_1 + \lambda_2 = a + b $  and 
$ a b - c^2 -1 = \lambda_1 \lambda_2 $.
Then for $ \lambda_1 = 0 $ it is $ \lambda_2 = a + b $ and thus  $ \det
\gamma = a b - c^2 = 1$ for all one-mode minimal uncertainty states.  
As we will see later (Section \ref{csts}) all states having this property are 
pure displaced squeezed vacua. 
\end{exa} \vspace{2ex}

We will need some basic lemmata of matrix analysis to cope with the
covariance matrices.  For an overwhelming variety of such little proofs
see Ref. \cite{B} and Ref. \cite{HJ}.  Most lemmata used in this thesis
are proved in the appendix.
\begin{lemma}\label{positiveGamma}
For every positive $N \times N$-matrix $A$ and every 
$M \times N$-matrix $C$:  $C A C^{\dag} $ is again positive.
\end{lemma}
\begin{proof}\vspace{2ex}
    It is  $\forall v \in \mathbbm{C}^{M}: (v, C \, A \, C^{\dag} \, v) =
    \sum_{l,m}^{N} \, \sum_{k,n}^{M} \, v_k^* \, C_{kl} \, A_{lm} \,
    C_{mn}^{\dag} \, v_n = \sum_{l,m}^{N} \, v'^*_l \, A_{lm} \, v'_m \ge
    0$,
 since $A$ is positive.
\end{proof}\vspace{2ex}

And finally, how do the moments of a state transform under symplectic
transformations $S \in Sp(2N, \mathbbm{R})$?
\begin{lemma}
    The covariance matrix $\gamma$ and the displacement
    $\vec d$ transform under symplectic transformations $S: \hat R \mapsto
    \hat R' = S \hat R$ according to \setlength{\mathindent}{2cm}
    \begin{eqnarray}\label{transgam}
	\vec d &\mapsto& S \, \vec d, \\
	\gamma &\mapsto& S \, \gamma \, S^{T}.
    \end{eqnarray}
\end{lemma}
This is easily seen by the definition of the displacement and
the covariance matrix. The resulting covariance matrix $\gamma' = S \, \gamma \, S^T$ still 
fulfills covariance matrix properties since $S \gamma S^T \ge 0$ for
$\gamma \ge 0$ and $S \gamma S^T + i \sigma = S ( \gamma + i\sigma )
S^T \ge 0$ for $\gamma + i \sigma \ge 0$.
\vspace{2ex}

We have done the first steps with covariance
matrices which will play an extraordinary role in this thesis.  We
will see in Section \ref{PoCN} how efficiently properties of Gaussian
states can be formulated on the level of covariance matrices.
Higher moments will not appear since Gaussian states are determined by
their first and second moments, so it is possible to calculate all
higher moments out of them.

\section{Characteristic Function}

The canonical operators  $\hat x$ and $\hat p$ have their drawbacks; 
especially the fact that they are unbounded operators on the Hilbert space 
having ``eigenvectors'' not belonging to it is unpleasant.  But there
is a possibility to avoid these technicalities.
\begin{defi}
We define the system of \emph{ Weyl operators}\index{$\hat
W_{\xi} $, Weyl operators elements of $\mathcal{B}(\mathcal{H})$}
:\vspace*{-0.3cm}
\begin{equation}
\hat W_{\xi}= e^{i \xi^T \sigma \hat R}
\end{equation}
for $ \xi \in \mathbbm{R}^{2N}$, $\sigma $ the symplectic matrix and
$\hat R $ defined in Section \ref{CR}.
\end{defi}
All $\hat W_{\xi} $ are unitary (bounded) operators on the Hilbert
space $\mathcal{H}$ fulfilling the following relations:

\begin{lemma} [Properties of the Weyl Operators]\label{Wr}\hspace{5cm}
    
    For $\xi, \eta \in \mathbbm{R}^{2N}$ and $ \xi^{j} \in  \mathbbm{R}^{2}$
    belonging to the $j-${th mode respectively it is:}
\begin{eqnarray*}
    \hat W_{\xi} &=& \bigotimes_{j=1}^{N} \hat W_{\xi^{j}}
    \qquad\qquad\quad\mbox{tensor product,} \nonumber \\
    \hat W_{\xi}^{\dag}&=& \hat W_{- \xi},\\
    \hat W_{\xi}\hat W_{\xi}^{\dag} &=&\hat W_{\xi}^{\dag} \hat W_{\xi} 
    = \mathbbm{1}  \qquad \quad\,\mbox{ unitarity,} \nonumber  \\ 
    \nonumber \hat W_{\xi} \hat W_{\eta} &=& e^{- { i \over 2} \xi^T \sigma
    \eta} \, \hat W_{\xi + \eta} \qquad \mbox{ Weyl relations,}\\ 
    \nonumber  tr[\hat W_{\xi} \, \hat W^{\dag}_{\eta}] &=& (2 \pi)^N \,
    \delta^{2 N}(\xi -\eta) \quad \mbox{orthogonality,}
\end{eqnarray*}
where the CCRs of the canonical operators have been translated to the Weyl relations 
of the Weyl operators.  The orthogonality should be taken with a grain of salt.
\end{lemma}
\begin{proof}  All properties are easily calculated except the
orthogonality being a bit tricky.  A physicists proof is given in the
appendix, Lemma \ref{traceweyl}.
\end{proof}\vspace{2ex}

\begin{exa} We construct the Weyl  operators for  a single mode.
With $\xi^T = (x, p)$ it follows: 
$ \hat W_{\xi}= \exp[ i (x \hat p - p \hat x)]$.
In the language of ladder operators, defined in Eq. \eqref{ladder}, we find with 
\begin{eqnarray*}
\alpha = \frac{x + ip}{\sqrt 2} &\mbox{ and }&
\alpha^* = \frac{x - i p}{\sqrt 2} 
\end{eqnarray*}
the \it displacement operator \rm $\hat D(\alpha)$\index{$\hat
D(\alpha)$, displacement operator} which is often used in quantum optics.
\begin{eqnarray}\label{4}
    \hat W_{\xi(\alpha)}= e^{\alpha^* \hat a - \alpha
    \,\hat a^{\dag}} = \hat D(- \alpha).
\end{eqnarray}
\end{exa}\vspace{-0.5cm}
The name ``displacement operator'' comes from the fact that the Weyl
operators generate translations in phase space :
\begin{eqnarray*}
\hat W_{\xi}\, \hat R \, \hat W^{\dag}_{\xi} &=& \sum_k \hat e_k ( e^{i
\xi^T \sigma \hat R}\, \hat R_k \,  e^{- i \xi^T \sigma \hat R})\\
&=& \sum_k \hat e_k 
(\hat R_k + i \sum_{i,j} \xi_i \sigma_{ij} [\hat R_j \,, \hat R_k] + 0) \\
&=& \sum_k \hat e_k ( \hat R_k +  \xi_k \hat{\mathbbm{1}}),
\end{eqnarray*}
where we have used the Baker-Campbell-Hausdorff formula, see Lemma \ref{BCH} 
and Ref. \cite{BR} for a proof.\\

The set of Weyl operators is furthermore a basis for the space of bounded linear operators 
$\mathcal{B}(\mathcal{H})$. Therefore it is possible to expand every linear 
operator from $\mathcal{B}(\mathcal{H})$  in that basis with a unique weighting function $\ f_A$:  
\begin{equation*}
A = {1 \over (2 \pi)^N} \int_{\mathbbm{R}^{2N}} 
\dif^{2N} {\xi}\: f_A(- \xi)\, \hat W_{\xi}. 
\end{equation*}
For every trace-class operator $ A \in \mathcal{T}^1(\mathcal{H}) $
(e.g., $||A ||_{1}= tr[ | A |] = tr [\sqrt{A^{\dag} \, A}]  <\infty$)
we can calculate the weighting function explicitly by using the orthogonality
of the Weyl operators.  We also use that for every bounded $ B \in
\mathcal{B} (\mathcal{H})$ the product $ A \, B $ is again in
$\mathcal{T}^1(\mathcal{H}) $,

\begin{eqnarray*}
tr [A \hat W_{\xi} ] &=& {1 \over (2 \pi)^N} \int_{\mathbbm{R}^{2N}} 
\dif^{2N}{\eta}\: f_A(- \eta)\, tr [\hat W_{\eta} \hat W_{\xi}] \\
 &=& {1 \over  (2 \pi)^N} \int_{\mathbbm{R}^{2N}} 
\dif^{2N}{\eta}\: f_A(-\eta)\,  (2 \pi)^N \, \delta^{2 N}(\xi + \eta)  \\
&=&  f_A(\xi).  
\end{eqnarray*}
From now on we will use only weighting functions of density matrices which we  call in analogy 
to classical probability theory \emph{ characteristic functions}. 
They are connected via:
\begin{defi} \label{character}
The \emph{characteristic function} of a state $\rho \in S(\mathcal{H})$ will be 
denoted by\footnote{We will drop the $\rho$ if confusions are
impossible.} 
{$\chi_{\rho}$} and is given by the expectation values of the Weyl
operators
\begin{equation*}
    \chi _{\rho}(\xi) = tr [ \rho ~\hat W_{\xi} ].
\end{equation*}
Then the state $\rho$ of $N$ modes can be expanded in the basis of
Weyl operators 
\begin{equation}\label{9}
    \rho = {1 \over (2 \pi)^N} \int_{\mathbbm{R}^{2N}} 
\dif^{2N} {\xi}~\chi_{\rho}(- \xi)~ \hat W_{\xi}.
\end{equation}
\end{defi}\vspace{1ex}

To be a proper characteristic function, $ \chi_{\rho}$ has to
fulfil the necessary conditions:
\begin{eqnarray*}
1.&& \, tr[ \rho ] = 1  \Rightarrow \chi_{\rho} (0) = 1 \\
2.&& \, \rho = \rho^{\dag} \quad \Rightarrow \chi_{\rho}(\xi) = \chi^*_{\rho} (- \xi)  \\
3.&& \, \rho \ge 0 \quad \; \, \Rightarrow tr[ \rho \, A \, A^{\dag}] \ge 0  
\mbox{ for all operators }  A \\
&& \mbox{especially for an arbitrary combination of Weyl  oper-}\\
&& \mbox{ators } ~ A = \sum_k c_k \,  \hat W_{\xi_k} 
\mbox { with } c_k \in \mathbbm{C} \mbox{ and } \xi_k \in \mathbbm{R}^{2 N}\\
&& \mbox{for } \chi_{\rho} \mbox{ follows } \sum_{k,l} c_k \,  c_l^*  \, \chi_{\rho}(\xi_k - \chi_l) \; 
e^{{ i \over 2} \xi_k^T \sigma \xi_l } \, \ge 0 \\
4.&& \, \chi_{\rho} \mbox{ is continuous at } \xi = 0
\end{eqnarray*}
For $3.$ we have used that the trace of a product of two positive operators 
is positive, see Lemma \ref{posmat} for a short proof. The third condition is named
\it $\sigma-$positive definiteness \rm and is the main condition 
$ \chi $ has to fulfil and surprisingly the above conditions are sufficient for $\chi $ 
(see Ref. \cite{H2} for a proof).\\

The moments of the state $ \rho$ can now be calculated with the help
of its characteristic function:
\begin{eqnarray*}
    \left[ {1 \over i}{\partial \over \partial \xi_k} \right] \chi_{\rho} (\xi) \Big|_{\xi = 0} \Big. 
    &=& \langle \hat R'_k \rangle_{\rho}\\
    \left[ - {\partial^2 \over \partial \xi_k \, \partial \xi_l} \right]
    \chi_{\rho} (\xi) \Big|_{\xi = 0} \Big.  &=& \langle \hat R'_k \, \hat
    R'_l\rangle_{\rho}\\
&\vdots&
\end{eqnarray*}
where $ \hat R' $ is the transformed vector of the canonical
coordinates $ \hat R' = \sigma \hat R $.
\section{Unitary and Symplectic Transformations}
\enlargethispage*{8pt}
We have seen that every state $\rho \in \mathcal{S}(\mathcal{H}) $ can
be expanded in the basis of the Weyl operators $\hat W_{\hat R,\xi} =
e^{i \xi^T \sigma \hat R} $ with a weighting function $ \chi(\xi)$:
\begin{equation}\label{four}
    \rho_{\hat R} = {1 \over (2 \pi)^N} \int_{\mathbbm{R}^{2N}} 
    \dif^{2N} {\xi}~~\, \chi(- \xi)~ \, \hat W_{\hat R, \xi}.
\end{equation}
Now the question arises how a symplectic transformation on the basis
operators $ \hat R $ change the state $\rho $?  To answer this
question we need an important theorem stated by  J. von Neumann, Ref. \cite{SN}, based 
on  the work of M.H. Stone, Ref. \cite{stone}.\newpage

\enlargethispage*{8pt}
\begin{theorem}[Stone-von Neumann]\label{SvN}\hspace{7cm}
 
     Let $\hat W ^{(1)} $ and $\hat W^{(2)}$ be two Weyl systems over a
    finite dimensional phase space ($N < \infty$), which obey the Weyl
    relations in Lemma \ref{Wr}.  \\
    If the two Weyl systems are \\
    1.  strongly continuous, i.e. $\forall \psi \in \mathcal{H} : \lim_{\xi
      \to 0} || \psi - \hat W_{\xi} \psi || = 0, $\\
    2. irreducible, 
    i.e. $\forall \xi \in \mathbbm{R}^{2N} : [\hat W_{\xi} ,  A ] = 0 
    \Rightarrow  A \propto {\mathbbm{1}}$ \\
    there exists a unitary operator $U$ such that 
\begin{equation*}
\forall \xi \in \mathbbm{R}^{2N} : \hat W ^{(1)} = U ~ \hat W^{(2)} ~  U^{\dag}. 
\end{equation*}
\end{theorem} 
When transforming the basis (\it passive\rm) with a symplectic
matrix $S$: $ \hat R \mapsto \hat R' = S ~ \hat R $, the state will be
expanded in the new Weyl system $\hat W_{S \hat R, \xi}$
\begin{eqnarray*}
\rho_{S \hat R} &=& {1 \over (2 \pi)^N} \int_{\mathbbm{R}^{2N}} 
\dif^{2N} {\xi}~~ \chi(- \xi)~ \hat W_{S \hat R, \xi}.
\end{eqnarray*}
Our systems of Weyl operators fulfill the necessary conditions for 
the Stone von Neumann theorem, so we know that the two Weyl systems 
are connected with a unitary transformation so that

\begin{eqnarray*}
\rho_{S \hat R} &\gleich&  {1 \over (2 \pi)^N} \int_{\mathbbm{R}^{2N}} 
\dif^{2N} {\xi}~~ \chi(- \xi)~ U^{\dag}(S)~ \hat W_{\hat R, \xi} ~ U(S)\\
&=& U^{\dag}(S)~ \rho_{\hat R}~ U(S).
\end{eqnarray*}
\begin{figure}[h]
\vspace{-0.3cm}
{\includegraphics[width=1\textwidth]{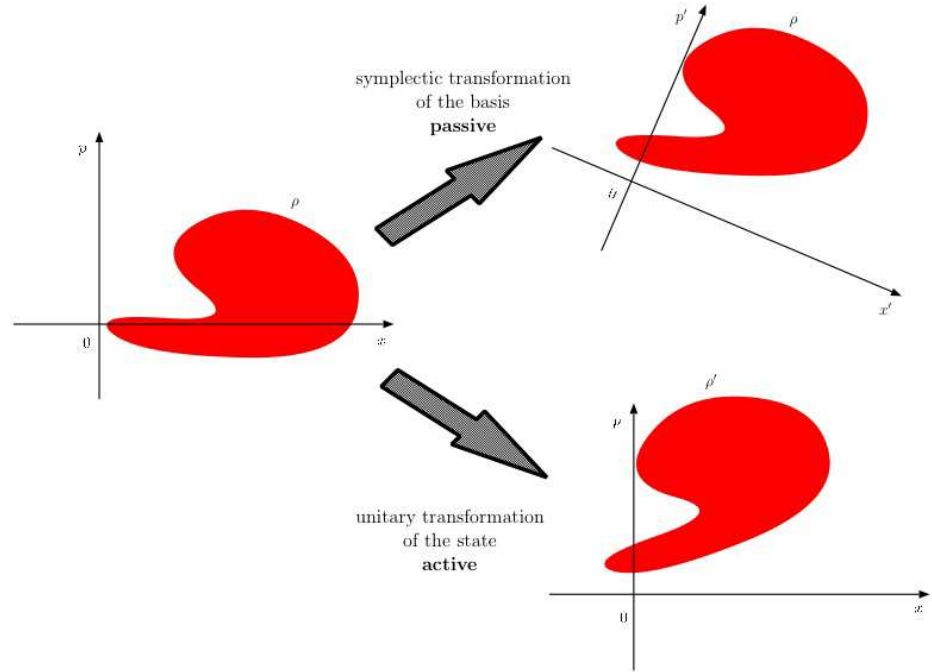}}
\vspace*{- 2.5 cm}\hspace*{0.5cm}\caption{\small Passive and active
transformations}\label{fertig}
\end{figure}\vspace{-1ex}

When comparing $\rho_{S \hat R}$ to $\rho_{\hat R}$ in  Eq. \eqref{four} 
we see that the above transformation could also be done by 
transforming the state (\it active\rm) while leaving the basis unchanged. 
In other words to every symplectic matrix $ S \in Sp(2N,
\mathbbm{R})$ exists a unitary operator $U(S)$ so, that $ \rho_{\hat R}
\mapsto \rho_{S \hat R} = U^{\dag}(S)~ \rho_{\hat R} ~ U(S)$.

\section{Bipartite Entanglement}\label{entanglement}

The crucial feature quantum mechanics provides for quantum information theory is \emph{entanglement}.
It is a purely quantum phenomenon with a lot of counterintuitive consequences. 
When measuring different parts of a composite system ($\mathcal{H} = 
\mathcal{H}_{A} \otimes \mathcal{H}_{B} $) the outcomes of the 
measurements may be correlated. This correlation could be generated 
by a preparation of the state using only  operations on each 
part of the system (local), classical communication and statistical 
mixing (LOCC). 
States which could be prepared like this will be called classically correlated or 
\emph{separable}. But this is not the whole story. In quantum theory a 
stronger kind of correlations can be observed. The so called 
\emph{entangled states} are non-local in the sense that they can not be prepared by LOCC 
and that both parts of an entangled pair do not have their own properties but contain 
only joint information. \\

The definition of separability shows the tensor product structure of 
locally produced states plus mixing.

\begin{defi}[Separability]\hspace{7cm}

    A state $\rho$ is \emph{ separable } with respect to the split $A|B$
    \emph{iff} it can be written (or approximated, e.g., in trace norm),
    with probabilities $p_{i} \ge 0 $ and $\sum_{i} p_{i} = 1 $ and proper
    density matrices $\rho_{i}^{(A)} $ and $\rho_{i}^{(B)} $ belonging to
    the parties $A $ and $B$ respectively, as
    \begin{equation}\label{separability}
	\rho = \sum_{i} p_{i} \, 
	\rho_{i}^{(A)} \otimes  \rho_{i}^{(B)}
    \end{equation} 
    that is, the closed convex hull of  product states.
    Otherwise the state is called  \emph{entangled}. The  set of separable density matrices  $\rho$ 
    on $\mathcal{H}_{A} \otimes \mathcal{H}_{B}$ ( with respect to the split $A|B$) will be 
    denoted by $S_{{A|B}}(\mathcal{H})$
    \index{$ S_{A"|B}(\mathcal{H})$, set of separable density matrices  $\rho$ 
    on $\mathcal{H}_{A} \otimes \mathcal{H}_{B}$,  } and is a closed convex subset of
    $S(\mathcal{H})$ defined in Def. \ref{states}.
\end{defi}

We see that this definition gives rise to complications since it is not easy to find out 
whether a given state $\rho$ can be written in the form \eqref{separability} or not. 
A significant amount of research  has been done  to find simpler criteria to  decide
whether a given state is entangled or not.  One of the most famous
criteria is the so called PPT-criterion. 
\begin{theorem}[PPT criterion for bipartite
systems]\label{peres} \hspace{2cm} 

Let $\rho$ be a state on a Hilbert space
$\mathcal{H} = \mathcal{H}_A \otimes \mathcal{H}_B$ which has a
\emph{non-positive partial transpose }then the state is entangled with
respect to the split $ A|B$.
\begin{equation*} 
\rho^{T_A} \not \ge 0 \Rightarrow \rho \mbox{ is entangled}
\end{equation*}
In the cases $1. \, \mathcal{H}_A= \mathcal{H}_B= \mathbbm{C}^2 $, 
$2.\, \mathcal{H}_A = \mathbbm{C}^2 $ and $ \mathcal{H}_B= \mathbbm{C}^3 $
and $3.$ for Gaussian states split into $1_A \times N_B$ modes this
criterion is also sufficient.
\end{theorem}\vspace{2ex}
\begin{minipage}[t]{13.2cm} 
    The necessary direction for entanglement was
    formulated by A. Peres in 1996 Ref. \cite{P}.  The sufficiency was
    shown by M.+ P.+R. Horodecki ($1$ and $2$) in 1996 Ref. \cite{HHH} and
    R.F. Werner and M.M. Wolf ($3$) in 2000  Ref. \cite{W3}.  The name PPT
    criterion is an abbreviation of positive partial transpose.
\end{minipage} \vspace{2ex}

The \emph{ partial (say $A$) transpose of a state
}\index{$\rho^{T_{A}}$, partial transpose of a state } is the state
transposed only on one of its subsystems (only $A$).  In formulae:
$\rho_{ \alpha, n; \beta, m}=(\rho^{T_A})_{\beta, n; \alpha, m} $
where $\alpha, \beta$ and $n, m$ belong to the parties $A$ and $ B$
respectively.  Note, that $\rho^{T_{A}}$ is still selfadjoint and has
trace one.  The partial transpose of a state is only a mathematical
tool and cannot perfectly  be  performed   by a physical transformation. 
Sometimes the partial transpose of a system's state is understood as a
time reversal in the transposed part only.  Phase conjugation in laser
beams actually realise such a time reversal but it can not be done
perfectly since there always comes a small amount of noise with it
(see Example 7 in the fourth chapter for a little
discussion).\vspace{1ex}

Almost every application in QIT uses non-classical correlations, e.g., 
teleportation \cite{1,1b,1c}, quantum cryptography \cite{2,2b,2c} and  dense coding  \cite{3}.  
Because all these funny
things strongly depend on the entanglement of the used states it is
often understood as a resource. Naturally one wants to know how
much of that resource one has. Due to its importance a great deal of 
effort has been invested in the last decade, to find a sensible 
 measure  quantifying entanglement.  
Several entanglement measures have been proposed, for
example the \emph{ entanglement of formation } in Ref.  \cite{4,4b} and
the \emph{ distillable entanglement } in Ref.  \cite{5} and further
discussions in Ref.  \cite{6,6b}.  We introduce two quantities
strongly related to entanglement measures, which we will need later
on.
The von Neumann entropy is the the quantum analogue of the Shannon
entropy Ref. \cite{7} and gives the degree of mixedness or impurity of a state.

\begin{defi}[von Neumann entropy] \label{vNe}\hspace{6cm}
    
The \emph{von Neumann entropy} of a mixed state $\rho \in S(\mathcal{H}) $ is defined as
\begin{equation*}
S(\rho) := - tr[ \rho \, \ln \rho] = - \sum_i \, p_i \ln p_i
\end{equation*}
where the $p_i$ are the eigenvalues of the state $\rho$.
\end{defi}

The only so far computable  entanglement measure is the
\emph{logarithmic negativity}, Ref. \cite{105,105b,105c,105d}.  
With  $||A||_1 := tr[\sqrt{A^{\dag} A}]$ we define the logarithmic 
negativity 
\begin{defi} \label{lognegrho} The \emph{logarithmic negativity} of a bipartite state
$\rho \in $ is given by
\begin{eqnarray*}
    E_{\mathcal N}(\rho) := \ln ||\rho^{T_A} ||_1.
\end{eqnarray*}
with respect to the split $A|B$. 
\end{defi}
If $\rho^{T_{A}}$ has negative eigenvalues $\lambda_{i}(\rho^{T_{A}})$
fulfilling $ 1 = tr[\rho^{T_{A}}] = \sum_{i}\lambda_{i}(\rho^{T_{A}})
$, the sum of the absolute values of the eigenvalues must be greater
than one: $||\rho^{T_{A}}||_{1} =\sum_{i}|\lambda_{i}(\rho^{T_{A}}) |
> 1 $.  Hence the logarithmic negativity of $\rho$ is greater than
$0$.  If $\rho$ has a positive partial transpose $\rho^{T_A} \ge 0$,
the logarithmic negativity vanishes since $tr[\sqrt{\rho^{T_A}
\rho^{T_A}}] = tr[ \rho^{T_{A}} ] = 1 $.  For all separable states
this is true.  But there may also exist entangled states with PPT
unless the PPT-criterion is also sufficient. In this thesis we will
mainly discuss the case where the PPT-criterion is also sufficient,
namely for Gaussian states of $1_{A} \times N_{B} $ modes.  Note, that
for all elements of $S(\mathcal{H})$ the logarithmic negativity is
positive definite.

\chapter{Gaussian States} \label{GS}

In quantum information theory more and more attention is paid to continuous variable states.
Gaussian states are a nice target to exploit since they are
available not only theoretically but can be observed and prepared in the
lab.  For example, any laser produces Gaussian states and most 
optical setup laser light can go through preserves this property. 
From the mathematical point of view they are the simplest case of
nontrivial CVS showing squeezing and entanglement.  To learn how one
can use continuous variable states for QIT purposes Gaussian states
play the key role and will be characterised in this chapter.

\begin{defi}
    A \emph{Gaussian $N-$mode state}  $\rho$  is a state,   whose
    characteristic function defined in Def. \ref{character} can be written
    as:
\begin{equation} 
    \chi_{\rho}(\xi) = \exp[-{1 \over 4}\xi^T \Gamma \xi + i \vec D^{T} \xi], 
    \qquad \xi \in \mathbbm{R}^{2N},
\end{equation}
where $\Gamma$ 
is the covariance matrix of the state and $\vec D $ the displacement
vector.
\end{defi}
No other parameters appear since in analogy to classical probability distributions 
the quantum Gaussian states are determined by their first and second
moments alone. When describing Gaussian states, we will often use 
only  their covariance matrices since they reflect all important 
properties  Gaussian states can have. 
We will investigate those properties in the subsequent sections after 
introducing some examples of Gaussian states. 

Unfortunately in the literature both, the matrix $\gamma$ defined in
Eq. \eqref{1} { and } the matrix $\Gamma$ appearing in the
characteristic function above, are called \emph{ covariance matrix}. 
Usually it is clear which one is meant and therefore we will not
distinguish them explicitly.  As a general rule in mathematical
formulae we will use small greek letters in the first case and capital
letters in the second.  The same holds for the displacements $\vec d$
and $\vec D$.
The two displacement vectors and the  two matrices
can be transformed into each other using the symplectic matrix defined
in Eq. \eqref{3} with the following transformation law:
\begin{eqnarray}\label{14}
\vec D &=& \sigma \, \vec d  \, \mbox{ and } \, \Gamma =\sigma  \, \gamma \, \sigma^T.
\end{eqnarray}
As one can easily check the $\Gamma$-matrix  again fulfills  covariance matrix  
properties.

\section{Coherent, Squeezed and Thermal States}\label{csts}

We introduce three important classes of pure one mode Gaussian states, 
namely the  coherent, the  squeezed and the  
thermal states of an electromagnetic field mode.  General reference for this
section is Ref.  \cite{BR} by S.M. Barnett and P.M. Radmore.

\subsubsection{Coherent States}

The coherent states  are defined as the eigenvectors of the non
hermitian annihilation operator $ \hat a $ : $\hat a | \alpha \rangle
= \alpha | \alpha \rangle,\index{$"|\alpha \rangle $, coherent states
with complex $\alpha$ } $ with $\alpha$ being a complex eigenvalue, and one
easily finds that those state vectors are  superpositions of number
state vectors of the form
\begin{equation}
 | \alpha \rangle = e^{- {| \alpha |^2 \over 2}} \,  
\sum_{n =0}^{\infty} \, {\alpha^n \over \sqrt{n!}} \,  |n \rangle.
\end{equation}

They can be  generated from the vacuum with the unitary  operators $\hat D(\alpha)$ 
introduced in Eq. \eqref{4} which induce displacements in phase space.
 
\begin{eqnarray*}
 \hat D( \alpha)& =& e^{\alpha \hat a^{\dag} - \alpha^* \hat a } 
= e^{\alpha \hat a^{\dag}} e^{- \alpha^* \hat a } e^{- {|\alpha |^2 
\over 2 }},\\
  | \alpha \rangle &:=& \hat D (\alpha) | 0 \rangle.
\end{eqnarray*}

They are an overcomplete nonorthogonal set of state vectors spanning
the entire Hilbert space resolving unity, 
\begin{equation*}
    \hat {\mathbbm{1}} = \int {\dif \Re(\alpha) \dif \Im(\alpha) \over \pi} 
    | \alpha \rangle \langle \alpha|,
\end{equation*}
and the scalar product between two coherent states 
\begin{equation*}
    \langle \alpha| \beta \rangle =e^{-{|\alpha|^2 + |\beta|^2 -2 \alpha^* \beta \over 2}}.
\end{equation*}

With $\alpha = {\eta_1 + i \eta_2 \over \sqrt{2} }$ and the
identification Eq. \eqref{4} we calculate the characteristic function of the
coherent states:
\begin{eqnarray*}
\chi_{\alpha} (\xi) &=& tr[ |\alpha \rangle \langle \alpha| \hat W_{\xi}] 
= tr[\hat D(\alpha)|0 \rangle \langle 0 | \hat D(- \alpha) \hat W_{\xi}]\\
&=&tr[ \hat W_{- \eta}|0 \rangle \langle 0 | \hat W_{ \eta} \hat W_{\xi}]
= e^{- {i \over 2 } \eta^T \sigma \xi} \, 
\langle 0 | \hat W_{\eta + \xi} \hat W_{- \eta} |0 \rangle\\
&=& e^{- {i \over 2 } \eta^T \sigma \xi} \, e^{ {i \over 2 } (\eta + 
\xi)^T \sigma \eta} \, 
\langle 0 | \hat W_{ \xi} |0 \rangle = e^{- { \xi_1^2 + \xi_2^2 \over
4}} \, e^{- {i} \eta^T \sigma \xi}\\
&=& e^{- { \xi_1^2 + \xi_2^2 \over 4}} \, e^{- {i} \sqrt{2}(\Re 
(\alpha), \Im (\alpha)) \sigma \xi},
\end{eqnarray*}
 where we used the cyclicity  of the trace, the Weyl relations from
 Lemma \ref{Wr} and $\langle 0 | \hat W_{\xi} |0 \rangle = e^{- {
 \xi_1^2 + \xi_2^2 \over 4}} $.  \\

The coherent states are thus of Gaussian type and furthermore all of 
them have the same covariance matrix: $\Gamma_{\alpha} = \mathbbm{1}$, while the
displacements depend on the value of $\alpha$: $ \vec D_{\alpha} =
\sqrt{2}\left(\begin{array}{c} \Im \, (\alpha) \\ - \Re \,(\alpha)
\end{array}\right).$ Similarly, we have  $\gamma_{\alpha}
= \sigma \Gamma_{\alpha} \sigma^T = \mathbbm{1}$ and $\vec d_{\alpha}
= - \sigma \vec D_{\alpha} = \sqrt{2} \left( \begin{array}{c} \Re \,
(\alpha) \\ \Im \,(\alpha) \end{array}\right)$.  The uncertainty relation
is exactly fulfilled with $ \Delta^{2} \hat x \cdot \Delta^{2} \hat p ={1 \over 4}
$ and is equally stretched in all directions in phase space. 
Therefore coherent states can be pictured as circles of diameter ${1
\over \sqrt{2}}$ around the points $(\sqrt{2} \Re(\alpha), \sqrt{2}
\Im(\alpha))$.  For $\alpha =0$ the coherent state vector $ | 0 \rangle$ is
just the vacuum with $\Gamma_{0} = \gamma_{0} =\mathbbm{1} $ and $\vec
D_{0} = \vec d_{0} = 0$.

\begin{figure}[h]
	\[{\includegraphics[width=0.75\textwidth]{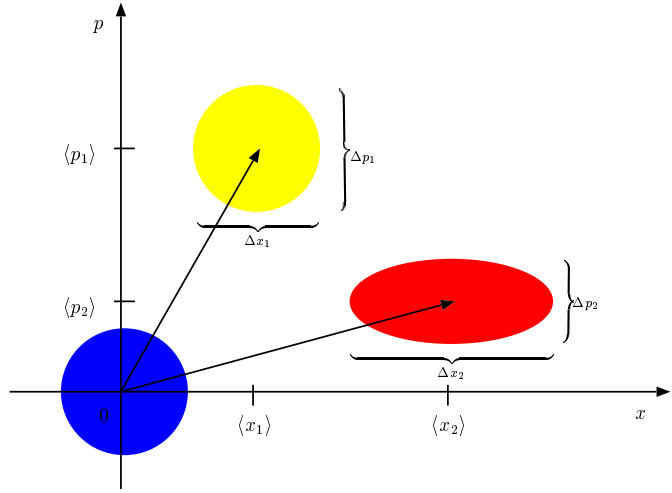}}\]
	\vspace*{-0.5cm}\hspace*{-0.5cm}\caption{\small States in phase
	space}\label{fig2}
\end{figure}
The picture shows the vacuum state (blue), a coherent state (yellow)
and a displaced squeezed vacuum  (red).  
 They all span the same area meaning that the product $\Delta^2 x \cdot \Delta^2 p $
is the same for all states, namely it is ${1 \over 4}$, the lower
bound of the uncertainty relation. Hence all these states are minimal
uncertainty states.

\subsubsection{Squeezed States}

Similarly one can produce states with non-energy conserving 
squeezing operators $\hat S(\zeta)$, though experimentally it is difficult
to realise such a transformation requiring strong Kerr media.
\begin{eqnarray*}
\hat S(\zeta) &:=& e^{-{\zeta \over 2}\hat a^{2 \dag} + {\zeta^* \over 
2} \hat a^2}\\
&=&  e^{-{1\over 2} \exp[i \phi]\tanh r \, \hat a^{2 \dag} }\, 
e^{-{1 \over 2} \ln (\cosh r)(\hat a^{\dag} \hat a + \hat a \hat 
a^{\dag})}\,
e^{{1\over 2} \exp[- i \phi]\tanh r \, \hat a^2}.
\end{eqnarray*} 
$\hat S $ is unitary since it is $\hat S (\zeta )= \hat S^{\dag}(-\zeta)$
with $ \zeta = r e^{i \phi}$.  The \emph{ squeezed vacuum} is prepared
by letting a squeezing operator act on the vacuum.
\begin{eqnarray*}
 | \zeta \rangle \index{$"|\zeta \rangle$, squeezed vacuum with
 complex $\zeta$ }& := & \hat S(\zeta) | 0 \rangle \\
& =& {1 \over \sqrt{\cosh r}} \sum_{n=0}^{\infty} {\sqrt{(2n)!} \over 
n!} \, 
(- {1 \over 2} \exp[i \phi ] \tanh r)^n \,  |2n \rangle. 
\end{eqnarray*}

The characteristic function of the squeezed vacuum with $\alpha = {\xi_1
+ i \xi_2 \over \sqrt{2}}$ is:
\begin{eqnarray*}
    \chi_{\zeta} (\xi) &=& tr[ |\zeta \rangle \langle \zeta| \hat W_{\xi}]\\
    &=&\langle 0 | \hat S(- \zeta) \hat D(-\alpha) \hat S(\zeta)|0 \rangle\\
    &=&\langle 0 | \hat S(- \zeta) \hat S(\zeta)  \hat D(-\alpha ') |0 \rangle 
    =  e^{- {|\alpha|^2 \over 2}}
\end{eqnarray*}
with $\alpha'$ determined by  commutator of $ \hat D(-\alpha)$ with $ \hat S(\zeta)$ and given by 
\begin{eqnarray*}
    \alpha' &=& { \xi_{1} [\cosh r + \cos \phi \sinh r] +  \xi_{2} \sin \phi \sinh r  \over \sqrt{2}}  \\
    &+& i \, { \xi_{1} \sin \phi \sinh r + \xi_{2} [\cosh r - \cos \phi \sinh r] \over \sqrt{2}}
\end{eqnarray*}
 we get
\begin{eqnarray*}
    \chi_{\zeta} (\xi)
    &=& e^{- {1 \over 4}  \xi_1^2 (\cosh 2r + \sinh 2r \cos \phi)} \, e^{- {1 \over 4}  \xi_2^2
    (\cosh 2r - \sinh 2r \cos \phi)  }\\
   && \times \, e^{- {1 \over 2} \xi_1 \xi_2 \sinh 2r \sin \phi},
\end{eqnarray*}
and hence all squeezed vacua are of Gaussian type.  They have vanishing 
displacements and covariance matrices 
\begin{equation*}
  \Gamma_{\zeta} = \left( 
  \begin{array}{cc}
    \cosh 2r + \sinh 2r \cos \phi & \sinh 2r \sin \phi \\
    \sinh 2r \sin \phi &\cosh 2r - \sinh 2r \cos \phi 
  \end{array}\right).
\end{equation*}
From the CM one can read off that the  variance of the state in one direction of 
the phase space is different than in the other; that is why the states
are called squeezed.  But the uncertainty relation is still fulfilled
as a short check shows.  \setlength{\mathindent}{0cm}
\begin{eqnarray*}
\Delta^2 \hat x \cdot  \Delta^2 \hat p &=& {1 \over 4}(\cosh 2r +\sinh 2r \cos 
\phi)(\cosh 2r -\sinh 2r \cos \phi)\\
&=& {1 \over 4}(\cosh^2 2r - \sinh^2 2r \cos \phi) = {1 \over 4} 
+ {\sinh^2 2r \sin^2 \phi \over 4} \ge {1 \over 4},
\end{eqnarray*}
where equality  (minimal uncertainty) is  reached for
$r= 0$ or $\phi =0$. In the first case there is no squeezing at all, 
$\hat S(0) = \mathbbm{1}$, and we again have the vacuum. In the second case
$\Gamma_{\zeta} $ has the form
\setlength{\mathindent}{2cm}
\begin{equation*}
  \Gamma_{\zeta=r} =\left(
  \begin{array}{cc}
    \cosh 2r + \sinh 2r  & 0 \\
    0  &\cosh 2r - \sinh 2r 
  \end{array}\right)
  =\left( 
  \begin{array}{cc}
    e^{ 2r} & 0 \\
    0  & e^{- 2r}
  \end{array}\right),
\end{equation*}
where we recognise that $\Gamma_{\zeta =r}$ could be prepared from the
vacuum with the symplectic squeezing transformation introduced in
Example 2 of the second chapter with the squeezing parameter $d=e^{r}$.
All \emph{ squeezed states } can be prepared from the vacuum by a
squeezing operation (squeezed vacuum) and an additional displacement in
phase space.

\newpage

\subsubsection{Thermal States}

The statistical concept of equilibrium states or Gibbs states is naturally
taken to the quantum world.  We define the thermal state of a system
described by a Hamiltonian $ \hat H $ as the exponential of the
Hamiltonian together with the factor $\beta = {1 \over k_B T}, $ where
$k_B $ is the Boltzmann constant and $ T $ is the temperature of the
system.
For a one mode harmonic oscillator the well known Hamiltonian is given
by $\hat H = \omega (\hat n + {1 \over 2})$, 
with $\hat n = \hat a^{\dag} \hat a $ being the number operator of the mode 
and  $\hbar=1$ throughout this thesis. The thermal state for this Hamiltonian is given by
\begin{eqnarray*}
\rho_{t}\index{$\rho_{t}$, thermal state of a harmonic oscillator } &=& {e^{- \beta
\hat H} \over tr[e^{- \beta \hat H}] }\\
&=& \sum_{n,m} | n \rangle \langle n | {e^{- \beta  \omega (\hat 
n + {1 \over 2}) } 
\over tr[e^{- \beta  \omega (\hat n + {1 \over 2})}] }| m 
\rangle \langle m |\\
&=& {1 \over tr[e^{- \beta  \omega (\hat n + {1 \over 2})}] } 
\sum_{n} e^{- \beta  \omega (n + {1 \over 2}) }  | n \rangle \langle n |, 
\end{eqnarray*}
with $tr[e^{- \beta  \omega (\hat n + {1 \over 2})}]  
= {1 \over 2 \sinh {\beta  \omega \over 2 }}$.

The characteristic function then is:
\begin{eqnarray*}
\chi_{t} (\xi) &=& 2 \,  \sinh {\beta  \omega \over 2 } \,\,\,  tr \left[\sum_{n}
| n \rangle \langle n | e^{- \beta  \omega (n + {1 \over 2}) }
\hat W_{\xi} \right]\\
\chi _{t}(\xi) &=& 2 \, \sinh {\beta  \omega \over 2 }\,\,\,   \sum_{n}
e^{- \beta  \omega (n + {1 \over 2}) } \langle n | \hat W_{\xi} |
n \rangle.
\end{eqnarray*}
The elements of the Weyl operators in the number basis with $\alpha =
{\xi_1 + i \xi_2 \over \sqrt{2} }$ are \setlength{\mathindent}{0cm}
\begin{eqnarray}\label{19}
\langle n | \hat W_{\xi} | m \rangle &=& \langle n | \hat D(- \alpha) |
m \rangle = \langle 0 | {\hat a^{n} \over \sqrt{n!}} \hat D(-\alpha) {\hat
a^{\dag \, m} \over \sqrt{m!}} |0 \rangle \nonumber \\
&=& 
{\begin{cases}
e^{- {| \alpha |^2 \over 2}} (- \alpha )^{n-m} \sqrt{m! n!} \sum_{l = 
0}^m  
{(- | \alpha |^2)^l  \over  l! (m - l)! (l + n - m)!}  
& \, m \le n \nonumber \\
e^{- {| \alpha |^2 \over 2}} (\alpha^* )^{m-n} \sqrt{m! n!} \sum_{l 
=0}^n  
{(- | \alpha |^2)^l  \over  l! (n - l)! (l + m - n)!} 
& \, m \ge n 
\end{cases}}\nonumber\\
&=& 
{\begin{cases}
e^{- {| \alpha |^2 \over 2}} (- \alpha )^{n-m} \sqrt{{m! \over n!}}
L_m^{n-m} (|\alpha|^2) & \, m \le n \\
e^{- {| \alpha |^2 \over 2}} (\alpha^* )^{m-n} \sqrt{{n! \over  m!}}
L_n^{m-n}( |\alpha |^2) & \, m \ge n,\nonumber\\
\end{cases}}
\end{eqnarray}\setlength{\mathindent}{2cm}

where $L_n^{k} (x)$ are the Laguerre polynomials;
so we get for $ n= m$ :
\begin{eqnarray*} 
\langle n | \hat W_{\xi} | n\rangle &=& \langle n | \hat D(- \alpha) | 
n \rangle = e^{- {| \alpha |^2 \over 2}} L_n^{0} (|\alpha|^2), 
\end{eqnarray*}
and $ \chi_{t} $ becomes
\begin{eqnarray*}
    \chi_{t} (\xi) &=& 2 \sinh {\beta  \omega \over 2 } e^{- {|
    \alpha |^2 \over 2}} \sum_{n} e^{- \beta  \omega (n + {1 \over
    2}) } L_n^{0} (|\alpha|^2). 
\end{eqnarray*}
With sum formula (8.975.1) in Ref. \cite{GR}  we finally get
\begin{eqnarray*}
   \chi_{t}(\xi) &=& 2 \sinh {\beta  \omega \over 2 } e^{- {| \alpha |^2
   \over 2}} e^{- {\beta  \omega \over 2} } { 1 \over 1- e^{- \beta
    \omega } } {e^{|\alpha|^2 e^{- \beta  \omega } \over e^{-
   \beta  \omega } -1}}\\
&=& e^{- {| \alpha |^2 \over 2}} 
e^{- |\alpha|^2   \over e^{ \beta  \omega } -1 }
=e^{-{\xi_1^2 + \xi_2^2  \over  4} 
{  e^{ \beta  \omega }  + 1 \over  e^{ \beta  \omega } -1 
}}\\
&=&e^{-{\xi_1^2 + \xi_2^2  \over  4 \tanh { \beta  \omega  \over 2}}}.
\end{eqnarray*} 
Hence the thermal state is a Gaussian state with vanishing
displacement and covariance matrix
\begin{eqnarray}
    \Gamma_{t} = { 1 \over \tanh { \beta  \omega \over 2}}
    \mathbbm{1} = \gamma_{t},
\end{eqnarray}
that is, a diagonal $2 \times 2$ matrix with identical  entries. 
We will meet  the thermal states again when discussing normal forms
of CMs in the next chapter. In the limit $ \beta = { 1 \over k_{B}T} \to \infty$ we get $\tanh { 
\beta  \omega  \over 2} \to 1$
and hence
\begin{eqnarray*}
\chi_{(T=0)} (\xi) =
e^{-{\xi_1^2 + \xi_2^2  \over  4}},
\end{eqnarray*} 
that is, the thermal state becomes the vacuum  in the limit of zero temperature.

\section{Normal Forms of Covariance Matrices}\label{16}

Covariance matrices are real symmetric matrices, hence it 
is always possible to diagonalise a CM with orthogonal matrices to the 
standard normal form which is the diagonal matrix of the eigenvalues of the CM. 
But other, symplectic, normal forms exist which are more convenient for covariance 
matrices because the resulting quantities are invariant under 
subgroups of the symplectic transformations.
We introduce two very important normal forms which are used 
extensively in the following chapters.

\subsection{Williamson Normal Form}

Surprisingly it is  possible to diagonalise real positive symmetric 
and even-dimensional matrices with  symplectic matrices. 
The resulting diagonal matrix is called \emph{Williamson normal form} 
and the proof of existence and uniqueness was given by J.Williamson 
in  1936, see Ref. \cite{JW}.

\begin{theorem}[Williamson normal form]\hspace{6cm}
    
    For every real strictly positive symmetric $2N \times 2N$ matrix $A$ there
    exists a matrix $S \in Sp(2N, \mathbbm R) $, so that 
    \begin{equation} 
        A^{WNF}\index{$A^{WNF} $, Williamson normal form of a positive, even
        dimensional matrix $A$ } =S A S^T = \left(
        \begin{array}{ccccc}                     
	a_1&0&&0&0\\
	0& a_1&&0&0\\
	&&\ddots&&\\
	0&0&&a_N&0\\
	0&0&&0&a_N                  
        \end{array}\right)
    \end{equation}
    with the \emph{ symplectic eigenvalues}\index{symplectic eigenvalues }
      $a_{i} > 0$ for all $i=1, \ldots, N$.  The symplectic
    matrix $S$ and the diagonal entries $a_{i}$ are unique up to
    permutations of the diagonal entries in $A^{WNF}$.
\end{theorem}
\bf Remark: \rm If $A$ had a vanishing eigenvalue it  would  not be possible to
bring it to the above form.  Instead, Williamson-diagonalise the regular, 
positive principal submatrices  of the  matrix $A$. Now transform the 
singular submatrix $A_0$ of $A$ to  its orthogonal diagonal form 
$D_{0} = O^{T} A_0 O = \left( 
\begin{array}{cc} 
  0&0 \\ 0 & a_{0} 
\end{array} \right)$ with $O \in SO(2,\mathbbm{R})
\subset Sp(2,\mathbbm{R})$.  The symplectic eigenvalue of $A_{0}$ is
counted as zero.\\

Note, that the determinant of $A$ is easily calculated from its
symplectic eigenvalues: $ \det A = \prod_{i=1}^{N} a_{i}^{2}$ and that
the symplectic eigenvalues of a matrix are invariant under symplectic
transformations.

Especially covariance matrices can be Williamson-diagonalised and as
we will see in the following discussion of the Williamson normal form
the transformed covariance matrix $\Gamma^{WNF} = \bigoplus_{i=1}^{N}
s_{i }\mathbbm{1}_{2 \times 2}$ describes the state of N uncoupled harmonic oscillators in a
heat bath with temperatures $T_i =T_{i}(s_i)$.

\subsubsection{Symplectic Eigenvalues }

In general it is difficult to find the symplectic transformation 
which brings a given matrix to Williamson normal form,  but the
eigenvalues are easily calculated with the following lemma
\begin{lemma}
    The symplectic eigenvalues of a real symmetric positive $2N \times 2N$
    matrix $A$ are the positive eigenvalues of the matrix $ i \sigma A $.
\end{lemma}
\begin{proof}
With $A = S^{-1 } A^{WNF} S^{-T}$
\begin{eqnarray*}
    \mbox{spec}(i \sigma A) 
    &=& \mbox{spec}(  i \sigma S^{-1 } A^{WNF} S^{-T} )
    = \mbox{spec}(  i S^{T} \sigma A^{WNF} S^{-T} )\\
    &=& \mbox{spec}( i S^{-T} S^{T} \sigma A^{WNF} )
    = \mbox{spec}( i \sigma  A^{WNF}) \\
    &=& \mbox{spec}\left( i \bigoplus_{j=1}^{N} \left[ a_{j} 
    \left( 
    \begin{array}{cc} 
        0&1\\ 
        -1&0 
    \end{array}  
    \right) \right] \right)
\end{eqnarray*}
were we used the properties of symplectic matrices and the cyclicity
of the spectrum, see Lemma \ref{8}.  The eigenvalues $\lambda_{i}$ of
$ i \sigma A^{WNF} $ are calculated via
\begin{eqnarray*}
    0 &\gleich& \det [i \sigma A^{WNF} - \lambda_{i} \mathbbm{1} ] 
    = \left|  \bigoplus_{j=1}^{N} \left[
    \begin{array}{cc}
       - \lambda_i & i a_{j}  \\
       - i a_{j} & - \lambda_{i}
    \end{array} \right] \right| \\
       &=& \prod^N_{j =1} \left |
    \begin{array}{cc}
        -\lambda_{i}&  i a_j \\
        - i a_j & - \lambda_{i}
    \end{array} \right| 
    = \prod^N_{j =1}(\lambda_{i}^{2} - a_{j}^{2}),
\end{eqnarray*}
we find the spectrum of $ i \sigma A $ to be $\mbox{spec}(i
\sigma A) = \{ a_{i}, -a_{i}\}_{i = 1}^{N}$ completing  the proof of the
lemma.
\end{proof}\vspace{1ex}

\begin{exa}
    Given a covariance matrix 
    $\Gamma = \left( 
    \begin{array}{cc}
        3&1 \\1 & 1
    \end{array} \right)
   $ then the symplectic eigenvalues of $\Gamma$ are the positive
   eigenvalues of $i \sigma \Gamma = \left(
    \begin{array}{cc}
        i&i \\ -3 i & -i
    \end{array} \right).  $ The eigenvalues  are $\lambda_{1,2} = \pm
    \sqrt{2}$, so the symplectic eigenvalue of $\Gamma$ is $\Gamma_{1} =
    \sqrt{2}$.  Even faster is $\det \Gamma = 2 = \Gamma_{1}^{2}$, thus
    $\Gamma_{1}= \sqrt{2}$.
\end{exa}

\subsubsection{Thermal States}
Question:
Given a covariance matrix in Williamson normal form and 
displacement $\vec D$,  what is the corresponding Gaussian state $\rho$ ?  
\\

We first see how block diagonal covariance matrices $\Gamma = 
\bigoplus_{i=1}^{N} \Gamma_{i}$ are  related to product states.
\begin{lemma}[Block diagonal CMs]\label{productCMs}
\hspace{4cm}
    
    A Gaussian state $\rho$ with  a covariance matrix in block diagonal 
    form is a product state, and conversely if $\rho$ is a  state of
    product form then its covariance matrix is of block diagonal form.
\end{lemma}
\begin{proof} We use the decomposition of $\rho$ in the Weyl system 
according to Eq.   \eqref{9} and insert the characteristic function
with the given block diagonal matrix $\Gamma = \bigoplus_{i=1}^{N}
\Gamma_{i}$ and displacement $\vec D$. For $\xi \in \mathbbm{R}^{2N}$ and 
$\xi_i \in \mathbbm{R}^2, \quad \forall i = 1,...N$, $\rho$ can be decomposed  
    \begin{eqnarray*}
	\rho &=& {1 \over (2 \pi )^{N}}\, \int \, \dif^{2N}
	\xi \; e^{- {\xi^{T} \Gamma \xi \over 4 } + i \vec D^{T} \xi }\, \hat
	W_{\xi}\\
	&=& {1 \over (2 \pi )} \, \int \, \prod_{i=1}^{N} (\dif^{2} \xi_{i}) \;  
	e^{- {\sum_{i=1}^{N}\xi_{i}^{T} \Gamma_{i} \xi_{i} \over 4 } } \, e^{ i
	\sum_{i=1}^{N}  \vec D_{i}^{T} \xi_{i} } \, 
	\bigotimes_{i=1}^{N} \hat W_{\xi_{i}} \\
	&=&\bigotimes_{i=1}^{N} \left[
	{1 \over (2 \pi )} \, \int \, \dif^{2} \xi_i \;
	e^{-  { \xi_i^{T} \Gamma_{i} \xi_i \over 4 } + i \vec D_{i}^{T} \xi_i }\, 
	\hat W_{\xi_{i}} \right] \\
	&=& \bigotimes_{i=1}^{N} \rho_{i},
    \end{eqnarray*}
    where the $\rho_{i} $ are one-mode Gaussian states with covariance
    matrices $\Gamma_{i} $ and displacements $\vec D_{i} $.
\end{proof}\vspace{2ex}

We learned in Section \ref{csts} that the one-mode thermal states of
the harmonic oscillator have the covariance matrix $\Gamma_{t} =
{1\over \tanh {\beta \omega \over 2 } } \mathbbm{1}$ and vanishing
displacement.  Since there is a one to one correspondence between the
pairs $\Gamma$ and $\vec D$ and the Gaussian states $\rho$, the
displaced thermal states are the only ones having such a diagonal
matrix.
The Gaussian state belonging to a CM in Williamson normal form $\Gamma
= \bigoplus_{i=1}^{N} s_{i} \mathbbm{1}$ with symplectic eigenvalues
$s_{i}$ and displacement $\vec D$ is a product state of displaced
thermal states with 
independent heat baths of temperature $T_{i}$.
\begin{equation*}
    \rho = \bigotimes_{i=1}^{N} \rho_{i} = \bigotimes_{i=1}^{N} 
    \hat W_{\vec D} \, \rho_{t}(T_{i}) \, \hat W^{\dag}_{\vec D}\,  .
\end{equation*}
The symplectic eigenvalues of the CM are then connected to the
temperature of the heat bath namely
\begin{equation*}
    s_{i}= {1 \over \tanh { \omega \over 2 k_{B} T_{i}}}.
\end{equation*}
    
\subsection{Simon Normal Form}

We introduce a normal form, as  proposed  in Ref. \cite{SCZ}, 
acting only locally on the first and second
mode respectively and therefore preserve the entanglement properties of a
given state $\rho$.
 \begin{theorem}[Simon normal form] \hspace{7cm}
    
     For every two-mode covariance matrix $\Gamma$ there exist
     matrices $S_1 \in Sp(2, \mathbbm R)$ and $S_2 \in Sp(2,\mathbbm R)$
     so that:
     \begin{equation*}
       \Gamma^{SNF}\index{$\Gamma^{SNF}$, Simon normal form of a
	 two-mode covariance matrix $\Gamma$ }	 = 
       (S_{1} \oplus S_{2})  \, \Gamma \, (S_{1}^{T} \oplus  
	 S_{2}^{T}) = \left(
	 \begin{array}{cccc}
	     a&0&c&0\\
	     0&a&0&d\\
	     c&0&b&0\\
	     0&d&0&b
	 \end{array}
	 \right).
     \end{equation*}
     This normal form is unique up to the little subtlety that only the
     relative sign of $c$ and $d $ is determined.
 \end{theorem}
 The transformations $S = S_{1} \oplus S_{2}$ are local transformations
 acting only on the first and second mode respectively, in particular they  
 do not change the entanglement properties of the covariance matrix they are acting on.

\subsubsection{Simon Invariants}

The determinants of the submatrices of $\Gamma$ and $\Gamma$ 
itself are preserved by the diagonalising symplectic transformation and appear as the 
\emph{ Simon invariants $a, b , c, d$  }\index{Simon invariants} with the
relations:
\setlength{\mathindent}{1cm}
\begin{eqnarray}\label{simonnumb}
\begin{array}{rclrcl}
\Gamma &=&\left( \begin{array}{cc} A & C \\ C^T & B \end{array} 
\right)
&a &=& \sqrt{\det A}, \\
&&& b &=& \sqrt{ \det B}, \\
\det \Gamma &=& a^2 b^2 + c^2 d^2 - ab (c^2 + d^2) \quad &c d &=& \; \;\,  \det C.
\end{array}
\end{eqnarray}

We hence reduce our ten parameter covariance matrix to four real
numbers still reflecting the entanglement properties of the original CM.
\setlength{\mathindent}{2cm} 
\section{Properties of Characteristic Numbers}{\label{PoCN}}

All derived properties of the covariance matrix $\gamma$ translate identically
to the matrix $\Gamma = \sigma \,  \gamma \, \sigma^{T}$.  Namely $\Gamma$ is
a real, symmetric, positive matrix fulfilling the uncertainty relation
and has the same determinant, eigenvalues and symplectic eigenvalues as
$\gamma$.

\subsubsection{Simon Invariants and Symplectic Eigenvalues}

In the last section we saw that there exist normal forms 
respecting the symplectic character of the canonical 
transformations on covariance matrices.  These forms were related to
invariant quantities whose properties we will discuss in this section. 
We will also see that properties of the states like squeezing and
entanglement are strongly connected to these numbers.  We denote the
eigenvalues of a given $N-$mode CM $\Gamma$ by $\lambda_i, $ for $i= 1,
..., 2N $, the symplectic eigenvalues by $\Gamma_j$, for $j=1, ...,N$
and the Simon invariants (only for two-mode systems) by $a, b, c, d$. 
\\

Assume we have a covariance matrix $\Gamma$ in Simon normal form and we want to 
know the eigenvalues and the symplectic eigenvalues of $\Gamma^{SNF}$. 
We find that the eigenvalues of $\Gamma^{SNF}$, expressed with the
Simon numbers, are given by
\begin{equation}\label{99}
  \begin{split}
    \lambda^{SNF}_{1, 2} &= \frac{a + b}{2}\pm \frac{1}{2}\sqrt{(a-b)^2 + 4 c^2},\\
    \lambda^{SNF}_{3, 4} &= \frac{a + b}{2}\pm \frac{1}{2}\sqrt{(a-b)^2 + 4 d^2}.
  \end{split}
\end{equation}

\bf{Remark}\rm: These numbers in general are not the eigenvalues of $\Gamma$
since in general the symplectic transformations $S_{1}$ and $S_{2}$
are active transformations (see Section \ref{12}).\\

Furthermore, the symplectic eigenvalues of a two mode system with
covariance matrix $\Gamma^{SNF}$ can be expressed by the Simon
invariants via
\begin{eqnarray}\label{sympsim}
    \Gamma_{1, 2} = \frac{1}{\sqrt 2}
\sqrt{a^2 + b^2 + 2 c d \pm \sqrt{(a^2 + b^2 + 2 c d )^2 - 4 \det
\gamma}} .
\end{eqnarray}
and are the same as the ones of $\Gamma$ itself.

\paragraph{Positivity }

From the positivity of $\Gamma$ it follows that
\begin{eqnarray}
\lambda_i&\ge& 0, \quad i= 1, \ldots, 2N, \nonumber\\
\Gamma_i& \ge & 0, \quad j= 1, \ldots, N,\\
a, b \ge 1 \, &\mbox{and }&\, c^2 \le a b , d^2 \le a b. \nonumber
\end{eqnarray}

\paragraph{Uncertainty Relation }

The Heisenberg uncertainty principle formulated in Section \ref{13}
is also true for the capital matrices  $\Gamma$
(see Eq. \eqref{14} and Lemma \ref{positiveGamma}).  Since the matrix
$\Gamma + i \sigma $ is positive it can be brought to Williamson
normal form by symplectic transformations $S$ without changing the
symplectic eigenvalues.  We have:
\begin{equation}\label{uncertainty}
    S \, [\Gamma + i\sigma ]\, S^{T} = S \, \Gamma \, S^{T} + i \sigma =
    \Gamma^{WNF} + i \sigma = \bigoplus_{j=1}^{N}\left (
    \begin{array}{cc}
	\Gamma_{j} & i \\
	-i& \Gamma_{j}
    \end{array}    \right ).
\end{equation}
To get a positive matrix all eigenvalues $\mu_i$, $i =1, ..., 2N$, 
of the above matrix have  to be positive. The equations determining 
these eigenvalues are 
\begin{eqnarray*}
\prod_{j=1}^{N} \,\,\, [(\Gamma_{j} - \mu_i)^{2 } -1 ] = 0, 
\quad \mbox{for } i=1,..., 2N,
\end{eqnarray*}
and hence 
\begin{equation*}
 \Gamma_{j} \pm 1 =   \mu_i  \ge 0, \quad \forall i=1, ..., N, \, \forall j=1, ..., 2N 
\end{equation*} 
 Thus the restrictions on the symplectic eigenvalues of $\Gamma$ are 
\begin{equation}\label{sym>1}
 \Gamma_j \ge 1, \quad \mbox{ for all } j = 1, \ldots, N. 
\end{equation}
Note, that the determinant of $\Gamma$ is therefore always greater than or 
equal to one.
For the Simon invariants we find the following, only necessary, criterion
\begin{equation}
\det \Gamma \ge \det A + \det B + 2\det C +1, 
\end{equation}
where $A, B, C$ are the submatrices of $\Gamma$ connected to the Simon
invariants via Eq. \eqref{simonnumb}
  
\paragraph{Pure States}
\begin{defi}
A state $\rho$ is called \emph{ pure } iff it can be written as a
projector  $ \rho = | \psi\rangle \langle \psi |$ whereas a \emph{ mixed
state } is a convex combinations $ \rho = \sum_{i} p_i | \psi_i
\rangle \langle \psi_i |$ of the states $| \psi_i\rangle \langle
\psi_i| $ with probabilities $p_i$ \, ($p_i \ge 0, \, \, \sum_i p_i = 1$) and
at least two $i$ so that $ p_i > 0$.
\end{defi}
It is well known that a state $\rho$ is pure \emph{ iff } $tr [\rho^{2}] =
 1$, but how can one express purity in covariance matrix language?
\begin{lemma}
    A Gaussian state $\rho$ with CM $\Gamma$ is pure if and only if
    \begin{eqnarray*}
        \det \Gamma = 1 \iff \forall i : \Gamma_i = 1 \iff ( i \sigma
        \Gamma)^2 = \mathbbm{1},
    \end{eqnarray*}
    where the $\Gamma_{i}$ are the symplectic eigenvalues of $\Gamma$.
    In the two-mode case the conditions on the Simon invariants are
    \begin{equation*}
        a = b= \cosh r  \, \mbox{ and } \, c = -d= \sinh r,
    \end{equation*}
    where $r$ is a \emph{squeezing parameter}  of the CM.
\end{lemma}

\begin{proof} We expand the state in the Weyl operator basis and
calculate the trace of its square
\begin{eqnarray*}
    \rho&=& {1 \over (2 \pi)^{N}} \int \dif^{2N}\xi \; \chi_{\rho}(-\xi)
    \, \hat W_{\xi} \\
    tr[ \rho^{2} ] &=& {1 \over (2 \pi)^{2N}} \int \dif^{2N}\xi  \int \dif^{2N}
    \eta \; \chi_{\rho}(-\xi) \, \chi_{\rho}(- \eta) \, tr [\hat W_{\xi} \, \hat
    W_{\eta}] \\
    &=& {1 \over (2 \pi)^{N}} \int \dif^{2N}\xi \int \dif^{2N}
    \eta \; \chi_{\rho}(-\xi) \, \chi_{\rho}(- \eta) \, \delta^{2N}(\xi + \eta) \\
    &=& {1 \over (2 \pi)^{N}} \int \dif^{2N}\xi \; \chi_{\rho}(-\xi) \,
    \chi_{\rho}( \xi).
\end{eqnarray*}
With $\rho$ being  a Gaussian state with covariance matrix $\Gamma $ and
arbitrary displacement we get
\begin{eqnarray*}
    tr[ \rho^{2} ] 
    &=& {1 \over (2 \pi)^{N}} \int \dif^{2N}\xi \; e^{-{\xi^{T} \Gamma \xi  \over 2} } 
    = {1 \over \sqrt{\det \Gamma }}.
\end{eqnarray*}
Hence $\rho$ is pure iff $\det \Gamma = 1$.  It follows directly that
the symplectic eigenvalues $\Gamma_{i}$, all necessarily greater than
or equal to one, just take their lower bound.  Furthermore half of the
eigenvalues of $ i \sigma \Gamma$ are one while the others are minus
one.  Hence $ (i \sigma \Gamma)^{2} $ has all eigenvalues equal to one
and is the identity matrix.  For the Simon invariants we easily get
the above relations by inserting the $\Gamma_{i}$ in
Eq. \eqref{sympsim}.
\end{proof}

\paragraph{Squeezed States}\label{GSPSQ}

If a (general) CV state has a smaller covariance in one direction of the phase
space than in another it is said to be squeezed.  For a basis
independent formulation one has to consider all those symplectic
transformations which do not change the degree of squeezing.  We
encountered those transformations when discussing normal forms of
symplectic matrices in Eq.  \eqref{passive}, where we learned that they
are in the intersection of the symplectic and the special orthogonal
group.
\begin{defi}
    An $N-$mode state $\rho$ with covariance matrix $\Gamma$ is 
    \emph{ squeezed } if and only if
    \begin{eqnarray*} 
        \exists \, S \in K(N), \; \exists k : \, (S \, \Gamma \, S^T)_{kk} < 1.    
    \end{eqnarray*} 
    Here $K(N) = Sp(2N, \mathbbm{R}) \cap SO(2N,\mathbbm{R})$ are the
passive symplectic transformations.
\end{defi}
We picked only those states which have a smaller covariance in at
least one phase space direction than the vacuum state.
Unfortunately, the  criterion is not practicable,
 but it can be formulated in terms of the eigenvalues of the CM as the
 following lemma shows:
\begin{lemma}  
    An $N-$mode state $\rho$ is squeezed if and only if the
    smallest eigenvalue $\lambda_{min}$ of the state's covariance matrix
    is smaller than one.
\end{lemma}
\begin{proof}
    First we show that the diagonal entries of a symmetric matrix $A \in
    \mbox{Mat}(N \times N, \mathbbm{R}) $ are always bigger or equal than
    the smallest eigenvalue $a_{min}$ of the matrix.  Let $\{ a_{k}
    \}_{k=1}^{N} $ be the orthonormal system of eigenvectors of $A $ and
    $\{ e_{j} \}_{j=1}^{N}$ an arbitrary orthonormal basis in
    $\mathbbm{R}^{N}$, then the diagonal entries of $A$ in that basis are
    $ A_{ii} = \langle e_{i} | A | e_{i} \rangle = \sum_{k,l =1}^{N} \,
    \langle e_{i}| a_{k} \rangle \langle a_{k} | A | a_{l} \rangle \langle
    a_{l}| e_{i} \rangle = \sum_{l =1}^{N} \, a_{l} |\langle a_{l}| e_{i}
    \rangle|^{2} \ge a_{min} \, \sum_{l =1}^{N} \, |\langle a_{l}| e_{i}
    \rangle|^{2} =  a_{min}, $ as stated.
        The passive transformations in $K(N)$ do not change the eigenvalues of the CM and
    since all diagonal entries are bigger or equal to the smallest
    eigenvalue we have $ \forall \, S \in K(N), \; \forall k : \,(S \,
    \gamma \, S^T)_{kk} \ge \lambda_{min} $ with equality when $S$
    diagonalises $\Gamma$ to its orthogonal normal form and the $k-$th
    entry is $\lambda_{min}$.  Hence the state can only be squeezed if and
    only if the smallest eigenvalue of the CM is smaller than one.
\end{proof}

\paragraph{Entangled States} 

In Section \ref{entanglement} we defined what entanglement is and how one can
decide whether a given state $\rho$ is entangled or not.  We introduced
the partial transpose of a bipartite system and stated the
PPT criterion for states $\rho \in S(\mathcal{H})$. 
For Gaussian states it is possible to formulate this criterion on
covariance matrix level.  We first see what effect the partial
transposition of a state has on the states covariance matrix.
\begin{lemma}\label{timerevers}
    The partially transposed covariance matrix (on part $A$) of a bipartite
    state $\rho$ of $N =N_{A} + N_{B}$ modes with covariance matrix
    $\Gamma$ is defined by the state $\rho^{T_{A}}$ with the covariance
    matrix $\Gamma^{T_{A}}$\index{$\Gamma^{T_{A}}$, partially transposed
    covariance matrix }, which is given by
    \begin{equation}
	\Gamma^{T_{A}} = (- M_{A} \oplus \mathbbm{1}_{B})  \,
	\Gamma \, (- M_{A} \oplus \mathbbm{1}_{B}),
    \end{equation}
    with $M_A  = M_A^{T} = \bigoplus_{i=1}^{N_A} \left(
    \begin{array}{cc}
	1&0\\ 
	0&-1 
    \end{array} \right)$. 
\end{lemma}
\begin{proof}
    Partial transposition is a basis dependent non-unitary operation  and since it 
    is not a completely positive map, it can  not be  realised perfectly by physical means.
    Mathematically the 
    transformation can be done, since the  spectrum of the partial transpose is not basis dependent,
    but we have to choose a basis for the calculation. 
    We denote the (multimode) number basis vectors by $|n_{A} \rangle$ and 
    $|n_{B} \rangle $ for $\mathcal{H}_{A}$ and $\mathcal{H}_{B}$ 
    respectively and calculate the characteristic function of the 
    passively transformed state $\rho \mapsto\rho^{T_{A}}$:
    \setlength{\mathindent}{0cm}
    \begin{eqnarray*} 
      \chi_{\rho} (\xi) &=& 
      tr [\rho \, \hat W_{\hat R, \xi}]
      = \sum_{n_A, n_B} \langle n_A n_B | \rho \hat W_{\hat R,\xi} | n_A n_B 
      \rangle\\
      &=& 
      \sum_{n_A, n_B, m_A, m_B} \langle n_A n_B | \rho |m_A m_B \rangle 
      \langle m_A |\hat W_{\hat R_{A},\xi_A} | n_A \rangle 
      \langle m_B |\hat W_{\hat R_{B},\xi_B} | n_B \rangle\\
      &\gleich& 
      \sum_{m_A, n_B, n_A, m_B} \langle n_A n_B | \, \rho \, |m_A m_B 
      \rangle 
      \langle n_A |\hat W_{\hat R'_{A},\xi_A} | m_A \rangle
      \langle m_B |\hat W_{\hat R'_{B},\xi_B} | n_B \rangle\\
      &=& tr [\rho^{T_A} \, \hat W_{\hat R', \xi}].
    \end{eqnarray*} 
    To determine the passive transformation on the basis operators $\hat R$,
    bringing $\rho$ to $\rho^{T_A}$, the above  expressions shall be equal.  
    We assume that on system $B$ no changes take place, $\hat R_{B} = \hat R'_{B}$, 
    while for system $A$ a linear homogenous transformation     
    \setlength{\mathindent}{2cm}
    \begin{equation*}
      M_{A} = \bigoplus_{i =1}^{N_A} M \quad \mbox{ fulfilling } 
\,  \hat R'_{A} = M_{A} \, \hat R_{A}
    \end{equation*}
    has to be determined.  
        For every single mode $i \in A$ we have by the
    definition of the Weyl operators: \setlength{\mathindent}{2cm}
    \begin{eqnarray}\label{express}
	\langle m_i |\hat W_{\hat R_{i},\xi_i} | n_i \rangle
	&= &\langle n_i |\hat W_{ M\hat R_{i},\xi_i} | m_i \rangle 
	=\langle n_i |e^{i \xi_{i}^{T} \sigma M \hat R_{i}} | m_i \rangle.
    \end{eqnarray}
    The $2 \times 2$-matrix $M$ is\emph{ not} a symplectic matrix because if it
    was one, partial transposition could in principle be   realised perfectly. 
    Additionally from symmetry considerations $M$ has to
    be selfinverse so that the determinant of $M$ has to be minus one. 
    Therefore it fulfills $M \sigma M^{T} = - \sigma$ and expression
    \eqref{express} becomes:
    \begin{eqnarray*}
	\langle n_i |e^{i \xi_{i}^{T} \sigma M \hat R_{i}}| m_i \rangle
	&=& \langle n_i |e^{- i ( M^{-1}\xi_{i})^{T} \sigma \hat R_{i}}| m_i \rangle\\
	&=&\langle n_i |\hat W_{ \hat R_{i}, -M^{-1} \xi_i} | m_i \rangle\\
	&=&\langle m_i |\hat W_{ \hat R_{i}, M^{-1} \xi_i} | n_i \rangle^{*}.
    \end{eqnarray*}\setlength{\mathindent}{0cm}
    
    With  the complex numbers $\alpha_{i} =  { \xi^{1}_{i} + i \xi^{2}_{i}\over 
    \sqrt{2}} $ and $\beta_{i} =  \sum_{j=1}^{2}{ M^{-1}_{1j}\xi^{j}_{i} 
    +i M^{-1}_{2j} \xi^{j}_{i}\over 
    \sqrt{2}} $ and with   Eq. \eqref{19} we get :  
    \begin{eqnarray*}
	\langle m_i |\hat W_{\hat R_{i},\xi_i} | n_i \rangle
	&=& \langle m_i |\hat D (- \alpha) | n_i \rangle \\
	&=& {
	\begin{cases}
	    e^{- {| \alpha_{i} |^2 \over 2}} 
	    ( -\alpha_{i} )^{m_{i}-n_{i}} \sqrt{{n_{i}! \over  m_{i}!}} 
	    L_{n_{i}}^{m_{i}-n_{i}} (|\alpha_{i}|^2) & \, n_{i} \le m_{i}, \\
	    e^{- {| \alpha_{i}|^2 \over 2}} 
	    (\alpha^{*}_{i})^{n_{i}-m_{i}} \sqrt{{m_{i}! \over  n_{i}!}}
	    L_{m_{i}}^{n_{i}-m_{i}}( |\alpha_{i}|^2) & \, n_{i} \ge m_{i},\\
	\end{cases}}\\ &\gleich&\\
	\langle m_i |\hat W_{ \hat R, M^{-1} \xi_i} | n_i \rangle^{*}	
	&=& {
	\begin{cases}
	    e^{- {| \beta_{i} |^2 \over 2}} 
	    ( -\beta^{*}_{i} )^{m_{i}-n_{i}} \sqrt{{n_{i}! \over  m_{i}!}} 
	    L_{n_{i}}^{m_{i}-n_{i}} (|\beta_{i}|^2) & \, n_{i} \le m_{i}, \\
	    e^{- {| \beta_{i}^2 |\over 2}} 
	    (\beta_{i})^{n_{i}-m_{i}} \sqrt{{m_{i}! \over  n_{i}!}}
	    L_{m_{i}}^{n_{i}-m_{i}}( |\beta_{i}|^2) & \, n_{i} \ge m_{i},
	\end{cases}}
    \end{eqnarray*} \setlength{\mathindent}{2cm}
    
    hence $\beta_{i} = \alpha^{*}_{i}$ and therefore $M^{-1} = M  = \left( 
    \begin{array}{cc}
	1&0\\
	0&-1
    \end{array}\right) =M^{T}$.  On the canonical operators, the
    displacement and the  covariance matrices, this transformation acts according
    to Eq. \eqref{1b} and Eq. \eqref{transgam} (even if the transformation
    is not symplectic but linear homogeneous):
\begin{eqnarray}\label{gTA}
    \hat R \mapsto \hat R' \; \; &=& \left(
    M_{A} \oplus \mathbbm{1}_{B} \right) \, \hat R, \nonumber\\
    \gamma \mapsto \gamma^{T_{A}} &=& (M_{A} \oplus \mathbbm{1}_{B})  \,
    \gamma \, (M_{A}\oplus \mathbbm{1}_{B}), \\
    \Gamma \mapsto \Gamma^{T_{A}} &=& ( - M_{A} \oplus \mathbbm{1}_{B})  \,
    \Gamma \, ( - M_{A} \oplus \mathbbm{1}_{B}),\nonumber
 \end{eqnarray}
 since $ \sigma M_{A} = - M_{A} \sigma$.
\end{proof}\vspace{2ex}

Note, that partial transposition leads to a reversal of
all momenta in the system $A$ while the positions stay unchanged and
the system $B$ is untouched.
We finally got a recipe of calculating the \emph{ partially transposed
covariance matrix}. 

\bf Attention: \rm There is a little inconsistency 
with the definiton of the partially transposed covariance matrix
since one does not transpose one  part of the matrix, as one does 
when calculating the partial transpose of a state. Instead, it is 
the covariance matrix \emph{associated } to the partial transpose of the 
state and is calculated by  appling a non-symplectic transformation 
 $T = -  M_{A} \oplus \mathbbm{1}_{B}$.\\

We easily see that $\Gamma^{T_{A}}$ is still a real, symmetric, 
positive matrix but if the partial
transpose of $\rho$ fails to be positive its covariance matrix
$\Gamma^{T_{A}}$ by Eq.  \eqref{15} does not fulfil the uncertainty
relation anymore and vice versa.  Finally the PPT 
criterion on covariance matrices is formulated in the following
theorem.
\begin{theorem}[PPT criterion for bipartite Gaussian systems]
    \label{peres2}\hspace{2cm}

    Let $\rho$ be a state on a Hilbert space $\mathcal{H} = \mathcal{H}_A
    \otimes \mathcal{H}_B$ with a partially transposed covariance matrix
    $\Gamma^{T_{A}}$ \emph{ not} fulfilling the uncertainty relation, then
    the state is entangled with respect to the split $ A|B$.
\begin{equation*} 
\Gamma^{T_A} + i \sigma\not \ge 0 \Rightarrow \rho \mbox{ is entangled.}
\end{equation*}
In case $\rho$ is a Gaussian $1_A \times N_B$-mode state this criterion
is also sufficient \cite{W3}.
\end{theorem}

As we saw in Eq. \eqref{uncertainty} the uncertainty relation for a matrix is
equivalent to restricting the symplectic eigenvalues of the matrix on
values greater than or equal to one.
We summarise that any state with covariance matrix $\Gamma$ is
entangled if its partially transposed matrix $\Gamma^{T_{A}}$ has an
eigenvalue smaller than one.
  \begin{eqnarray}
      \Gamma^{T_{A}} + i \sigma \not \ge 0 \,  \iff  \, \exists j : \Gamma^{T_{A}}_j & < & 1,  
  \end{eqnarray}
with $\Gamma^{T_{A}}_j$ the symplectic  eigenvalues of $\Gamma^{T_A}$.
It follows immediately that at least one of the (usual) eigenvalues of
$\Gamma^{T_{A}}$ is smaller than one and since
$\mbox{spec}(\Gamma^{T_{A}} ) = \mbox{spec}(\Gamma)$ every entangled
state is necessarily squeezed.\vspace{2ex}

\begin{lemma}\label{25}
The logarithmic negativity, defined in Def. \ref{lognegrho}, of a
Gaussian state $\rho$ and CM $\Gamma$  can be calculated with the help of its
partially transposed covariance matrix $\Gamma^{T_{A}}$ having the
symplectic eigenvalues $\Gamma^{T_{A}}_{i}$:
\begin{equation}\label{logneg}
   	E_{\mathcal{N}}(\rho) = - \sum_{i} \; \min \left( \ln
   	[\Gamma^{T_{A}}_{i} ], 0 \right) 
	=: E_{\mathcal{N}}(\Gamma)
	=: E_{\mathcal{N}}(\gamma).
\end{equation}
\end{lemma}
\begin{proof}
    As we saw in Section \ref{16},  every Gaussian state can be brought to
    a product of thermal states with different temperatures by applying an
    appropriate displacement and symplectic transformation.  The state
     was then  written as 
     \begin{equation*}
       \rho = \bigotimes_{i} \left[
	 \sum_{n_{i}=0}^{\infty} p_{n_{i}} \, |n_{i } \rangle \langle n_{i}|
	 \right]
     \end{equation*} 
     where the probabilities are given by $p_{n_{i}} = 2 \sinh
     {\beta_{i}  \omega \over 2} \, e^{-\beta_{i}  \omega (n_{i}
       + {1 \over2} )}$ which are connected to the symplectic eigenvalues
     $\Gamma_{i}$ of the CM of $\rho$ via $\Gamma_{i} = {1 \over \tanh
       {\beta_{i}  \omega \over 2}}$ or after a little calculation $
     p_{n_{i}} = {2 \over \Gamma_{i} + 1}\left( {\Gamma_{i} - 1
       \over \Gamma_{i} +1} \right)^{n_{i}}$.  The construction also applies
     for the partial transpose of a state but since $\rho^{T_{A}}$ is not
     necessarily positive the prefactors in $p^{T_{A}}_{n_{i}}$ are not
     probabilities anymore, since they can be negative although they still
     sum up to one.  The relation 
     \begin{equation*}
       p^{T_{A}}_{n_{i}} = {2 \over \Gamma^{T_A}_{i} + 1}
\left( {\Gamma^{T_{A}}_{i} - 1 \over \Gamma^{T_{A}}_{i} +1}
       \right)^{n_{i}}
     \end{equation*}
     still holds, but it is not possible to assign a sensible temperature to
     the state $\rho^{T_{A}}$.   
     We calculate the logarithmic negativity of a Gaussian state $\rho$:
     \begin{eqnarray*}
       E_{\mathcal{N}}(\rho) = \ln ||\rho^{T_{A}}||_{1} = \ln
       \left\|\bigotimes_{i} \left( \sum_{n_{i}=0}^{\infty} p^{T_{A}}_{n_{i}}
       \, |n_{i } \rangle \langle n_{i}| \right) \right\|_{1},
     \end{eqnarray*}
     where we used that the diagonalising unitary transformation on  
     $\rho^{T_{A}}$ leaves the trace norm invariant.
\newpage
Applying the
definition of the trace norm gives: \setlength{\mathindent}{0cm}
     \begin{eqnarray*}
       E_{\mathcal{N}}(\rho) 
       & =& \ln \; tr\left [\sqrt{\bigotimes_{i} \left(
	   \sum_{n_{i}=0}^{\infty} (p^{T_{A}}_{n_{i}})^{2} \, |n_{i } \rangle
	   \langle n_{i}| \right)} \: \right] \\
       &=& \ln \; tr\left [\bigotimes_{i} \left( \sum_{n_{i}=0}^{\infty}
	|p^{T_{A}}_{n_{i}} | \, |n_{i } \rangle \langle n_{i}| \right)
	\right] 
       = \ln \; \left( \prod_{i} \left[ \sum_{n_{i}=0}^{\infty} \,
	 |p^{T_{A}}_{n_{i}} | \right] \right)
     \end{eqnarray*}
    and inserting the $p^{T_{A}}_{n_{i}}$ gives
    \begin{eqnarray*}
      E_{\mathcal{N}}(\rho) &=& \sum_{i} \;\left( \ln \left[
	\sum_{n_{i}=0}^{\infty} \, {2 \over {\Gamma^{T_{A}}_{i} + 1} } \left|
	    {\Gamma^{T_{A}}_{i} - 1 \over \Gamma^{T_{A}}_{i} +1} \right|^{n_{i}}
	    \right] \right)\\
      &=& \sum_{i} \;\left( \ln \left[ {2 \over \Gamma^{T_{A}}_{i} + 1 -
	  |\Gamma^{T_{A}}_{i} - 1|}  \right] \right)\\
      &=& - \sum_{i} \;
      \begin{cases}
	0 & \Gamma^{T_{A}}_{i} \ge  1 \\
	\ln [ {\Gamma^{T_{A}}_{i}} ]& \Gamma^{T_{A}}_{i} \le 1 \end{cases}
      = - \sum_{i} \; \min \left( \ln [\Gamma^{T_{A}}_{i} ], 0 \right).
    \end{eqnarray*}
    \setlength{\mathindent}{2cm}
    
The same logarithmic negativity results when using $\gamma^{T_{A}}$,
having the same symplectic eigenvalues as $\Gamma^{T_A}$.
\end{proof}

\section{States of Maximal Entropy}

\begin{defi}\label{lgrho}
    The logarithm of a state $\rho \in S(\mathcal{H})$ is
    defined via its diagonalised normal form $\mbox{\rm diag
    \it}(p_{1}, p_{2}, ... ) = U^{\dag} \, \rho \, U $ according to
    \vspace{-0.4cm}
    
    \begin{equation*}
        \ln (\rho) := U \, \mbox{\rm diag}(\ln p_{1}, \ln p{2} \ldots ) \,
        U^{\dag}
    \end{equation*}
    where $p_{1}, p_{2}, ... $ are the positive eigenvalues of $\rho$ and
    $\ln p_{i} $ the logarithms of the eigenvalues.
\end{defi}

\begin{lemma}\label{lgtensor} The logarithm of a tensor product of two states
$\rho_{A}$ and $\rho_{B}$ is given by \vspace{-0.6cm}
    
    \begin{equation*}        
        \ln (\rho_{A} \otimes \rho_{B} ) = \ln (\rho_{A}) \otimes \hat
        {\mathbbm{1}}_{B} + \hat {\mathbbm{1}}_{A} \otimes \ln(\rho_{B})
    \end{equation*}
\end{lemma}\setlength{\mathindent}{0cm}
\begin{proof}
    The tensor product of two states can be diagonalised locally with
    \begin{eqnarray*}
        && (U_{A} \otimes U_{B})\, (\rho_{A} \otimes \rho_{B})\, (U_{A}^{\dag}
        \otimes U_{B}^{\dag}) = \mbox{diag}(p_{1}^{A},  p_{2}^{A}, ...)
        \otimes \mbox{diag}(p_{1}^{B}, p_{2}^{B}, ...)\\
        && = \left(
        \begin{array}{cccccc}
	p^{A}_{1} p^{B}_{1}&&&&&\\
	& \negthinspace  \negthinspace  \negthinspace  \negthinspace  \negthinspace 
	p^{A}_{1} p^{B}_{2}&&&&\\
	&& \negthinspace  \negthinspace  \negthinspace  \negthinspace  \negthinspace 
	\ddots &&&\\
	&&&  \negthinspace  \negthinspace  \negthinspace  \negthinspace  \negthinspace 
	p^{A}_{2} p^{B}_{1}&&\\
	&&&&  \negthinspace  \negthinspace  \negthinspace  \negthinspace  \negthinspace 
	p^{A}_{2} p^{B}_{2}&\\
	&&&&&  \negthinspace  \negthinspace  \negthinspace  \negthinspace  \negthinspace 
	\ddots  
        \end{array}\right)
        =\bigoplus_{i}^{\dim \mathcal{H}_{A}} \left[p^{A}_{i}
        \mbox{diag}(p_{1}^{B}, p_{2}^{B}, ...) \right].
    \end{eqnarray*}
    where $p_{i}^{A}$ and $p_{j}^{B}$, are the of eigenvalues of
    $\rho_{A}$ and $\rho_{B}$ respectively.  The logarithm of the tensor
    product is by Def.  \ref{lgrho} given by
    \begin{eqnarray*}
        \ln (\rho_{A} \otimes \rho_{B})&=&  (U_{A} \otimes U_{B})\, \left( 
        \begin{array}{cccccc}
	\ln p^{A}_{1} + \ln p^{B}_{1}&&&&&\\
	& \negthinspace  \negthinspace  \negthinspace  \negthinspace  \negthinspace 
	\negthinspace  \negthinspace  \negthinspace  \negthinspace  \negthinspace 
	\negthinspace  \negthinspace  \negthinspace  \negthinspace  \negthinspace 
	\negthinspace  \negthinspace  \negthinspace  \negthinspace  \negthinspace
	\negthinspace  \negthinspace  \negthinspace  \negthinspace  \negthinspace
	\ln p^{A}_{1} + \ln p^{B}_{2}&&&&\\
	&&  \negthinspace  \negthinspace  \negthinspace  \negthinspace  \negthinspace
	\negthinspace  \negthinspace  \negthinspace  \negthinspace  \negthinspace 
	\negthinspace  \negthinspace  \negthinspace  \negthinspace  \negthinspace 
	\negthinspace  \negthinspace  \negthinspace  \negthinspace  \negthinspace
	\negthinspace  \negthinspace  \negthinspace  \negthinspace  \negthinspace
	\ddots &&&\\
	&&& \negthinspace  \negthinspace  \negthinspace  \negthinspace  \negthinspace
	\negthinspace  \negthinspace  \negthinspace  \negthinspace  \negthinspace 
	\negthinspace  \negthinspace  \negthinspace  \negthinspace  \negthinspace 
	\negthinspace  \negthinspace  \negthinspace  \negthinspace  \negthinspace
	\negthinspace  \negthinspace  \negthinspace  \negthinspace  \negthinspace
	\ln p^{A}_{2} + \ln p^{B}_{1}&&\\
	&&&&  \negthinspace  \negthinspace  \negthinspace  \negthinspace  \negthinspace
	\negthinspace  \negthinspace  \negthinspace  \negthinspace  \negthinspace 
	\negthinspace  \negthinspace  \negthinspace  \negthinspace  \negthinspace 
	\negthinspace  \negthinspace  \negthinspace  \negthinspace  \negthinspace
	\negthinspace  \negthinspace  \negthinspace  \negthinspace  \negthinspace
	\ln p^{A}_{2} + \ln p^{B}_{2}&\\
	&&&&&  \negthinspace  \negthinspace  \negthinspace  \negthinspace  \negthinspace
	\negthinspace  \negthinspace  \negthinspace  \negthinspace  \negthinspace
	\negthinspace  \negthinspace  \negthinspace  \negthinspace  \negthinspace 
	\negthinspace  \negthinspace  \negthinspace  \negthinspace  \negthinspace 
	\negthinspace  \negthinspace  \negthinspace  \negthinspace  \negthinspace 
	\ddots  
        \end{array}\right) \, (U_{A}^{\dag}
        \otimes U_{B}^{\dag}) \\
        &=& (U_{A} \otimes U_{B})\,\left( 
        \begin{array}{ccc}
	\ln p^{A}_{1} \hat {\mathbbm{1}}_{B}&&\\
	&	\negthinspace  \negthinspace  \negthinspace  \negthinspace  \negthinspace 
	\negthinspace  \negthinspace  \negthinspace  \negthinspace  \negthinspace 
	\ln p^{A}_{2} \hat {\mathbbm{1}}_{B}&\\
	&& 	\negthinspace  \negthinspace  \negthinspace  \negthinspace  \negthinspace 
	\negthinspace  \negthinspace  \negthinspace  \negthinspace  \negthinspace 
	\ddots 
        \end{array}\right) \, (U_{A}^{\dag}\otimes U_{B}^{\dag})\\
        & + & (U_{A} \otimes U_{B})\,\left( 
        \begin{array}{cccccc}
	 \ln p^{B}_{1}&&&&&\\
	&
	\negthinspace  \negthinspace  \negthinspace  \negthinspace  \negthinspace 
	\ln p^{B}_{2}&&&&\\
	&&  
	\negthinspace  \negthinspace  \negthinspace  \negthinspace  \negthinspace 
	\ddots &&&\\
	&&& 
	\negthinspace  \negthinspace  \negthinspace  \negthinspace  \negthinspace 
	\ln p^{B}_{1}&&\\
	&&&& 
	\negthinspace  \negthinspace  \negthinspace  \negthinspace  \negthinspace  
	\ln p^{B}_{2}&\\
	&&&&& 
	\negthinspace  \negthinspace  \negthinspace  \negthinspace  \negthinspace 
	\ddots 
        \end{array}\right)\, (U_{A}^{\dag}
        \otimes U_{B}^{\dag})\\
        &=& U_{A} \, \mbox{diag}(\ln p_{1}^{A}, \ln p_{2}^{A}, ... ) \,
        U_{A}^{\dag} \otimes \hat {\mathbbm{1}}_{B} \\
	&& \hfill + \hat {\mathbbm{1}}_{A} \otimes U_{B} \, \mbox{diag}(\ln
	p_{1}^{B}, \ln p_{M_{B}}^{B}, ... )\,U_{B}^{\dag}\\
        &=& \ln (\rho_{A}) \otimes \hat {\mathbbm{1}}_{B} + \hat
        {\mathbbm{1}}_{A} \otimes \ln (\rho_{B})
    \end{eqnarray*}
    \normalsize
    as stated in the lemma.
\end{proof}
\setlength{\mathindent}{2cm}

\begin{theorem}[States of Maximal Entropy]\label{17} \hspace{6cm}
   
    Of all states with fixed first and second moments, the Gaussian state
    maximise the von-Neumann entropy.
\end{theorem}

\begin{proof} The definition of the von Neumann entropy was given in
Def. \ref{vNe}. We again transform the Gaussian state $\rho$ to a
product of thermal states
\begin{equation*}
    \rho' = U^{\dag } \rho \, U = \bigotimes_{i=1}^{N}{e^{-
    \beta_{i} \hat H_{i}} \over tr [e^{- \beta_{i} \hat H_{i}}]},
\end{equation*}
 where $\hat H_{i} = \omega {\hat x^{2}_{i} + \hat p^{2}_{i}
    \over 2 }$.  The logarithm of $\rho'$ is with Lemma \ref{lgtensor}
    given by
    \begin{equation*}
	\ln \rho' =\sum_{i=1}^{N} \hat {\mathbbm{1}}_{i-1}\, \otimes \left[ -
	\beta_{i}  \omega {\hat x^{2}_{i} + \hat p^{2}_{i} \over 2 } -
	\ln \left( tr [e^{- \beta_{i} \hat H_{i}}] \right) \hat{ \mathbbm{1}}
	\right] \, \otimes \hat {\mathbbm{1}}_{N-i},
    \end{equation*}
    where all addends shall be understood as identity
    operators on all $N-1$ modes except the i-th mode where the above
    operators are inserted.
        For the Gaussian state $\rho$ and an arbitrary state $\delta$ with the
    same first and second moments $\vec D$ and $\Gamma$ we calculate the
    difference of their entropies
    \begin{eqnarray*} 
        S(\rho ) - S(\delta) &=& S(\delta|| \rho) + tr [( \delta - \rho ) \ln  \rho] \\
        &\ge&   0 + tr [ U \, ( \delta - \rho ) \, U^{\dag} \, \ln \rho'] \\
        &=& tr [ ( \delta' - \rho' )  \ln \rho']. 
    \end{eqnarray*}
    Here $\delta'$ and $\rho'$ are the transformed states with vanishing
    displacement and covariance matrix $\Gamma^{WNF}$ and $S( \rho'||
    \rho)$ the relative entropy of $\rho$ and $\rho'$ which is a positive
    quantity, see proof in Ref. \cite{NC}.   Finally $tr [ (
    \delta' - \rho' ) \ln \rho'] = 0$ since from the above discussion we
    see that $\ln \rho'$ is a polynomial of second degree in the canonical
    operators $\hat x_{i}$ and $\hat p_{i} \; (i = 1, \ldots , N) $, and
    hence picks out only the first and second moments of the difference,
    and because $\delta'$ and $\rho'$ have identical first and second
    moments, they vanish. This completes the proof of the theorem.
\end{proof}

\chapter{Gaussian Operations}

\section{General Gaussian Operations}
Quantum operations preserving the Gaussian character of all Gaussian
states are naturally called \emph{ Gaussian operations}, 
see Ref. \cite{103,103b,103c} and Ref. \cite{104,104b,104c} for Gaussian channels.  
As we want to stay in the set of Gaussian states we have to take care that the
operations we apply do not drive us out of that set. In the following  we
will briefly discuss the classes of Gaussian operations.

\paragraph{Gaussian Unitary Operations} The unitary evolutions $\hat U = e^{i \hat
H t }$ with the Hamiltonian $\hat H$ which preserve the Gaussian
character of all Gaussian states are those where $\hat H$ is a
polynomial of second degree in the canonical operators $\hat x_{i}$
and $\hat p_{i}$ for $i = 1, \ldots, N$.  One understands that by
recalling that a Gaussian state $\rho$ can be brought to an
exponential form were the exponent is a polynomial of second degree in
the basis operators (see the proof of Theorem \ref{17}).  Only a
Hamiltonian which is again such a polynomial can transform all
Gaussian states to Gaussian states.  We immediately see that these
Hamiltonians can only generate translations in phase space and
symplectic transformations $S \in Sp(2N, \mathbbm{R})$.  The most
relevant operations which we can implement experimentally are of
Gaussian type.  An example of a non-Gaussian unitary operation is the
realisation of the Kerr effect whose Hamiltonian is proportional to
third powers of the ladder operators.

\paragraph{Gaussian Dilation /Channels}

Consider an $N-$mode Gaussian system coupled to the environment it is
living in.  In covariance matrix language the CMs of the system and
the environment sum to the covariance matrix of the whole, according to
\begin{equation*}
   \gamma_{w} = \gamma_{s} \oplus \gamma_{e}. 
\end{equation*}
When applying a transformation on the system we have to take into
account that the system and the environment always interact with each
other and the environment may evolve during this process.  We assume
that this interaction is of symplectic type.  Hence the CM of the
whole gets transformed by
\begin{equation*}
    \gamma_{w} \mapsto S \, \gamma_{w}\, S^{T} 
    \, \mbox{ with } 
    S = \left(
    \begin{array}{cc} 
        S_{s} & S_{i, 1}\\ S_{i, 2} & S_{e} 
    \end{array} \right), 
\end{equation*}
with $S_{s}$ ($S_{e}$) the part of the transformation belonging to the system
(environment) only and the $S_{i}$ describing the interaction between
the system and the environment.  Note, that the submatrices of 
$S$ are not necessarily symplectic.

Since we only observe the system the behaviour of the environment is
neglected.  In mathematical formulae we take the average over all
possible configurations of the environment, that is on states,  we trace out the
environment.
 \begin{equation*}
    \rho'_s = tr_{e} [ \hat U (S)\, \rho_{w }\, \hat U^{\dag}(S)]. 
\end{equation*} 
with $ \hat U(S)$ being the unitary transformation associated to 
the symplectic transformation $S$. The corresponding mathematical 
operation  for the covariance matrices of the states is to take 
the  principal submatrix of the covariance matrix of the whole,  belonging 
only to the system. This reduction is denoted by brackets with an index $[\,\, ]_s$.
\begin{equation*}
    \gamma'_{s} = [ S \, \gamma_{w }\,  S^{T}]_s. 
\end{equation*}
\enlargethispage*{10pt}
The resulting covariance matrix for the system alone is then
\begin{eqnarray*}
    \gamma'_{s} &=& \left[ \left(
    \begin{array}{cc} 
        S_{s} & S_{i, 1}\\ S_{i, 2} & S_{e} 
    \end{array} \right) \left(
    \begin{array}{cc}
        \gamma_{s} & 0 \\ 0& \gamma_{e}
    \end{array} \right) \left(
    \begin{array}{cc} 
        S^{T}_{s} & S^{T}_{i, 2}\\ S^{T}_{i, 1} & S^{T}_{e} 
    \end{array} \right)\right]_s\\
        &=&  S_{s} \, \gamma_{s} \,  S^{T}_{s} +   S_{i, 1} \,  \gamma_{e} \,  S^{T}_{i, 1}. 
\end{eqnarray*}
With the coupling we do not get only the desired transformation $S_{s} \,
\gamma_{s}  \, S^{T}_{s}$ but also an additional part depending on the
state the environment is in and the interaction between system and
environment. 
Formally the symplectic transformation can be done on the composite 
of the system and the environment.  In reality that happens automatically 
and  with our ignorance of the environments degrees of freedom we only  can
see the effects on the system, mathematically formulated by taking
 the reduction on the system while neglecting the environment. 
We recognise that the interaction 
between system and environment introduces some noise in the system 
denoted by $ G = S_{i, 1}\,  \gamma_{e}\,  S^{T}_{i, 1} $.
 Thus, the coupling leads to a  decoherence process in the system.\\

But can we not prepare everything so that the noise vanishes?
We have to make sure that $S$ is a symplectic matrix and that the
resulting $\gamma_{s}$ has still covariance matrix properties.  So
first of all $G$ has to be symmetric what it absolutely is.  Secondly
we have $S_{s} \, \sigma_{s} \,  S_{s} + S_{i, 1} \,  \sigma_{e} \, S^{T}_{i, 1} =
\sigma_{s}$ and thirdly the uncertainty relation has to be fulfilled
by $\gamma_{e}$:
\begin{eqnarray*}
    \gamma_{e} + i \sigma_{e} \ge 0 &\Rightarrow&
    S_{i, 1} \, \gamma_{e} \,  S^{T}_{i, 1} + i \,  S_{i, 1} \,  \sigma_{e} \,  S^{T}_{i, 1} \ge 0\\
    &\Rightarrow & G = S_{i, 1} \, \gamma_{e} \,  S^{T}_{i, 1} \ge - i \,  S_{i, 1} \,
    \sigma_{e} \, S^{T}_{i, 1} \\
    &\Rightarrow& G  + i \, \sigma_{s} -  i \,  S_{s} \, \sigma_{s} \,  S^{T}_{s} \ge 0.
\end{eqnarray*}
We see that the noise $G$ has to fulfil a kind of uncertainty relation as well,
depending on the transformation $S_{s}$ we want to implement on the system.
Especially the case $G = 0$ is allowed only in case $S_{s}$ was
itself a symplectic transformation. 
We conclude that all experimentally available operations, not only those  
of symplectic type for the system, can be done but there is always  a  
quantum lower bound for the precision of the transformation which can not 
be beaten.

\begin{exa} We determine the minimum noise for a time reversal, 
e.g., phase conjugation of laser light. The transformation  on the covariance matrix of
the system is done with $M_A$, the matrix for partial 
transposition or time reversal introduced in Lemma \ref{timerevers}
\begin{equation*}
    \gamma \mapsto M_A \, \gamma \, M_A =  \left(
    \begin{array}{cc}
	1&0\\ 0&-1 
    \end{array} \right) \gamma \left(
    \begin{array}{cc}
	1&0\\ 0&-1 
    \end{array} \right).
\end{equation*}
Thus the noise $G$ has to fulfil the uncertainty relation
\begin{eqnarray*}
    0 &\le& G + i \sigma - i \left( \begin{array}{cc} 1&0\\ 0&-1
\end{array} \right) \sigma \left( \begin{array}{cc}
    1&0\\ 0&-1 
\end{array} \right) \\
&\le& G + 2 i \sigma.
\end{eqnarray*}
As we learned in Section \ref{PoCN} such an uncertainty relation
requires symplectic eigenvalues of $G$ greater or equal to two.
A perfect phase conjugation is thus not possible, 
the precision can be arbitrary small when using, e.g., strong 
laser pulses, but we can not reduce the noise completely since we have to fulfil a
minimal quantum limit.
\end{exa}

\paragraph{Adding of Classical Noise} 

Adding classical noise to a Gaussian state $\rho$ is described  by a 
convex combination of  random displacements of the state $\rho$, distributed with a 
normalised Gaussian distribution 
\begin{equation*}
\lambda(\xi) = \mathcal{N} e^{- \xi^T {1 \over \Delta } \xi}
\end{equation*} 
with a positive real symmetric $\Delta$.
The resulting $\rho'$ can then be written as the integral
\begin{equation*}
  \rho \mapsto \rho' = \int \dif^{2N}\xi \:\, \lambda(\xi) \,\, \hat W^{\dag}_{\xi}\, \rho \, \hat W_{\xi},
\end{equation*}
and  the covariance matrix $\Gamma_{\rho}$ of $\rho$ then changes according to 
\begin{equation*}
\Gamma_{\rho} \mapsto \Gamma_{\rho'} = \Gamma_{\rho} + \Delta
\end{equation*}
This process is always allowed since adding a positive matrix $\Delta$
to a covariance matrix $\Gamma$  gives  a proper covariance matrix $\Gamma'$.
For proofs of  these statements, please see the proof of Theorem \ref{24} and Lemma \ref{absorb}.

\paragraph{Measurements} 
The measurements on parts of multi-mode Gaussian states resulting again
in a Gaussian state are exactly those, which can be described as
projections on other Gaussian states.  We will exploit this a bit when
calculating the  Schur complement  in the next section.
In the following we will discuss  homodyne measurements.

\newpage

\section{Homodyne Measurements}
 
 One of the experimentally well realisable measurements is the homodyne 
 detection, where one mode of the measured $N-$mode state $ \rho$ is coupled to
a local oscillator mode and measured together.
The local oscillator is usually prepared in a coherent state with
state vector $|\alpha_{p} \rangle $ and the annihilation operator
belonging to the local oscillator will be denoted by $ \hat a_{p}$
(with index $p$ for pump mode).
\begin{figure}[h]
	{\includegraphics[width=1\textwidth]{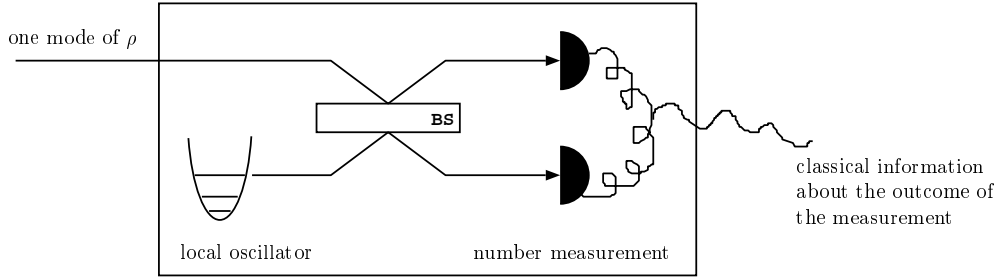}}
	\vspace*{- 0.5 cm}\hspace*{0.5cm}
	\caption{\small Homodyne measurement box}\label{fig7}
\end{figure}

With a homodyne measurement setup we are able to measure the expectation 
values of the quadratures of the measured mode
\begin{equation*}
    \hat x(\phi ) = \frac{\hat a e^{i\phi}+ \hat a^{\dag}e^{-i
    \phi}}{\sqrt 2}.
\end{equation*}
We present a small calculation without taking imperfect
detectors into account.
Behind the beam splitter we find the annihilation and creation operators of 
the modes $\hat b$ and $\hat b_p$ which are composed of the incoming modes
$\hat a$ belonging to $\rho$ and $\hat a_p$ belonging to the local oscillator
with the transmission and reflection coefficients playing the role of
the weighting prefactors.  We get:
\begin{eqnarray*}
  \hat b_p &=& T^* \, \hat a_p - R^* \, \hat a,\\
   \hat b&=& T \, \hat a + R \, \hat a_p,
\end{eqnarray*}
with the commutator relations $[\hat b ,\hat b^{\dag}] = | T |^2 + | R |^2$
when assuming $[\hat a,\hat a_p] =[\hat a, \hat a^{\dag}_p]= [\hat
a^{\dag}, \hat a_p] = [\hat a^{\dag},\hat a^{\dag}_p]=0$. 
\setlength{\mathindent}{4cm}
\begin{figure}[h] 
	\[{\includegraphics[width=0.5\textwidth]{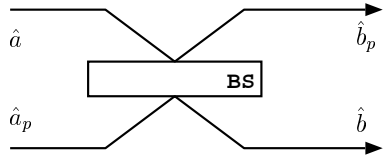}}\]
	\vspace*{- 0.5cm}\hspace*{0.5cm}\caption{\small Beam splitter}\label{fig8}
\end{figure}
\setlength{\mathindent}{2cm}

To set $ | T |^2 + | R |^2 =1 $ therefore gives correct annihilation  and 
creation operators and 
is an energy conservation restriction for a non-absorbing  mirror. 
The expectation values of the number operators after passing the beam splitter are:
\begin{eqnarray*}
   \langle \hat n_{\hat b} \rangle &=& |T|^2 \langle \hat n_{\hat a} \rangle + 
   |R|^2 |\alpha_p|^2 +|T R \alpha_p| \cdot
   \langle \hat a e^{i \phi} + \hat a^{\dag} e^{-i \phi}\rangle,  \\
   \langle \hat n_{\hat b_p} \rangle &=& |R|^2 \langle \hat n_{\hat a} \rangle 
   + |T|^2 |\alpha_p|^2 -|T R \alpha_p| \cdot
   \langle \hat a e^{i \phi} + \hat a^{\dag} e^{-i \phi}\rangle,
\end{eqnarray*}
with $\alpha_p$ the coherent state the pump  mode $\hat a_p$ was prepared in, 
and $\phi$ a combination, $\phi= \phi_T - \phi_R - \phi_{\alpha_p}$, of  
the phases of the transmission and  reflection coefficient and the complex number 
$\alpha_{p}$ respectively.
If we first measure the number of photons coming out in the two modes 
behind the beam spitter and then subtract them from each other 
we find the number difference to be:
\begin{eqnarray*}
   \langle \hat n_{\hat b} \rangle - \langle \hat n_{\hat b_p} \rangle &=&
   2 |T R \alpha_p| \langle \hat a e^{i \phi} + \hat a^{\dag} e^{-i
   \phi}\rangle \\
   &=& \sqrt 2 |T R \alpha_p| \langle \hat x(\phi)\rangle 
   = \frac{ |\alpha_p|}{\sqrt 2} \langle \hat x(\phi)\rangle,
\end{eqnarray*} with, e.g., $|T | = |R| =\frac{1}{\sqrt 2}$. \\
For a more accurate calculation involving imperfect photon detectors
see Ref. \cite{VW},  where it is shown that even in this case one can do
precise measurements when using a strong local oscillator, e.g.,
$\alpha_p$ is large.
The output state and classical information after the homodyne detection is
hence a projection of the $N$-mode state on the localised states
of the measured mode and the output is a Gaussian $(N-1)$-mode state. 
To go on we will need the covariance matrix and the displacement
vectors of the, about a position $x$,  localised state vectors.

\subsection{Moments of the Localised States}

We define the  state vectors  $| \psi_{x ,\epsilon}\rangle$ localised 
at $x$ with width $\epsilon$.
\begin{eqnarray}\label{xpsi}
   \langle x' | \psi_{x, \epsilon}\rangle 
    := \frac{\exp\left( \frac{-\left(x-x'\right)^2}{ 2 \epsilon^2}\right)}
    {\sqrt {2 \pi \epsilon^2}},
\end{eqnarray} 
with 
\begin{eqnarray*}
   tr[|\psi_{x, \epsilon} \rangle\langle \psi_{x, \epsilon}|]
   = \int_{\mathbbm{R}} \dif x'  |\langle x' | \psi_{x, \epsilon} \rangle |^2 
   = \frac{1}{2 \epsilon \sqrt \pi }.
\end{eqnarray*}
So $|\psi_{x, \epsilon}\rangle$ is not normalised but we find
\begin{eqnarray*}
    \int_{\mathbbm{R}} \dif x'  \langle x' | \psi_{x, \epsilon} \rangle  = 1.
\end{eqnarray*}

\setlength{\mathindent}{2cm}
Let us now calculate the first and second moments of the state vector $ |
\psi_{x, \epsilon} \rangle$
\begin{eqnarray*} 
   \langle \hat x \rangle \, \; \,
   &=& \frac{1}{tr[|\psi_{x, \epsilon} \rangle\langle \psi_{x, 
   \epsilon}|]}
   \int_{\mathbbm{R}}  \dif x' \langle x' | \hat x |\psi_{x, \epsilon} \rangle 
   \langle \psi_{x, \epsilon} | x'\rangle  \nonumber \\
  & =&  2 \epsilon \sqrt\pi  \int_{\mathbbm{R}}  \dif x' x' 
   \frac{\exp\left( \frac{-\left(x-x'\right)^2}{ \epsilon^2}\right) }
   {2 \pi \epsilon^2}\\
   &=& \nonumber  \frac{1}{\epsilon \sqrt\pi}  
   \int_{\mathbbm{R}} \dif x'' \left(x''+x\right)  
   \exp\left( \frac{-x''^2}{ \epsilon^2}\right) \nonumber
   =  x  
\end{eqnarray*} 
because of the symmetry of the Gaussian.

\begin{eqnarray*}  
   \langle \hat x^2 \rangle \,
   &=& 2 \epsilon \sqrt\pi  \int_{\mathbbm{R}}  \dif x' x'^2 
   \frac{\exp\left( \frac{-\left(x-x'\right)^2}{ \epsilon^2}\right)}
   {2 \pi \epsilon^2}\\ 
   &=& \frac{1}{\epsilon \sqrt \pi} \left[  2 \int_0^{\infty}\dif x'' x''^2  
   \exp\left( \frac{-x''^2}{ \epsilon^2}\right) + x^2 \epsilon \sqrt \pi \right]  \\
   &=& \frac{1}{\epsilon \sqrt \pi} \left[ \frac{\epsilon^3\sqrt \pi}{2} + 
   x^2 \epsilon \sqrt \pi \right]  
   = x^2 + {\epsilon^2 \over 2},\nonumber
\end{eqnarray*} 

\begin{eqnarray*}
   \langle \hat p \rangle \, \; \,
   &=& \frac{1}{tr[|\psi_{x, \epsilon} \rangle\langle \psi_{x, 
   \epsilon}|]}
   \int_{\mathbbm{R}}  \dif x' \langle x' | \hat p |\psi_{x, \epsilon} \rangle 
   \langle \psi_{x, \epsilon} | x'\rangle \\
   &=&  2 \epsilon \sqrt\pi  \int_{\mathbbm{R}}  \dif x' 
   \left(- i \frac{\partial \langle x' | \psi_{x, \epsilon} \rangle  } {\partial x'}\right)
   \langle \psi_{x, \epsilon} | x' \rangle \\
   &=&   - \frac{2i \sqrt\pi}{\epsilon}  \int_{\mathbbm{R}}  \dif x' 
   \left(x - x'\right)|\langle \psi_{x, \epsilon} | x' \rangle|^2 
   = 0, 
\end{eqnarray*}

\begin{eqnarray*}
   \langle \hat p^2 \,\rangle 
   &=&  2 \epsilon \sqrt\pi  \int_{\mathbbm{R}}  \dif x' \left((- i)^2 
   \frac{\partial ^2\langle x' | \psi_{x, \epsilon} \rangle  }
   {\partial x'^2}\right)
   \langle \psi_{x, \epsilon} | x' \rangle\\
   &=& \frac{2 \sqrt \pi}{\epsilon}  \int_{\mathbbm{R}}  \dif x' 
   \left(1 - \frac{\left(x - x'\right)^2}{\epsilon^2}\right) 
   |\langle \psi_{x, \epsilon} | x' \rangle|^2 \\
   \\
   &=& \frac{1}{2 \epsilon^2},\nonumber 
\end{eqnarray*}

\begin{eqnarray*}
   \langle \hat x \hat p \rangle 
   &=& \frac{1}{tr[|\psi_{x, \epsilon} \rangle
   \langle \psi_{x, \epsilon}|]} \int_{\mathbbm{R}}  \dif x' 
   \, x ' \left(-i \frac{\partial \langle x' | \psi_{x, \epsilon} \rangle } 
   {\partial x'}\right) \langle \psi_{x, \epsilon} | x'\rangle \nonumber\\ 
   &=&  2 i \epsilon \sqrt \pi \int_{\mathbbm{R}}  \dif x'  
   \frac{\partial \left(x '\langle x' | \psi_{x, \epsilon} \rangle \right)} 
   {\partial x'} \langle \psi_{x, \epsilon} | x'\rangle    
  \\& =& i \epsilon \sqrt \pi \int_{\mathbbm{R}}  \dif x'  
   |\langle x' | \psi_{x, \epsilon} \rangle|^2 
   =   i \epsilon \sqrt \pi \frac{1}{2 \epsilon \sqrt\pi} 
   = \frac{i}{2}\nonumber\\
   \langle \hat p \hat x \rangle & =& \langle \hat x \hat p\rangle^* 
   =\frac{- i }{2}.
\end{eqnarray*} 
The displacement vector and the covariance matrix of the state vector  $
|\psi_{x, \epsilon} \rangle $ are:
\begin{eqnarray*}
   \vec d_{x,\epsilon} &=&
   \left(\begin{array}{c}\langle \hat x\rangle \\ 
   \langle \hat p \rangle \end{array}\right) 
   = \left(\begin{array}{c}x \\    0 \end{array}\right),\\
   \gamma_{x, \epsilon} &=& \left(\begin{array}{cc}2 \left(\langle \hat x^2 \rangle 
   - \langle \hat x \rangle^2\right) & 2 \langle \hat x \hat p \rangle - i \\ 
   2 \langle \hat p \hat x \rangle + i &2 \left(\langle \hat p^2 \rangle 
   - \langle \hat p \rangle^2\right) \end{array}\right)
   =\left(\begin{array}{cc}\epsilon ^2&0 \\ 0 & \frac{1}{\epsilon^2} \end{array}\right).
\end{eqnarray*}

This result is very convincing since for a well localised state vector  
$ |\psi_{x, \epsilon}\rangle$ with $\epsilon \to 0$, the mean value should  
be $ x$ and the variance of the position should be small.  On the other hand 
we expect the variance of the momentum to grow, following the uncertainty
relation.  The uncertainty relation can be expressed in terms of the
covariance matrix : $ \det \gamma \ge 1$ which is in our case
fulfilled for every $\epsilon$ : $\det \gamma_{x, \epsilon} =
\frac{\epsilon^2}{\epsilon^2}= 1$.  To go on we will use the
covariance matrix $ \Gamma_{x, \epsilon} = \sigma \, \gamma_{x, \epsilon}
\, \sigma^T =\left(
\begin{array}{cc}
    {1 \over \epsilon ^2}&0 \\
    0 & \epsilon^2 
\end{array}\right)$
and the displacement $\vec D_{x, \epsilon } = \sigma \,  \vec d_{x, \epsilon} = 
\left( \begin{array}{c} 0\\ -x \end{array} \right).$\\

In the real world we can not do precise position measurements since our detectors 
have imperfect efficiency rates and  always  some dark counts.
If we assume that in the limit of large numbers  the measurement outcomes 
are approximately Gaussian distributed 
the following argument is true, even when the width $\epsilon$ of the distribution  is nonzero.

\subsubsection{Schur Complement}

We may now calculate what the outcoming state's covariance matrix is, after realising 
a projection on an arbitrary Gaussian one-mode state.  For $ \rho_{AB}$ a  Gaussian two-mode state
with covariance matrix $\Gamma= \left( \begin{array}{cc}A&C\\C^{T}&B
\end{array} \right)$ and displacement $\vec D$ we have
\setlength{\mathindent}{0cm}
\begin{eqnarray*} 
   \chi_{\rho} \left(\xi_A, \xi_B \right) := tr[\hat W _{\xi} \,
   \rho_{AB} ] = \exp \left(-\frac{\xi^T \Gamma \xi}{4} +i \vec D^T \xi\right),
   \, \mbox{ with } \, \xi^{T} = (\xi_{A}^{T} \xi_{B}^{T}) \in \mathbbm{R}^4.
\end{eqnarray*}
Similarly we have for the Gaussian one-mode state $\omega$ on whom we project a
covariance matrix $\Gamma_{\omega} $ and displacement $\vec D_{\omega} $
\begin{eqnarray*}
   \chi_{\omega} \left( \eta_B\right) := tr[\hat W _{ \eta} \omega ]
 = \exp\left(- \frac{\eta^T \Gamma_{\omega} \eta}{4}\right), \, \mbox{
 with } \, \eta^{T}= \eta_{B}^{T} \in \mathbbm{R}^4.. 
\end{eqnarray*} 

With the expansion in Eq. \eqref{9} the state $ \rho' $ can be
written as:

\begin{eqnarray*}
   \rho' &=& tr_{B} [ \rho_{AB} \, (\hat {\mathbbm{1}} \otimes \omega_{B} )] \\
   &=& tr_{B} \left[{1 \over (2\pi)^2} \int\dif^{4}\xi \;\; \chi_{\rho} (- \xi) \hat
   W_{\xi} \;\, 
   \left(\hat {\mathbbm{1}} \otimes  {1 \over 2 \pi} \int \dif^{2} \eta_{B} \;\; \chi_{\omega} (-
   \eta_{B}) \hat W_{\eta_{B}} \right) \right]\\
   &=& {1 \over (2\pi)^3} \int\dif^{4}\xi \int\dif^{2} \eta_{B} \;\;\chi_{\rho}
   (- \xi) \chi_{\omega} (- \eta_{B}) \hat W_{\xi_{A}} \; tr_{B} [ \hat
   W_{\xi_{B}} \hat W_{\eta_{B}} ],
\end{eqnarray*}
and with the trace formula for the Weyl operators in Lemma \ref{Wr} we
have
\begin{eqnarray*}
    \rho' &=& {1 \over 2\pi} \int\dif \xi_A \; 
   \left( {1 \over 2\pi}\int \dif \xi_B\;\chi_{\rho}(-\xi)\;
    \chi_{\omega}(\xi_{B}) \right) \;\, \hat W_{\xi_A}.
\end{eqnarray*}
We  find the characteristic function of the output state $\rho' $ to be:
\begin{eqnarray*} 
   \chi_{\rho'} (-\xi_A) &=& {1\over 2\pi}\int \dif \xi_B \;\chi_{\rho} (-\xi)\;
   \chi_{\omega}(\xi_{B})\\
   &=& {1\over 2 \pi}\int \dif \xi_2 \; 
   \exp \left(-\frac{\xi^T \Gamma  \xi}{4} \right) 
   \exp\left(- \frac{\xi_B^T \Gamma_{\omega}\xi_B}{4}\right) \\
   && \times \exp \left(-i \vec D^{T} \xi \right) \; \exp \left(+ i \vec
   D^{T}_{\omega} \xi_{B} \right) \\
   &=&    {1\over 2\pi}\exp\left(-{\xi_A^T A  \xi_A \over 4}-i \vec D_A^T 
   \xi_A\right) \;  \int \dif \xi_B \exp\left(- {\xi_B^T
   \left(\Gamma_{\omega} + B \right) \xi_B \over 4}\right)\\
   && \times\exp \left(-{\xi_A^T C \xi_B \over 2}\right) \exp\left(-i
   (\vec D^{T}_B - \vec D^{T}_{\omega}) \xi_{B} \right).
\end{eqnarray*} 
Here we realise that  only projections on Gaussian states can transform 
all Gaussian states again to Gaussian states. If the characteristic 
function of  $\omega$ had third powers of the variable $\xi$ 
in its exponential, the characteristic function of $\rho'$ could 
take a non-Gaussian form as well, for some Gaussian states $\rho$. 
We go on and  transform the integration variables $ \xi_B \rightarrow \eta +
C ^{-1}\xi_B$ and include a transformation constant $\mathcal{N}$
and get:

\begin{eqnarray*} 
   \chi_{\rho'}(-\xi_A) &=& {\mathcal{N}\over 2\pi}\exp\left(-{\xi_A^T A \xi_A
   \over 4}-i \vec D_A^T \xi_A\right)\exp\left(- {\eta^T \left(\Gamma_{\omega}
   + B \right) \eta \over 4}\right)\\
   &&\times\exp \left(-{\xi_A^T C \eta \over 2}\right) \exp\left(-i (\vec
   D^{T}_B - \vec D^{T}_{\omega}) \eta \right)\\
   && \times \int \dif \xi_B \exp\left(- {\xi_B^T C ^{-1T}
   \left(\Gamma_{\omega} + B \right) C ^{-1}\xi_B \over 4}\right)\\
      && \times  \exp\left( - {\eta^T \left(\Gamma_{\omega} + B \right)
      C^{-1}\xi_B \over 2}\right)
     \\&&\times \exp \left(-{\xi_A^T \xi_B \over 2}\right) \exp\left(-i
     (\vec D^{T}_B - \vec D^{T}_{\omega}) C ^{-1}\xi_B \right)
\end{eqnarray*} 
and set $\eta = - [\Gamma_{\omega} + B ]^{-1} \left( C^{T} \xi_{A} + 2
i (\vec D_{B} - \vec D_{\omega})\right)$ so that linear terms of $\xi_B $ in the
integral vanish and the integral shortens to
\setlength{\mathindent}{2cm}
\begin{eqnarray*}
    I = \int \dif \xi_B 
        \exp\left(- {\xi_B^T C ^{-1T} \left(\Gamma_{\omega} + B \right)
       C ^{-1}\xi_B \over 4}\right)
\end{eqnarray*} 
with $I$ being a real number.  Using that $B $ and $\Gamma_{\omega}$ are
symmetric we finally get:
\setlength{\mathindent}{0cm}
\begin{eqnarray*} \chi_{\rho'}(-\xi_A) &=& {I
\mathcal{N}\over 2\pi} \exp\left(-{\xi_A^T A \xi_A \over 4} -i \vec D_A^T
\xi_A\right)\\
       && \times \exp\left( { \xi_A^T C [\Gamma_{\omega} + B ]^{-1}
       C^{T}\xi_{A}\over 4}\right) \exp \left(i{\xi_A^T C [\Gamma_{\omega} +
       B ]^{-1} (\vec D_{B} - \vec D_{\omega})}\right)\\
       && \times \exp\left({- \left(\vec D_B^T - \vec D_{\omega}^T\right)
       [\Gamma_{\omega} + B ]^{-1} (\vec D_{B} - \vec D_{\omega})}\right).
\end{eqnarray*}
\setlength{\mathindent}{0cm}
The $\xi_{A}-$independent factors in $\chi_{\rho'}$ have to factor to one since
the normalisation condition gives $\chi_{\rho'}(0) = 1$.  Hence 
\begin{eqnarray*} 
    \chi_{\rho'}(\xi_A)  
          &=&  \exp\left(-{\xi_A^T A \xi_A \over 4} +i \vec D_A^T  \xi_A\right)
          \exp\left( { \xi_A^T C [\Gamma_{\omega} + B ]^{-1} 
          C^{T}\xi_{A}\over 4}\right)\\
          && \times \exp \left(-i{\xi_A^T C [\Gamma_{\omega} + B ]^{-1} 
          (\vec D_{B} - \vec D_{\omega})}\right).
\end{eqnarray*} 
We find the displacement of the state after the projection  to be\setlength{\mathindent}{2cm}
\begin{equation}\label{Dstrich}
    \vec D' =\vec D_A - C \left(\Gamma_{\omega} +B \right)^{-1} (\vec D_B - \vec D_{\omega})
\end{equation}
and the covariance matrix  
\begin{equation}\label{Gammastrich}
    \Gamma' = A - C \left(\Gamma_{\omega} +B \right)^{-1}C ^T.
\end{equation}

The new $\Gamma'$ is a combination of submatrices of $\Gamma$  and is
known as the \emph{Schur complement of the matrix $ B + \Gamma_{\omega} $ 
in  $\tilde \Gamma = \Gamma + \left( 
\begin{array}{cc}  
  0&0\\0& \Gamma_{\omega} 
\end{array} \right),$ } 
see Ref. \cite{HJ}.  We will need it in the next chapter when discussing about
measuring the degree of entanglement of an unknown state.\\

\begin{exa} 
  Let us calculate $\Gamma'$ for the special covariance matrix 
  $\Gamma_{x, \epsilon}=  \left(
  \begin{array}{cc} 
    {1 \over  \epsilon^2} & 0  \\ 0  &  \epsilon^2 
  \end{array}\right) $ 
of the localised states.\\  We find with 
$B = \left( 
\begin{array}{cc} 
  b_1 & b_2 \\ b_2 & b_3 
\end{array} \right)$ the inverse
\begin{eqnarray*}
  \left(\Gamma_{x, \epsilon} + B \right)^{-1}
   &=& \left( \begin{array}{cc}                              
     b_1 + {1 \over \epsilon^2} & b_2 \\ 
     b_2 & b_3 + {\epsilon^2} 
   \end{array}\right)^{-1}\\
   &=& { 
   \left(\begin{array}{cc}  
     b_3 \epsilon^2 + \epsilon^4 & - b_2 \epsilon^2\\
     - b_2 \epsilon^2& b_1 \epsilon^2 + 1
   \end{array}\right) \over (b_1 \epsilon^2 + 1)(b_3 + \epsilon^2)-\epsilon^2 b_2^2}.
 \end{eqnarray*}
We finally get the covariance matrix after the projection
\begin{equation*}
  \Gamma' = A - C \, {\left( 
    \begin{array}{cc}  
      b_3 \epsilon^2 + \epsilon^4 & - b_2 \epsilon^2\\
      - b_2 \epsilon^2& b_1 \epsilon^2 + 1
    \end{array}\right) 
    \over (b_1 \epsilon^2 + 1)(b_3 + \epsilon^2)-\epsilon^2 b_2^2}\, C^T.
 \end{equation*} 
Since the $\Gamma_{x, \epsilon}$ do not really depend on $x$,
$\Gamma'$ is the same for every projection on any localised state, 
while the displacements are different for different projections. 
For $\epsilon \to 0 $ the above inverse becomes the  Moore-Penrose 
inverse, generalising  matrix inversion to singular matrices. 
\end{exa}
 
\newpage
\subsection{Projection on a Coherent State by Homodyne Measurement}
\label{procoh}

In the next chapter we will need an experimentally realisable setup to project one mode of a 
 Gaussian multimode photon state on a coherent state.  To do this
 projection we will use homodyne detection because experimentally it
 can often be achieved perfectly while other projection strategies are
 not as efficient.

We discuss the case where the state $\rho$ to be measured has two modes and
we want to project it on a coherent state of its second mode. 
To realise this projection with homodyne measurements we have to
find a scheme allowing us to do so. We will show that the following setup is convenient.

\setlength{\mathindent}{1cm}
 \begin{figure}[h]
\[	{\includegraphics[width=1\textwidth]{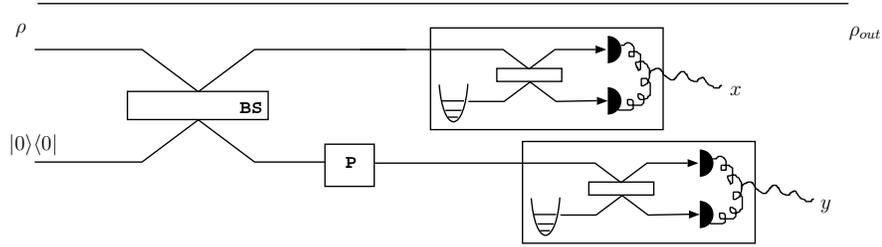}}\]
	\vspace*{- 1cm}\hspace*{-2.0cm}
	\caption{\small Measurement setup}\label{procohe}
\end{figure}
\setlength{\mathindent}{2cm}

The mathematical description of this processing is done by unitaries
followed by homodyne measurements in the second and third mode. 
\begin{eqnarray*}
 \langle \psi_{y, \epsilon} |_{C} \, \langle
    \psi_{x, \epsilon} |_{B} \, \hat U^{P}_{C} \,\hat U^{BS}_{BC} \;
    (\rho_{AB} \, \otimes |0\rangle \langle 0 |_{C}) \; \hat U^{\dag
    \,BS}_{BC} \, \hat U^{\dag \, P}_{C} \, |\psi_{x, \epsilon}
    \rangle_{B} \, | \psi_{y, \epsilon} \rangle_{C}. 
 \end{eqnarray*}
With all of these ingredients being Gaussian we can identify the
outcoming state by using only the covariance matrix representation of
all operations.
We start with a two mode  state $\rho$ having an arbitrary covariance matrix
$\Gamma = \left( \begin{array}{cc} A & C \\C^{T} &B \end{array}
\right)$ and displacement $\vec D = \left( \begin{array}{c} \vec D_{A} \\ \vec D_{B}
\end{array}\right)$, the vacuum $ |0 \rangle \langle 0|$ with CM
$\Gamma_{0} = \mathbbm{1}$ and vanishing displacement (see Section
\ref{csts}). We will also need the localised states, with width $\epsilon$, 
realised by the homodyne measurements all having the covariance matrix
$\Gamma_{\epsilon} = \left(
\begin{array}{cc} 
    {1 \over \epsilon^{2}} & 0 \\ 0 & \epsilon^{2}
\end{array} \right)$ 
and different displacements 
$\vec D_{x} = \left( 
\begin{array}{c} 
    0\\-x 
\end{array} \right). $   We apply a $50:50$ beam splitter described
by the symplectic transformation 
\begin{equation*}
    S^{BS} = {1 \over \sqrt{2}} \left(  
    \begin{array}{cc}
	\mathbbm{1} & - \mathbbm{1} \\
	\mathbbm{1} & \mathbbm{1} 
    \end{array} \right),
\end{equation*}
a one mode ${\pi \over 2}-$phase shifter 
\begin{equation*} S^{P}_{\pi \over 2} = \left(
    \begin{array}{cc} 
	0& 1\\ -1 & 0
    \end{array} \right) = \sigma
\end{equation*}
and send two modes to homodyne measurement boxes.  We
calculate step by step what happens on the covariance matrix and the
displacement of $\rho$ passing this setup.  \newpage
\setlength{\mathindent}{0cm} \begin{tabular}{p{2.2cm}|p{9.9cm}}\hline
operation & displacement and covariance matrix \\ \hline \hline
\vspace{1cm}
    
    {adding the} vacuum as a third mode &
    \vspace{-0.4cm}
    
    \begin{eqnarray*}
        \vec D \mapsto \vec D' = \left(
        \begin{array}{c}
	\vec D_{A}\\\vec D_{B}\\ 0 \end{array} \right)
    \end{eqnarray*}
    \begin{eqnarray*}
        \Gamma \mapsto \Gamma' = \left(
        \begin{array}{ccc}
	A&C & 0 \\
	C^{T}&B& 0\\
	0 &0& \mathbbm{1} 
        \end{array}\right) 
    \end{eqnarray*}\\ \hline
    \vspace{1cm} 
    
    beam splitter operation on the second and third mode&
    \vspace{-0.4cm}
    
    \begin{eqnarray*}
        \vec D' \mapsto &\vec D''& = S^{BS} \vec D' = \left(
        \begin{array}{c}
	\vec D_{A} \\
	{1 \over \sqrt{2}} \vec D_{B} \\
	{1 \over \sqrt{2}} \vec D_{B} 
        \end{array} \right)
    \end{eqnarray*} 
    \begin{eqnarray*}
        \Gamma' \mapsto &\Gamma''& = S^{BS} \,  \Gamma' \, S^{T \, BS}
         =\left(
        \begin{array}{ccc}
	A & {C\over\sqrt{2}} & {C \over\sqrt{2}} \\
	{C^{T}\over \sqrt{2}} &{B +\mathbbm{1} \over 2} & {B- \mathbbm{1} \over 2} \\
	{C^{T}\over\sqrt{2}} & {B -\mathbbm{1} \over 2} & {B + \mathbbm{1} \over 2}
	\end{array}\right) \end{eqnarray*}\\ \hline
    \vspace{1cm} 
    
    phase shifter operation on the third mode &
    \vspace{-0.4cm}
    
    \begin{eqnarray*}
        \vec D'' &\mapsto &\vec D''' = S^{P}_{\pi \over 2} \, \vec D'' 
        = \left(
        \begin{array}{c}
	\vec D_{A} \\
	\! \! \! \! {1 \over \sqrt{2}} \vec D_{B} \\
	{1 \over \sqrt{2}} \sigma \vec D_{B} 
        \end{array} \right) 
    \end{eqnarray*} 
    \begin{eqnarray*}
        \Gamma'' &\mapsto &\Gamma''' = S^{P}_{\pi \over 2} \, \Gamma'' \, S^{P\, T}_{\pi \over 2}
         =\left(
        \begin{array}{ccc}
	A 
	& {C \over \sqrt{2}} 
	& {C \sigma^{T} \over \sqrt{2}} \\
	{C^{T} \over \sqrt{2}} &\; \; {B +\mathbbm{1} \over 2} & \; \; {B- \mathbbm{1}
	\over 2} \sigma^{T} \\
	{\sigma C^{T} \over \sqrt{2}} 
	& \sigma {B -\mathbbm{1} \over 2} 
	& \sigma {B +\mathbbm{1} \over 2} \sigma^{T} 
        \end{array}\right) 
    \end{eqnarray*}\\
	\hline
    \vspace{1cm}
    
    homodyne measurement on the second mode with the outcome
    $x$ &
    \vspace{-0.4cm}
    
    \begin{eqnarray*}
    \vec D''' &\mapsto &\vec D^{(4)}  = \left(
    \begin{array}{c}
        \vec D_{A} \\
        {1 \over \sqrt{2}} \sigma \vec D_{B} 
    \end{array} \right) 
    \\&& \quad - \left(
    \begin{array}{c}
        {C \over \sqrt{2}} \\
        \sigma{B -\mathbbm{1} \over 2} 
        \end{array}\right) \left [{B+ \mathbbm{1} \over 2}+\Gamma_{\epsilon}
        \right]^{-1} \left[{1 \over \sqrt{2}} \vec D_{B} - \vec D_{x} \right ]
\end{eqnarray*} 
\begin{eqnarray*}
    \Gamma''' &\mapsto& \Gamma^{(4)}= \left(
    \begin{array}{cc}
        A & {C \sigma^{T} \over \sqrt{2}} \\
        {\sigma C^{T} \over \sqrt{2}} & \sigma {B + \mathbbm{1} \over 2} \sigma^{T}
        \end{array}\right)\\
        &&\quad - \left(
    \begin{array}{c}
        {C \over \sqrt{2}} \\
        \sigma {B -\mathbbm{1} \over 2} 
    \end{array}\right) \left[{B + \mathbbm{1} \over 2} +
    \Gamma_{\epsilon} \right]^{-1} \left(
    \begin{array}{cc}
        {C^{T}\over \sqrt{2}} & {B -\mathbbm{1} \over 2} \sigma^{T}
        \end{array}\right)
\end{eqnarray*}
\end{tabular} 

We abbreviate\setlength{\mathindent}{2cm}
\begin{eqnarray*}
    M&=&M^{T}=
    {\left [{B+ \mathbbm{1} \over 2}+\Gamma_{\epsilon} \right]^{-1}\over 2} \\
    &=& { \left(
    \begin{array}{cc} 
        1+b_{3} +2 \epsilon^{2} & - b_{2}\\
        - b_{2} & 1 + b_{1} +{2 \over \epsilon^{2}}
    \end{array} \right)  \over \det B +b_{3} (1 + {2 \over \epsilon^{2}}) 
    + b_{1} (1 +2 \epsilon^{2}) 
    + 2 (\epsilon +{1 \over \epsilon})^{2} +1 }. 
\end{eqnarray*}
And get for $\vec D^{(4)}$:
\begin{equation*}
    \vec D^{(4)} = \left(
    \begin{array}{c}
        \vec D_{A} -{\sqrt{2} C M } \left[{1 \over \sqrt{2}} \vec D_{B} - \vec D_{x} \right
        ] \\
        {1 \over \sqrt{2}} \sigma \vec D_{B} -    
        \sigma (B -\mathbbm{1}) M \left[{1 \over \sqrt{2}} \vec D_{B} - \vec D_{x}
        \right ] \end{array} \right)
\end{equation*}
and for $\Gamma^{(4)}$ we find:
\begin{eqnarray*}
          \Gamma^{(4)} &=& \left(
    \begin{array}{cc}
        A - {C M } {C^{T}}
        & {C \over \sqrt{2}}\left(\mathbbm{1} 
        - { M } (B -\mathbbm{1}) \right)\sigma^{T}\\
        \sigma \left( \mathbbm{1} 
        - (B -\mathbbm{1}) M \right) {C^{T}\over \sqrt{2}} 
        & { \sigma \left[ (B +\mathbbm{1})  -  (B-\mathbbm{1}) M (B
        -\mathbbm{1}) \right] \sigma^{T} \over 2} 
    \end{array}\right).
\end{eqnarray*}
As the last step we implement a homodyne measurement on the former third
mode with the outcome $y$.  The final displacement
and covariance matrix are given by \setlength{\mathindent}{0cm}
\begin{eqnarray*}
    \vec D^{(4)} &\mapsto& \vec D^{(5)} =
    \vec D_{A} -{C M \sqrt{2}} \left[{1 \over \sqrt{2}} \vec D_{B} - \vec D_{x} \right ]
     - {C \over \sqrt{2}}\left( \mathbbm{1} - M (B -\mathbbm{1}) \right)
    \sigma^{T} \\
    && \qquad  \qquad\quad\times \left[ { \sigma \left[ (B +\mathbbm{1})
    - (B-\mathbbm{1}) M (B -\mathbbm{1}) \right] \sigma^{T} \over 2}
    +\Gamma_{\epsilon} \right]^{-1} \\
    && \qquad \qquad \qquad\quad\times({1 \over \sqrt{2}} \sigma \vec
    D_{B} - \sigma (B -\mathbbm{1}) M \left[{1 \over \sqrt{2}} \vec D_{B}
    - \vec D_{x} \right ] -\vec D_{y}),
\end{eqnarray*}
\begin{eqnarray*}
    \Gamma^{(4)} &\mapsto& \Gamma^{(5)}= 
    A - C M C^{T} - {C \over \sqrt{2}}\left( \mathbbm{1} - M 
    (B -\mathbbm{1}) \right) \sigma^{T}  \\
    && \qquad\qquad\qquad \qquad \times \left[ { \sigma \left[ (B
    +\mathbbm{1}) - (B-\mathbbm{1}) M (B -\mathbbm{1}) \right] \sigma^{T}
    \over 2} +\Gamma_{\epsilon} \right]^{-1} \\
    && \qquad \qquad \qquad\qquad\qquad  \times \sigma
    \left( \mathbbm{1} - (B -\mathbbm{1}) M \right) {C^{T}\over \sqrt{2}}
\end{eqnarray*}
Surprisingly this expression reduces to
\begin{eqnarray}\label{schurg}
  \Gamma^{(5)} &=&  A - C \left[ B+ \mathbbm{1} \right]^{-1} C^{T},
\end{eqnarray}
where no $\epsilon$ appear anymore, meaning that  even a huge width of a Gaussian 
distributed measurement outcome is not relevant for the resulting covariance matrix.
From the covariance matrix and the displacement of the resulting state $\rho'$
we can read off that it would have been the same to project the initial
state $\rho$ on coherent state vectors $|\alpha(x,y,\epsilon) \rangle$.  The
complex number $\alpha$ depends on the outcomes of the homodyne
measurements: $x$ and $y$, and on the Gaussian detection distribution
with width $\epsilon$, with all those quantities appearing in $\vec D^{(5)}$.

\chapter{Measuring  Entanglement of Two-Mode States}

To determine whether a given unknown Gaussian two-mode state is entangled or not 
and how much, one  can measure all entries of the covariance matrix, i.e., ten real numbers.  
But since we only want to know a special property of the state it could be 
that fewer measurements are necessary to answer the same question. We propose  a scheme 
which makes it possible to determine the symplectic eigenvalues of the
partially transposed covariance matrix with nine measurements.  In the
two-mode case, where the PPT criterion is necessary and sufficient,
these symplectic eigenvalues are adequate to find out the degree of
entanglement, measured in the logarithmic negativity.  

\section{Measurement Scheme}
Given an unknown covariance matrix\footnote{We will use again 
the small covariance matrices $\gamma$. Keep in mind that the 
connection between the capital and the small CM is given by 
$\Gamma = \sigma \gamma \sigma^T$.}
$\gamma$ of a Gaussian two-mode state  $\gamma
= \left( \begin{array}{cc} A & C \\ C^{T } & B \end{array} \right)$
with $A, B$ real symmetric and $C$ real $2 \times 2$ matrices.

\paragraph{Step 1}
Measure all entries of 
$A = \left( \begin{array}{cc} a_1& a_2 \\ a_2 & a_3 \end{array}\right)$ 
and determine $a = \sqrt{\det A}= a_1 a_3 - a_2^2$. Experimentally
this can be done by simple position and momentum measurements with
homodyne detection.  The measurement outcomes of a position/momentum
measurement will approximately be Gaussian distributed.  The variance
of the measured distribution is an estimator  for the covariance 
matrix element $a_{1}$/$a_{3}$.  To measure the off-diagonal 
element $a_{2}$ we apply a ${ \pi \over 4 }-$phase shifter 
$S^P_{\pi \over 4 } = {1 \over \sqrt{2} } \left(
\begin{array}{cc} 
    1 & -1\\ 1&1
\end{array}\right)$ so that  the diagonal element $A_{11}$  transforms to a
linear combination of all elements of $A$ namely $A'_{11} = a_{1} +
a_{3} - 2 a_{2} $.  When measuring the position of the transformed
state one will approximately get a Gaussian distribution with
variance $A'_{11}$ from which we can determine the value of $a_{2}$.

\paragraph{Step 2}
Measure all entries of $B= \left( \begin{array}{cc} b_1& b_2 \\ b_2 &
b_3 \end{array}\right)$ and determine $b = \sqrt{\det B}= b_1 b_3 -
b_2^2. $
\paragraph{Step 3}
Calculate and implement the local symplectic transformation which
brings $A$ and $B$ to their diagonal form.\setlength{\mathindent}{2cm}
\begin{equation*}
S = S_A \oplus S_B : \gamma \mapsto \gamma' = 
S \gamma S^T = \left( \begin{array}{cc}
a \mathbbm{1} & C' \\ C'^{T } & b \mathbbm{1}
\end{array} \right)
\end{equation*}
with $C'= S_A C S^T_B$.  The experimental setup depends on the
matrices $A$ and $B$ but the transformations $S_{A}$ and $S_{B}$ are generally 
phase shifters with variable phase $\phi$ : \, $P_{\phi} = \left(
\begin{array}{cc} 
    \cos \phi & -\sin \phi\\ \sin \phi &\cos \phi 
\end{array}\right)$ which should be available for every $\phi$.

\paragraph{Step 4}
As we saw in Subsection \ref{procoh} it is possible to project a two
(or more) mode system on coherent states of its second mode using
homodyne measurements.  Although the projections done with the
proposed setup change with the measurement outcomes of the homodyne
boxes, it is always a projection on a coherent state.  The CM of the
measured state changes in every case according to
\begin{equation*}
    \gamma' \mapsto \gamma'' = a \mathbbm{1} -(b+1)^{-1} C'C'^T. 
\end{equation*}
giving a proper one-mode covariance matrix.
Apply the projection physically. \\

\bf Remark: \rm  The different displacements of the coherent states have
no effect on the covariance matrix but only on the displacement the
resulting state has.  

\paragraph{Step 5}
Measure all entries of 
$\gamma'' = \left( \begin{array}{cc} g_1& g_2 \\ g_2 & g_3 \end{array}\right)$. 
Since $\gamma''$ is symmetric we only have to measure three times, 
one less as when measuring the non-symmetric matrix $C$. 
Calculate the matrix $ C' C'^T= (b+1)(-\gamma'' + a \mathbbm{1}) $ and the
absolute value of $\det C$: 
\begin{eqnarray*}
    \det (C' C'^T) &=& (\det C')^2 = (\det S_A C S^T_B)^2\\
    &=&(\det C)^2 = (b+1)^2 [(a-g_1 )(a-g_3) - g_2^2].
\end{eqnarray*}

\paragraph{Step 6}
With the following calculation we determine the determinant of the
initial $\gamma$:
\begin{eqnarray*}
&\det & \left[\left( 
\begin{array}{cc}
    \mathbbm{1}& -{1 \over b} C' \\ 
    0 & \mathbbm{1} 
\end{array}\right) \left( 
\begin{array}{cc} 
    a\mathbbm{1}& C' \\ 
    C'^T & b \mathbbm{1} 
\end{array}\right) \left(
\begin{array}{cc} 
    \mathbbm{1}& 0 \\
    - {1 \over b} C'^T & \mathbbm{1} 
\end{array}\right)\right] \\
&=& \det \left[ 
\left( \begin{array}{cc} a \mathbbm{1} - {1 \over b} C'  C'^T & 0 \\ 
0 & b \mathbbm{1} \end{array}\right)
\right] =b^2   \det \left[  a \mathbbm{1} - {1 \over b} C'  C'^T \right]\\
&=&
   \det \left[   (b + 1)  \gamma'' - a  \mathbbm{1} \right] 
\gleich \det \gamma' = \det \gamma. 
\end{eqnarray*}
\paragraph{Step 7} 
We now know the Simon normal form of the matrix $\gamma$ 
namely 
\begin{equation*}
\gamma^{SNF} =\left( \begin{array}{cccc}
a  & 0 & c& 0 \\ 0&a&0&d \\ c&0&b&0\\ 0&d&0&b\end{array} \right)
\end{equation*}
with $a$ from \bf Step 1\rm, 
$b$ (\bf 2\rm) and $c, d$ determined by the
equations $cd = \det C$ (\bf 5\rm) and $\det \gamma = a^2 b^2 + c^2
d^2 - ab (c^2 + d^2)$ (\bf 6\rm).  There is a little ambiguity in \bf
5 \rm since the sign of the determinant of $C$ is not fixed.

\section{Determining the Degree of Entanglement}

We remember that every two-mode covariance matrix
can be brought to Simon normal form only by local symplectic
transformations which do not change the degree of entanglement.
The symplectic eigenvalues of the covariance matrix in Simon normal form are
\setlength{\mathindent}{0cm}
\begin{equation}\label{SNFG}
    (\gamma^{SNF})_{1,2} = 
{ 1 \over \sqrt 2} \sqrt{a^2 + b^2 + 2 c d \pm 
\sqrt{(a^2 + b^2 + 2 c d )^2 - 4 \det \gamma}}.
\end{equation}

To decide whether the given state $\gamma$ is entangled or not we use
the PPT-criterion stating that an entangled Gaussian state must have a partial
transpose which violates the uncertainty relation for covariance
matrices.  The symplectic eigenvalues of the partially transposed CM can
be calculated using again the Simon invariants via : 
\begin{equation}\label{SNFGTA}
  (\gamma^{SNF})^{T_A}_{1,2} = { 1\over \sqrt 2} \sqrt{a^2 + b^2 - 2 c d 
    \pm \sqrt{(a^2 + b^2 - 2 c d)^2 - 4 \det \gamma}}.
\end{equation} 
From the preceding steps we have all ingredients to determine the values of $ 
(\gamma^{SNF})_{1,2} $ and $ (\gamma^{SNF})^{T_A}_{1,2} $ 
with the little subtlety that we do not know the sign of $cd$. 
We first take the positive sign for $\det C$ and calculate all eigenvalues. 
If all eigenvalues in Eq.   \eqref{SNFG} and \eqref{SNFGTA} are greater
than one, the state was not entangled.  This does not change when
taking the negative sign of $\det C$ since the eigenvalues of
$(\gamma^{SNF})^{T_{A}} $ and $\gamma^{SNF}$ then just interchange. 
We hence know that the state is not entangled but can not decide if
the state has Simon normal form \setlength{\mathindent}{2cm}
\begin{equation*}
    \gamma^{SNF_{1}} =
    \left( \begin{array}{cccc}
    a  & 0 & c& 0 \\ 0&a&0&d \\ c&0&b&0\\ 0&d&0&b\end{array} \right)
    \mbox{ or } \gamma^{SNF_{2}} 
    =\left( \begin{array}{cccc}
    a  & 0 & -c& 0 \\ 0&a&0&d \\ -c&0&b&0\\ 0&d&0&b\end{array} \right).
\end{equation*}

If one of the eigenvalues was smaller than one then this eigenvalue
belongs to $ (\gamma^{SNF})^{T_A} $ and the state is entangled.
In fact we are able to determine the degree of entanglement the state 
posses with only nine measurements instead of ten when measuring the 
whole state.  The exact degree of entanglement can  be calculated with
Eq. \eqref{logneg}
\begin{eqnarray*}
    E_{\mathcal{N}}(\gamma) = - \sum_{i} \min(\ln \gamma_{i}^{T_{A}}, 0).
\end{eqnarray*}
and the symplectic eigenvalues from above.

\section{Discussion of the Measurement Strategies}

To experimentally realise a measurement is often a really 
expensive adventure, but the costs could be reduced when performing less measurements.  
The easiest way to determine the symplectic eigenvalues of a covariance matrix
 would be to measure just all elements of the CM; that makes 
ten different kinds of measurements. 
But why not learn from the available results of previous measurements? 
As we showed in the previous steps, it is possible to get the 
symplectic eigenvalues  with less queries, when adjusting the 
strategy dependent on the information we already gained.
For future work, a challenging task would be the optimisation of such strategies.
One could even try to  proof how many kinds of measurements 
one  necessarily has to perform to determine the symplectic eigenvalues 
of the covariance matrix of a given state, when it is allowed to 
learn during the process.  \\

But maybe our strategy is not that good because the variances of the
measured covariance matrix entries could be worse as in the ten number
case.  It could be, that when measuring all entries of the CM  one has to measure
every entry  $N_{10}$ times to get the same results and the same variances as
with our nine number idea for  $N_9 >> N_{10}$ measurements.  The energy saved
when only measuring nine instead of ten  entries is then spend on
more tries for every element.\\

We simulated both strategies, assuming that $\gamma$ is the matrix
\begin{equation*}
\gamma = \left( 
\begin{array}{cccc}
  3.5 & 0& 2.5 &0\\
  0&3&0&-2.5\\
  2.5&0&3.5&0\\
  0&-2.5&0&3
\end{array}\right).
\end{equation*}
The calculated symplectic eigenvalues of the given $\gamma$ are
\begin{eqnarray}\label{gam}
    \gamma_{1}=  2.345 \quad \gamma_{2} = 1.732 
\end{eqnarray}
and the symplectic eigenvalues of $\gamma^{T_A} $ are given by
\begin{eqnarray}\label{gamTA}
    \gamma^{T_{A}}_{1} = 0.707\quad \gamma^{T_{A}}_{2} = 5.745,
\end{eqnarray}
thus $\gamma$ is entangled.
To determine the entries of the covariance matrix we simulated  
 $N= 10^2, 10^3, 10^4, 10^5, 10^6$  measurements in total. 
For the usual ten number strategy we measured every entry $N_{10}= {N \over 10}$ 
times, while for the nine number strategy we measured every single quantity $N_9 = {N \over 9}$ times.
The estimations of the entries of the covariance matrix where done using 
ten (nine)  Gaussian probability distributions, centered about zero. 
The variances of the first four Gaussians where the  diagonal
entries of the covariance matrix $\gamma$, which one can experimentally determine by  
position  or momentum measurements. For the off-diagonal entries one first 
has to apply phase shifters, to bring them on the diagonal. Measuring  again 
the position of such a transformed state gave values for those entries.
For every expectation value we took the average over all $N$ measurements 
and calculated the variance of the measurement outcomes. These results are 
the estimators for the covariance matrix entries.
The whole procedure was done $M$ times for $M = 1000 $  measurements. 
From the estimated covariance matrix we calculated the symplectic eigenvalues of $\gamma$
and $\gamma^{T_{A}}$, each $M$ times, using  Eq. \eqref{SNFG} and Eq. \eqref{SNFGTA}
respectively and got an estimator  for each of them.
We compared  the average of these measurement outcomes to the exact values 
and found the following result.\\

With the usual ten measurements the estimation of the smallest symplectic eigenvalue 
$\gamma_1^{T_A}$, from which one can determine the degree of entanglement of the Gaussian state,
 was a bit better. The decrease of the abberation of the estimated symplectic eigenvalue in comparison to 
the exact value was approximatly equal in both strategies. The standard deviation of the measured 
$ \gamma^{T_A}_1$ from the exact one, is  calculated using the $M$ estimated values of $\gamma^{T_A}_1$
\setlength{\mathindent}{2cm}
\begin{equation}
\Delta \gamma_1^{T_A} = \sqrt{{1 \over M - 1} \, \sum_{i=1}^{M} \, 
\left[ (\gamma^{T_A}_1)_i -(\gamma^{T_A}_1)_{exact} \right]^2 }
\end{equation}

\setlength{\mathindent}{1.3cm}
 \begin{figure}[h]\label{estim}
\[	{\includegraphics[width=0.7\textwidth]{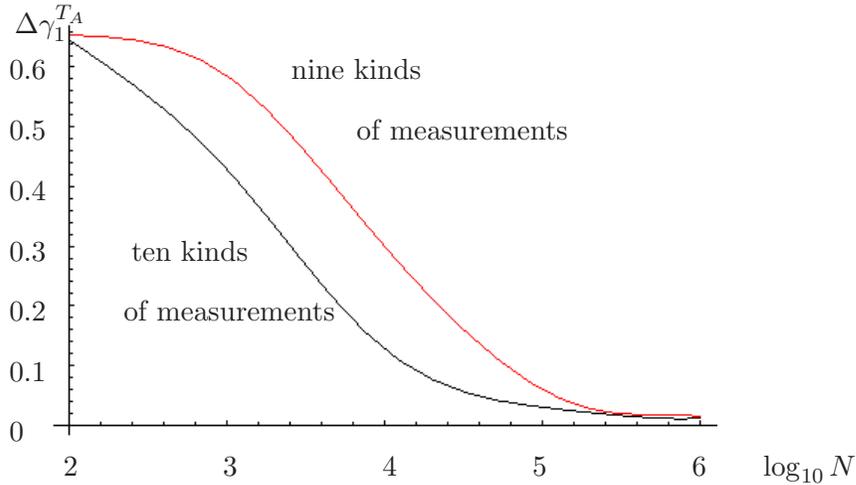}}\]
\vspace{-6.95cm}

\hspace{1.2cm}$\Delta \gamma_1^{T_A}$\\\vspace{-0.35cm}

\hspace{1cm} 0.6 \hspace{3cm} nine kinds \\\vspace{-0.15cm}

\hspace{1cm} 0.5 \hspace{4cm}of measurements\\\vspace{-0.15cm}

\hspace{1cm} 0.4\\\vspace{-0.15cm}

\hspace{1cm} 0.3 \hspace{1cm}ten  kinds\\\vspace{-0.16cm}

\hspace{1cm} 0.2 \hspace{0.9cm}of measurements\\\vspace{-0.15cm}

\hspace{1cm} 0.1\\\vspace{-0.15cm}

\hspace{1cm} 0\\\vspace{-0.5cm}

\hspace{1.85cm}2 \hspace{1.8cm}3 \hspace{1.8cm}4\hspace{1.8cm}5 \hspace{1.8cm}6 \hspace{0.5cm} $\log_{10}N$

\vspace{-2ex}\hspace{-4cm}\caption{\small Ten Measurements versus Nine}
\end{figure}
\vspace{0.4cm}

The figure shows the standard deviation $\Delta \gamma_1^{T_A}$ 
for the ten measurement strategy (lower curve) and the 
nine measurement strategy (upper curve). The  behaviour of the 
variances for the total  number of  measurements $N= 10^2, 10^3, 10^4, 10^5, 10^6$ is  shown on a logarithmic scale.
 Unfortunately our nine measurement strategy gives worse results  
than  the ten measurement strategy. With increasing $N$ the variances of the 
symplectic eigenvalue $\gamma_1^{T_A}$ estimated with  the two strategies 
decreases, as it should be.

\newpage

\section{Collective Measurements}

Finally we note, that it is by no means clear that single mode  measurements, 
as we employed in the proposed setup,  turn out to give better measurement 
strategies than a measurement using more copies of a state. 
Sometimes it can be useful to implement collective measurements, meaning that 
not only one copy of a state is measured but two or more copies are merged 
and processed together. The measurements of those multimode states could 
allow to further reduce the number of queries.  We often want to know 
only a special quantity composed of entries of the covariance matrix. 
In this case, with collective measurements it is indeed possible to 
reduce the number of  measurements, as the following example shows. 
\vspace{2ex}
\renewenvironment*{exa}{
   \begin{sloppypar} \hspace{1.4cm} \stepcounter{exa} 
      \begin{minipage}[t]{11.5cm}  \small \bf{Example} \arabic{exa} : \rm \normalsize
\setlength{\mathindent}{1cm}}{\\ \end{minipage}   \end{sloppypar} }

\begin{exa}
  Let $\gamma$ be a two-mode covariance matrix of the form 
  \begin{eqnarray*}
    \gamma = \left(
    \begin{array}{cc}
      A & C \\
      C^T & B 
    \end{array} \right).
  \end{eqnarray*}
  We are interested in the determinant of $A + B$ and would naively measure 
  the three entries of each matrix $A$ and $B$. 
  Employing  collective measurements instead could give the desired 
  determinant with only three measurements, as the following strategy shows.
  Take two copies of a state with covariance matrix $\gamma$ and apply a beam splitter \\\vspace{1ex}

$\left( 
    \begin{array}{cccc}
      \mathbbm{1}&0&0&0\\
      0&{\mathbbm{1} \over \sqrt{2}} & - {\mathbbm{1} \over \sqrt{2}}& 0\\
      0&{\mathbbm{1} \over \sqrt{2}} &{\mathbbm{1} \over \sqrt{2}}& 0\\
      0&0&0&\mathbbm{1}
    \end{array}\right)\left(
    \begin{array}{cccc}
      A & C &0&0\\
      C^T & B&0&0\\
      0&0&A&C\\
      0&0&C^T&B 
    \end{array}\right)\left( 
    \begin{array}{cccc}
      \mathbbm{1}&0&0&0\\
      0&{\mathbbm{1} \over \sqrt{2}} & {\mathbbm{1} \over \sqrt{2}}& 0\\
      0&-{\mathbbm{1} \over \sqrt{2}} &{\mathbbm{1} \over \sqrt{2}}& 0\\
      0&0&0&\mathbbm{1}
    \end{array}\right)$\vspace{1ex}

$  =  {1 \over 2 } \left(
    \begin{array}{cccc}
      2~ A & \sqrt{2} ~C & \sqrt{2}~C & 0\\
      \sqrt{2} ~C^T & B + A & B - A & - \sqrt{2}~ C\\
      \sqrt{2}~ C & B - A & B + A & \sqrt{2}~ C\\
      0 & - \sqrt{2}~ C^T & \sqrt{2}~ C^T & 2~ B 
    \end{array}\right).$\\\vspace{1ex}

We can just measure the three entries of the principal  matrix $B+A$, where 
for the offdiagonal entry of $B + A$ one  has to implement an additional  
${\pi \over 4}-$phase shifter on the second mode. 
With these three measurements we can calculate the determinant without 
knowing what the independent  values of $A$ and $B$ are. We hence saved 
half of the costs with just a little trick.
\end{exa}  

\renewenvironment*{exa}{
   \begin{sloppypar} \hspace{2cm} \stepcounter{exa} 
      \begin{minipage}[t]{10.9cm}  \small \bf{Example} \arabic{exa} : \rm \normalsize
\setlength{\mathindent}{1cm}}{\\ \end{minipage}   \end{sloppypar} }
 
  \chapter{Entanglement Witnesses}
We explain how separability can be formulated  on covariance matrix
level and introduce the advantageous concept of entanglement
witnesses (EW)\index{EW, entanglement witness}.  We propose a scheme to
efficiently estimate the degree of entanglement of an \emph{ unknown }
state by realisable experimental measurements.  Global reference is
the script on Gaussian states, Ref. \cite{RWJE}, to be published.  
\setlength{\mathindent}{2cm}

\section{Separability}
We review some basic properties of covariance matrices and formulate
another separability criterion on the CMs.  First,  we recall the definition
of separability of states and extend it to $M$ parties.
\begin{defi}[Separability]
    A state $\rho$ is separable
with respect to $M$ parties \rm iff \it it can be written (or
approximated, e.g., in trace norm), with probabilities $p_{i} \ge 0 $
and $\sum_{i} p_{i} = 1 $ and proper density matrices $\rho_{i}^{(j)}
$ belonging to the $j$th party, as
    \begin{equation*}
	\rho = \sum_{i} p_{i} \, 
	\rho_{i}^{(1)} \otimes \ldots \otimes \rho_{i}^{(M)},
    \end{equation*} 
    that is, the closed convex hull of the  $M$-product states. 
\end{defi}
Fortunately the separability of states can be formulated on the 
level of their covariance matrices with the following theorem, proved  in Ref. \cite{W2}.
\begin{theorem}[Separability of CMs]\label{24}
  Let $\gamma $ be the covariance matrix of a state $\rho$, which is
  separable with respect to $M$ parties.  Then there exist proper
  covariance matrices $\gamma^{(1)}, \ldots, \gamma^{(M)} $
  corresponding to the $M$ parties, such that
  \begin{equation*}
    \gamma \ge \gamma^{(1)} \oplus \ldots \oplus \gamma^{(M)}.
  \end{equation*}
  Conversely, if this condition is satisfied, then the Gaussian state
  with covariance matrix $\gamma$ is separable.  If the stated 
  relation is fulfilled by a CM we will name the covariance
matrix itself \emph{ separable with respect to the $M$ parties}.
\end{theorem}

\begin{proof}
    For the first statement let the state $\rho$ be decomposed  
    $\rho =  \sum_{i} p_{i} \, \rho^{i}$ where 
    all $\rho^{i}$ are $M$-product states and with probabilities
    $p_{i} \ge 0, \, \sum_{i} p_{i} = 1$. For $\gamma$ covariance matrix 
    and $\vec d$ displacement vector  of the state $\rho$ it is 
    \begin{equation*}
	tr[\rho \, \{\hat R_k - \langle \hat R_k \rangle , 
	\hat R_l - \langle \hat R_{l} \rangle \}_+] 
	= \gamma_{kl} + 2 d_{k} d_{l} = \sum_{i} p_{i} (\gamma_{kl}^{i} + 2 d_{k}^{i} d_{l}^{i}), 
    \end{equation*}  
    where the $\gamma^{i}$ are the block diagonal covariance  matrices 
    (Lemma \ref{productCMs})
    $\gamma^{i} = \bigoplus_{j=1}^{M} (\gamma^{i})^{(j)}$ and  $\vec d^{i}$ 
    the first moments of the $\rho^{i}$. \\

    The difference $ \Delta_{kl} = \gamma_{kl} - \sum_{i} p_{i} \gamma_{kl}^{i} 
    = 2 \sum_{i} p_{i} \, d_{k}^{i} d_{l}^{i} - 2 d_{k} d_{l} $ is a
    positive definite matrix, since for every $\vec v \in \mathbbm{R}^{2N}
    $ it is
    \begin{eqnarray*}
	(\vec v, \Delta \vec v) &=& 2 \sum_{i} p_{i} \, \sum_{kl}  v_{k}
	d_{k}^{i} d_{l}^{i} v_{l} - 2 \sum_{kl}  v_{k} d_{k} d_{l} v_{l} \\
	&=& 2 \sum_{i} p_{i} \, |(\vec v, \vec d^{i})|^{2} - 2 |(\vec v, \vec d)|^{2}
    \end{eqnarray*} 
    and with $\vec d= \sum_{i} p_{i} \vec d^{i} $ we get
    \begin{eqnarray*}
	(\vec v, \Delta \vec v) &=& 2 \sum_{i} p_{i} \, |(\vec v, \vec
	d^{i})|^{2} - 2 \left|\sum_{i} p_{i} (\vec v, \vec d^{i}) \right|^{2} \\
	&\ge& 2 \sum_{i} p_{i} \, |(\vec v, \vec d^{i})|^{2} - 2 \sum_{i}
	p_{i}^{2} |(\vec v, \vec d^{i})|^{2} \\
	&=& 2 \sum_{i} p_{i} \, |(\vec v, \vec d^{i})|^{2} (1- p_{i}) \ge 0.
    \end{eqnarray*}

    Thus   it is indeed possible to find a direct sum of proper 
    covariance matrices $\gamma^{(1)} \oplus \ldots \oplus \gamma^{(M)}$, 
    namely the $\gamma^{(j)} :=\left( \sum_i p_i \gamma^i \right)^{(j)}$, that fulfil 
    the stated relation  since 
    \begin{eqnarray*}
	\gamma &=& \sum_{i} p_{i} \gamma^{i} + \Delta 
	\ge  \sum_{i} p_{i} \left[ \bigoplus_{j=1}^{M} (\gamma^{i})^{(j)}\right] 
	= \bigoplus_{j=1}^{M}\left[ \left( \sum_{i} p_{i} \gamma^{i} \right)^{(j)} \right].
    \end{eqnarray*}

    In order to proof the second part of the theorem let $\gamma$ be a 
    covariance matrix 
    \begin{eqnarray*}
	\gamma 
	=  \bigoplus_{j=1}^{M}  \gamma^{(j)} + \Delta, 
    \end{eqnarray*} with an arbitrary positive  matrix $\Delta$ and
    $\lambda$ a (classical) normalised, about zero centred Gaussian probability
    distribution in $\mathbbm{R}^{2N}$ with covariance matrix ${\Delta
    \over 2}$ :
    \begin{equation*}
	\lambda(\eta) = \mathcal{N} \, e^{- {\eta^{T}  {1\over \Delta} \eta }},
    \end{equation*}  
    let $\omega$ be a density operator describing a Gaussian 
    state with block diagonal covariance matrix $\bigoplus_{j=1}^{M} \gamma^{(j)}$ and 
    vanishing displacement. The expansion of $\omega$ then reads
    \begin{equation*}
	\omega ={1 \over (2\pi)^{N}} \int \dif^{2N}\xi \;  e^{- { \xi^{T} \sigma 
	\left[ \bigoplus_{j=1}^{M} \gamma^{(j)} \right] \sigma^T \xi \over 4} } \;
	\hat W_{\xi}
    \end{equation*}
    where we have used the relation $\Gamma=\sigma \gamma \sigma^T$. We
    construct the state $\rho$ with $\lambda$ and $\omega$ :
    \begin{equation*}
	\rho := \int \dif^{2N} \! \!\eta \,\, \lambda(\eta) \, \hat 
	W^{\dag}_{\eta} \, \omega \, \hat W_{\eta},
    \end{equation*}
    and find that it is a Gaussian state with covariance matrix $\gamma$
    since its characteristic function is of the form :
    \begin{eqnarray*}
	\chi_{\rho}(\xi) &=& tr[\rho \, \hat W_{\xi}] 
	= \int \dif^{2N} \! \!\eta \,\, \lambda(\eta) \, 
	tr[ \hat W^{\dag}_{\eta} \, \omega \, 	\hat W_{\eta} \hat W_{\xi}] \\
	&=& tr[  \omega \,  \hat W_{\xi}] \int \dif^{2N} \! \!\eta \,\, 
	\lambda(\eta) \, e^{- i \eta^{T} \sigma \xi}  \\     
	&=& e^{- {\xi^{T}\sigma [\bigoplus_{j=1}^{M} \gamma^{(j)}] 
	\sigma^{T} \xi \over 4}} \, e^{- {\xi^{T} \sigma \Delta  \sigma^{T} \xi \over 4}}   
	= e^{- {\xi^{T}\sigma \gamma \sigma^{T} \xi \over 4}}.    
    \end{eqnarray*} 

    We now have to proof that  $\rho$ is separable with respect to the $M$ 
    parties. We will need the following lemma to complete the proof:
    \begin{lemma}\label{aconv}
	Let $C$ be a closed convex set, $\rho$ be a $C$ valued function and 
	$g$ 	be a continuous strictly positive function 
	$g: \mathbbm{R} \to \mathbbm{R}^{+} $ with 
	$ \int_{- \infty}^{+ \infty} \dif  t \, g(t) = 1 $. 
	Then the integral
	\begin{equation} \label{a}
	    \int_{- \infty} ^{+\infty} \dif  t \, g(t) \, \rho(t) 
	\end{equation}
	is again an element of $C$.
    \end{lemma}
    \begin{proof}
	Let  $G$ be the integral of $g$, that is $G' = g $, 
	which is strictly monotonically increasing and $G: \mathbbm{R} \to
	[0,1]$.  We observe that the integral Eq.   \eqref{a} can be
	substituted to the form
	\begin{equation*}
	    \int_{- \infty}^{+ \infty} \dif t \, g(t) \, \rho(t) =  \int_{- \infty}^{+ 
	    \infty} \dif t \, G'(t) \, \rho(t) =
	    \int_{0}^{1} \dif x \, \rho(G^{-1}(x)). 
	\end{equation*}
	By definition of the Riemann integral that is equal to
	\begin{equation}\label{sum}
	    \lim_{
	    \begin{subarray}{c} 
		N \to \infty \\ \Delta x_{i}\to 0 
	    \end{subarray}} \quad 
	    \sum_{i=1}^{N} \rho(G^{-1}(x_{i}) ) \Delta x_{i},
	\end{equation}
	with $0 = x_{1} \le x_{2} \le \ldots x_{N+1} = 1 $ and $\Delta x_{i} = 
	x_{i+1} -x_{i} \ge 0 $ and hence  $\sum_{i=1}^{N} \Delta x_{i} = 1$.
	The sum in Eq. \eqref{sum} is  a convex combination 
	and thus an element of C. Since C is closed the limit is reached in
	$C$.
    \end{proof}
    \vspace{2ex}
    In our problem the multi-dimensional function $\lambda $ can be 
    brought to product form by orthogonal diagonalization. 
    Its  factors   are then indeed  continuous,  normalised and strictly positive  functions. 
    Furthermore all  $\omega_{\eta} := \hat W^{\dag}_{\eta} \, \omega \, \hat W_{\eta} $ 
    are product states, because the covariance matrix of $\omega$ is of 
    block diagonal form, and hence separable, while the operators $\hat W_{\eta}$ 
    only displace $\omega$ without affecting its covariance matrix. 
    We  use the previous  lemma  iteratively  and find that the state 
    $ \rho = \int \dif^{2N} \! \!\eta \,\, \lambda(\eta) \,\omega_{\eta} $  is separable.
    This completes the proof of the theorem.
\end{proof}

\begin{defi}
  We  define the closed convex sets \\
  \begin{tabular}{lcp{10.0cm}}
    $\Gamma(\mathbbm{R}^{2N})$\index{$\Gamma(\mathbbm{R}^{2N})$, closed
    convex set of $2N \times 2N$ covariance matrices} &=& Set of covariance
    matrices $\gamma$, elements of $Sym(2N, \mathbbm{R}) $ with the
    properties $\gamma \ge 0$ and $\gamma\pm i \sigma \ge 0$.\\
    $\Gamma_{A|B}(\mathbbm{R}^{2N})
    $\index{$\Gamma_{A"|B}(\mathbbm{R}^{2N})$, closed convex set of $2N
    \times 2N$ covariance matrices, separable with respect to the split
   $A"|B$ } &=& Set of separable covariance matrices $\gamma\in
   \Gamma(\mathbbm{R}^{2N}) $ ( with respect to the split $A|B$), with
   $\gamma \ge \gamma_{A} \oplus \gamma_{B}$ for some $\gamma_{A} \in
   \Gamma(\mathbbm{R}^{2N_{A}}), \gamma_{B} \in
   \Gamma(\mathbbm{R}^{2N_{B}})$ with $N =N_{A} + N_{B}$ .
  \end{tabular}
\end{defi}
\begin{defi}\label{princsub} 
The\emph{ principal submatrices} $A_{r}$ of a square matrix $A$ are the
matrices which result when one neglects one or more columns and rows
always of the same number.
\end{defi}

\begin{lemma}\label{absorb} 
    Let $\gamma$ be a covariance matrix, element of 
    $\Gamma(\mathbbm{R}^{2N})$    $\left( \Gamma_{A|B}(\mathbbm{R}^{2N}) \right)$ , 
    $P$ a positive matrix and 
    $\alpha \ge 1$  then 
    \begin{eqnarray*}
	1. &\,& \gamma' = \gamma + P  \\
	2. &\,& \gamma'' = \alpha \gamma
    \end{eqnarray*}
    are still  elements of $\Gamma(\mathbbm{R}^{2N})$     
    $\left( \Gamma_{A|B}(\mathbbm{R}^{2N})\right)$ and 
    \begin{eqnarray*}
	3.  && \mbox{ all principal submatrices of } \gamma \mbox{ containing }
	 N_r \mbox{ modes}
     \end{eqnarray*}
     are elements of $\Gamma(\mathbbm{R}^{2N_r})$    $\left( \Gamma_{A|B}(\mathbbm{R}^{2N_{r}})\right)$.
 \end{lemma}
 \begin{proof} 
    $1. $ Is obvious.  \\
    $2. \, \gamma''$ is still a real positive matrix. 
    The uncertainty relation is fulfilled for $\alpha \ge 1 $ since 
    $\alpha \gamma + i \sigma 
    = (\alpha -1 ) \gamma + (\gamma + i \sigma) \ge 0 $.  If $\gamma$ was
    separable with $\gamma \ge \gamma_{A} \oplus \gamma_{B}$, $\gamma''$ is
    still separable since $\gamma'' \ge \alpha \gamma_{A} \oplus \alpha
    \gamma_{B}$ with proper covariance matrices $\alpha \gamma_{A}$ and
    $\alpha \gamma_{B}$.  \\
    $3.  $ All principal
    submatrices of a positive matrix $A$ are always positive since for all
    $v \in \mathbbm{R}^{2N} : \langle v | A | v \rangle \ge 0$. 
    Especially for all $v_{r} $ in the reduced spaces $
    \mathbbm{R}^{2N_{r}} $ with $N_{r} \le N$.  Then $ 0 \le \langle v_{r}
    | A | v_{r} \rangle = \langle v_{r} | A_{r} | v_{r} \rangle $ where
    the $A_{r}$ are the principal submatrices of $A$ depending on the
    reduction one chose.  That applies for $\gamma_r$ and also for
    $\gamma_r + i \sigma_{r}$, so that $\gamma_r$ is again a CM on
    $\Gamma(\mathbbm{R}^{2N_r})$.  Additionally the positive matrix
    $\gamma - \gamma_{A} \oplus \gamma_{B}$ stays positive for reductions
    on $N_{r}$ modes.
\end{proof}
\vspace{2ex}
\section{Entanglement Witnesses}

A relatively new method to
distinguish separable from entangled states results from the insight
that the set of covariance matrices and the set of separable CMs form
closed convex subsets of the space of real symmetric $2N \times 2N$
matrices.  The convexity allows to describe these subsets completely
by the intersection of closed half-spaces or equivalently by a family
of linear inequalities.  One family of such inequalities is given by
the so called \emph{ entanglement witnesses.}  They were first
proposed in Ref. \cite{EW}, further investigations can be found
in Ref. \cite{40} on discrete states and on continuous states in
Ref. \cite{RWJE}.  We consider bipartite CVS systems and start with the
definition of appropriate entanglement witnesses.
\begin{defi}\label{entwit}\index{EW, entanglement witness}
We define the following sets of real symmetric $2N \times 2N$-matrices
\begin{eqnarray*} 
   \mathcal{Z}(\mathbbm{R}^{2N})
   \index{$\mathcal{Z}(\mathbbm{R}^{2N})$, set of real symmetric
   matrices,  with $Z \in \mathcal{Z}(\mathbbm{R}^{2N}) \, \iff Z \ge 0 $
   and $str[Z] \ge{1 \over 2} $} &=& \{Z |\, \forall \gamma \in
   \Gamma(\mathbbm{R}^{2N}): \, tr [\gamma \, Z] \ge 1 \}\\
   \mathcal{Z}_{A|B}(\mathbbm{R}^{2N}) 
   \index{$\mathcal{Z}_{A"|B}(\mathbbm{R}^{2N})$, set of real symmetric
   matrices,  with $Z \in \mathcal{Z}_{A"|B}(\mathbbm{R}^{2N}) \, \iff Z \ge 0 $
   and $str[Z_{A}] + str[Z_{B}] \ge{1 \over 2} $} &=& \{Z | \forall
   \gamma \in \Gamma_{A|B}(\mathbbm{R}^{2N}): \, tr [\gamma \, Z] \ge 1, \}
\end{eqnarray*}  
where $A|B$ denotes the split of a bipartite system with $N = N_A +N_B $ modes and  
$\gamma$ covariance matrix. All matrices $Z \in \mathcal{Z}_{A|B}(\mathbbm{R}^{2N}) $ 
will be called \rm entanglement witnesses \it for the split $A|B$.
\end{defi}
The defined sets are again closed convex subsets with $ \mathcal{Z}(\mathbbm{R}^{2N})  \subset 
\mathcal{Z}_{A|B}(\mathbbm{R}^{2N}) $. 
Furthermore it is possible to assign to every
entanglement witness a quadratic Hamiltonian. 
 \begin{defi} Let $Z$ be
   an element of $\mathcal{Z}_{A|B}(\mathbbm{R}^{2N}) $ and $ \hat R =
   (\hat x_{1}, \ldots,  \hat p_{N}).^{T}$
   We define the Hamiltonian 
   \begin{equation*}
     \hat Z[\hat R]\index{$\hat Z[\hat R]$, denotes the  Hamiltonian defined
       by a matrix $Z \in \mathcal{Z}_{A"|B}(\mathbbm{R}^{2N}) $}: = 2
     \sum_{k,l = 1}^{N} Z_{kl} \hat R_k \hat R_l.
   \end{equation*}
 \end{defi}
 
 With the help of the trace we see how the matrix $Z$ is related to the 
 expectation value of the operator $\hat Z$. Let  $\gamma$ be the covariance matrix 
 and $\vec d$ the displacement of a state $\rho$.
 The expectation value of any $\hat Z$ can be expressed in terms of the moments of the state:
 \begin{eqnarray}\label{27}
   \langle \hat Z[\hat R] \rangle_{\rho} &=& tr[\hat
     Z[\hat R] \, \rho] = 2 \sum_{k,l = 1}^{N} Z_{kl} \, tr[\hat R_k \hat R_l
     \, \rho] \nonumber\\
   &=&  \sum_{k,l = 1}^{N} Z_{kl} (\gamma_{kl} + 2 d_{k} d_{l})
   =  tr[Z \gamma] + 2 \vec d^{T} Z \vec d.
 \end{eqnarray}
 
 In the following lemma we see how entanglement
 witnesses detect entangled states. 
 \begin{lemma}\label{gammazab}
   For $\gamma$ covariance matrix of a bipartite system of $N_A + N_B =
   N$ modes, it is:
   \begin{eqnarray*} 
     \gamma \,\mbox{ is separable } &\iff& \forall Z \in \mathcal{Z}_{A|B}(\mathbbm{R}^{2N}) 
      \,  : tr[Z \gamma ] \ge 1
   \end{eqnarray*}
   with respect to the split $A|B$.
 \end{lemma} 
For Gaussian states only the covariance matrix determines the
entanglement properties, thus a separable covariance matrix is
necessary and sufficient for a Gaussian state $\rho$ to be separable.
For non-Gaussian states it is necessary to have a separable covariance
matrix but not sufficient since higher moments could show
entanglement.  The lemma for non-Gaussian states is thus
\setlength{\mathindent}{0cm}
\begin{eqnarray*} 
    \exists Z \in  \mathcal{Z}_{A|B}(\mathbbm{R}^{2N}) : \, tr[Z \gamma ] <
    1&\iff& \gamma \,\mbox{ is entangled  }\\
    &\Rightarrow &\mbox{all CV states having the covariance } \\
    &&\mbox{ matrix } \gamma \mbox{ are entangled.}
\end{eqnarray*}
\begin{proof}
    $\Rightarrow$ is clear from the definition of 
     entanglement witnesses in Def. \ref{entwit}. 
     In order to prove the $\Leftarrow$
    direction we will need an important theorem from the theory of 
    convex sets which can be found in \cite{stanford}  given here without proof:

    \begin{theorem}[Separation of point from convex set]\label{Sopf}
	\hspace{2cm} 
	    
	Let $Y$ be a closed, convex, nonempty set of matrices, $Y \subset
	Sym(2N, \mathbbm{R})\footnote{The theorem holds for more general sets,
	we state only what we will need.}$, let $x \in Sym(2N, \mathbbm{R})
	\backslash Y$.  Then there exists a matrix $M \in Sym(2N, \mathbbm{R})
	, \, M \not = 0 $ with \setlength{\mathindent}{2cm}
	\begin{equation*}
	    tr[M x] < \min_{y \in Y} \, tr[M y].  
	\end{equation*}
    \end{theorem}
    \vspace{2ex}
    The set of separable covariance matrices $ \Gamma_{A|B}(\mathbbm{R}^{2N})$ 
    is such a closed convex set.  In order to prove Lemma \ref{gammazab} we
    assume to have an entangled $x$ and show that there has to exist a $Z
    \in \mathcal{Z}_{A|B}(\mathbbm{R}^{2N}) $ which then detects it and
    gives $tr[Z \, x] < 1 $.
    
\begin{figure}[h]
    \hspace*{3cm}
{\includegraphics[width=0.5\textwidth]{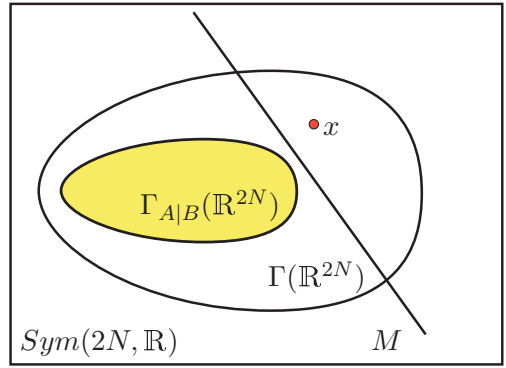}}

\vspace*{-3.3cm}

\hspace*{3.0cm}\begin{minipage}[b]{6.5cm}
\hspace*{4.2cm} $x$ \\ \vspace{0.1cm}

\hspace*{1.9cm}$\Gamma_{A|B}(\mathbbm{R}^{2N})$\\ 
                                   
\hspace{3.6cm}\vspace{0.3cm}$\Gamma(\mathbbm{R}^{2N})$ \\ \vspace{-0.4cm}

\hspace{0.3cm}$Sym(2N,\mathbbm{R})$ \hspace{2.3cm} $M$
\end{minipage}

\caption{\small Separation of point from convex set}\label{hyper}
\end{figure}
      
    With Theorem \ref{Sopf} we find a matrix $M \in Sym(2N, \mathbbm{R})$
    so that \setlength{\mathindent}{0cm}
    \begin{equation}\label{negc}
	tr[M \, x ] < c = \min_{\gamma' \in
    \Gamma_{A|B}(\mathbbm{R}^{2N})} \, tr[M \, \gamma'] \le tr[M \,
    \gamma] \quad \forall \, \gamma \in \Gamma_{A|B}(\mathbbm{R}^{2N}).
    \end{equation}
     \\
    Let us assume that the real number $c$ is strictly positive. 
    Then we get $tr[{M \over c} \, x] < 1 \le tr[ {M \over c} \, \gamma ]
    \quad \forall \,  \gamma \, \Gamma_{A|B}(\mathbbm{R}^{2N})$ 
    but if this is true $ {M \over c} = Z$ has to 
    be an element of $\mathcal{Z}_{A|B}(\mathbbm{R}^{2N})  $, since it is a real 
    symmetric matrix with the defining condition in Def. \ref{entwit}.
    We have found that if we take an $x \not \in \Gamma_{A|B}(\mathbbm{R}^{2N})$ 
    we will find an $Z \in \mathcal{Z}_{A|B}(\mathbbm{R}^{2N})  $ which gives $tr[ Z 
    \,x ] < 1 $ or equivalently $ \forall Z \in \mathcal{Z}_{A|B}(\mathbbm{R}^{2N}) : tr[ Z 
    \,\gamma ] \ge  1  \Rightarrow \gamma \in 
    \Gamma_{A|B}(\mathbbm{R}^{2N})$.     \\
    
    We still have to discuss the case $ c \le 0$.  Let us show that $M$ is
    positive by contradiction.  Suppose $M$ were not positive and had an
    eigenvalue $m_0 < 0$ with the corresponding eigenvector $\mu \in
    \mathbbm{R}^{2N}$.  We construct the matrix $\gamma '_{kl} =
    \gamma_{kl} + \lambda \mu_k \mu_l $ for an arbitrary $\lambda \ge 0$
    and a separable matrix $\gamma$, which by Lemma \ref{absorb} is itself
    an element of $\Gamma_{A|B}(\mathbbm{R}^{2N})$.
    We find that $tr [M \gamma ']= tr[M \gamma ] + \lambda\, m_0 \, \mu^2 \to - \infty$ 
    for $\lambda \to \infty$. With this construction we see that $M$ cannot have any eigenvalue 
    smaller than zero because Eq. \eqref{negc} has to hold and 
    $tr[M \, \gamma'] = tr[M \, \gamma] + \lambda  m_{0}  \mu^{2} $ would 
    reach values smaller than $tr[M x] $ for sufficiently large $\lambda$.  Hence $M$
    is positive.\\
    But then the following equation holds for every $\gamma
    \in \Gamma_{A|B}(\mathbbm{R}^{2N})$: \setlength{\mathindent}{2cm}
    \begin{eqnarray*}
      tr[M \gamma] &=& tr[M \, S \, \gamma^{WNF} \, S^{T} ] = 
      tr[ S^{T} \, M \, S \, \gamma^{WNF} ] \\
      &=& tr[ S^{T} \, M \, S \, (\gamma^{WNF} -\mathbbm{1})] +  
      tr[S^{T} \, M \, S ]\\
      &>& 0,
    \end{eqnarray*}
    since the first addend is again the trace of a (semi-)positive matrix,
    see Eq. \eqref{sym>1} and Lemma \ref{positiveGamma}, while the second
    addend is the trace of a positive matrix which is only zero if all
    eigenvalues of it were zero that is the matrix $ S^{T} \, M \, S $ and
    hence the matrix $M$ itself was the zero matrix.  But Theorem
    \ref{Sopf} excluded the $M = 0 $ case.  We conclude that there exists
    at least one nonzero $M \in Sym(2N, \mathbbm{R}) $ so that for all
    $\gamma \in \Gamma_{A|B}(\mathbbm{R}^{2N}) $ we find a strictly
    positive $c$.  This completes the proof of Lemma \ref{gammazab}.
\end{proof}

\begin{defi}
    The \emph{symplectic trace} (denoted as $str[A]$\index{$str[A]$,
    symplectic trace of a positive, symmetric, even dimensional matrix $A$
    } ) of an even dimensional, positive, symmetric matrix $A$ is defined
    as the sum of the symplectic eigenvalues (counted once).
\end{defi}

\begin{lemma}
Let $A = \left(  
\begin{array}{cc}
A_1 & A_2 \\ 
A_2^T & A_3 
\end{array} \right)$ denote a real positive $2N \times 2N$  matrix, then the symplectic trace 
has the following  properties: \setlength{\mathindent}{2cm}
\begin{eqnarray*}
str[S \, A \, S^T] &=& str[A] \quad \forall S \in Sp(2N,\mathbbm{R})\quad \mbox{invariance }\\
str[p A] &=& p \, str[A] \quad p \ge 0 \quad \mbox{multiplicity} \\
str[A] &\le& str[A_1] + str [A_3]
\end{eqnarray*}
\end{lemma}
\begin{proof}
  The invariance is clear since symplectic transformations preserve all
  symplectic eigenvalues, additionally the symplectic eigenvalues just
  all multiply with the factor $p$ so that multiplicity holds. The last 
  property can be derived as a consequence of Theorem \ref{47}.
\end{proof}\vspace{2ex}

But given a real symmetric $2N \times 2N$ matrix, is it
an entanglement witness or not?  The following theorem answers the
question providing a check, based on the symplectic eigenvalues the
given matrix possesses.  
\begin{theorem}\label{47} Let $Z$ be a real
  symmetric $2N \times 2N$ matrix on a phase space of $N = N_A +N_B$
  modes.  Let $Z_A$ and $Z_B$ denote the principal submatrices of $Z$
  belonging to the subsystems $A$ and $B $ respectively.  Then :\\
  $\begin{array}{lcl}
    \qquad \qquad 1.\, Z \in \mathcal{Z}(\mathbbm{R}^{2N}) 
    \index{$\mathcal{Z}(\mathbbm{R}^{2N})$,   set of real symmetric
      matrices,  with $Z \in \mathcal{Z}(\mathbbm{R}^{2N}) \, \iff Z \ge 0 $
      and $str[Z] \ge{1 \over 2} $}   
    &\iff& Z \ge 0 \; \mbox{and } \:
    str[Z] \ge {1 \over 2}\\
    \qquad \qquad  2.\,  Z \in \mathcal{Z}_{A|B}(\mathbbm{R}^{2N})
    \index{$\mathcal{Z}_{A"|B}(\mathbbm{R}^{2N})$, set of real symmetric
      matrices,  with $Z \in \mathcal{Z}_{A"|B}(\mathbbm{R}^{2N}) \, \iff Z \ge 0 $
      and $str[Z_{A}] + str[Z_{B}] \ge{1 \over 2} $}  
    &\iff &  Z \ge 0 \; \mbox{and } \: str [Z_A] + str[Z_B] \ge {1 \over 2}. 
\end{array}$ 
\end{theorem}
Note that $str [Z] \ge{1 \over 2}$ is stronger than $str[Z_A] + str[B] \ge {1 \over 2}$.\\

\begin{proof} (following reference \cite{RWJE})\\
$1.  \Rightarrow$ : With the same argument as in the proof of Lemma
\ref{gammazab} we find that $Z \in \mathcal{Z}(\mathbbm{R}^{2N}) $ has
to be positive.  Since $Z \ge 0$ we can find a symplectic
transformation to bring $Z$ to Williamson normal form:
\setlength{\mathindent}{2cm}
\begin{equation*}
Z^{WNF}= S_{z}\,Z\,S_{z}^T= Z_0 \oplus \bigoplus_i \left(\begin{array}{cc}
z_i&0\\0&z_i \end{array}\right)
\end{equation*}
where   $Z_0$ is a matrix with determinant zero and zero symplectic eigenvalues.  
We assume that the particular $Z$ has a vanishing $Z_{0}$ part an go on.  The
symplectic trace of $Z$ is now easily calculated: $ str[Z] =
str[Z^{WNF}] = \sum_i z_i$.  Furthermore we have with the basis
transformation $\hat R' \mapsto \hat R = S_{z} \hat R'$
\begin{eqnarray}\ tr[\hat Z[\hat R'] \rho' ]&=& tr[\hat Z^{WNF}[\hat R]
    \rho ]= 2 \sum_{i} \left(\langle Q_i^2 \rangle_{\rho} + \langle
    P_i^2 \rangle_{\rho} \right) z_i \nonumber \\
    \label{haros}
    &=&  2 \sum_{i} \left( 2 \langle \hat n_i \rangle_{\rho} + 1 \right) z_i \,
    \ge 2 \, str[Z], 
\end{eqnarray} where $\hat n_i = \hat a_i^{\dag} \hat a_i $.
The right hand side is the expectation value of the
Hamiltonian of harmonic oscillators which is bounded from below by the
ground state energy reached when $\rho$ in Eq.   \eqref{haros} was the
ground state.

On the other hand the expectation value of $\hat Z$ is related to the
state's moments via $ tr[\hat Z[\hat R'] \rho'] = tr [Z \gamma'] + 2
\, \vec d'^{T} \,  Z  \, \vec d' \ge tr[Z \gamma'] \ge 1 $ for all $\gamma' \in
\Gamma(\mathbbm{R}^{2N}), \, \rho' \in S(\mathcal{H})$ by Def.  
\ref{entwit}.  Since the  boundary in Eq.   \eqref{haros} is reached,
we conclude the stated property of $Z$, namely $ 2 \, str[Z] \ge 1 $. 
Also for all $Z'$ identical to $Z$ but with a nonzero $Z_{0}$ part it
follows $str[Z'] = str[Z] \ge {1 \over 2}$.  \vspace{1.5ex}

$1. \Leftarrow$ :  
  For all $\gamma \in \Gamma (\mathbbm{R}^{2N})$ and with $Z^{WNF}$ again Williamson
  normal form of $Z$, including a $Z_{0}$ part, we calculate:
\begin{eqnarray*}
tr[Z \, \gamma] &=& tr[S^T \, Z^{WNF} S \gamma] 
= tr[ Z_0 \gamma'_0] + \sum_i  tr[z_i \mathbbm{1} \gamma'_i]\\
&\ge&   \sum_i z_i (\lambda_i^1 + \lambda_i^2),
\end{eqnarray*}
where $\gamma'_0$ and $\gamma'_i$ are the principal submatrices of
$\gamma'$.  $\gamma'_{0}$ has the size of $Z_{0}$ while all
$\gamma'_{i}$ are $2 \times 2$ covariance matrices with the positive
eigenvalues $\lambda_{i}^{1}$ and $\lambda_{i}^{2}$ which fulfil
$\lambda_i^1 \cdot \lambda_i^2 \ge 1 $ for every submatrix since the
determinant of $\gamma'_{i}$ is necessarily greater or equal to one. 
And since for every $\lambda \ge 0$ the relation $0 \le
(\sqrt{\lambda} - {1 \over \sqrt{\lambda}})^2 = \lambda + {1 \over
\lambda} - 2 $ holds, it follows that
\begin{eqnarray*}
\forall \gamma \in \Gamma(\mathbbm{R}^{2N}): \, tr[Z \, \gamma] \ge \sum_i z_i
(\lambda_i^1 + {1 \over \lambda_i^1}) \ge 2 \sum_i z_i = 2 \, str[Z] \ge
1.
\end{eqnarray*} 
Thus $Z$ is an element of $\mathcal{Z}(\mathbbm{R}^{2N}) $ as stated. 
\vspace{1.5ex}

$2. \Rightarrow$ : For $Z \in \mathcal{Z}_{A|B}(\mathbbm{R}^{2N}) $ we run the above argument 
(proof of Lemma \ref{gammazab}) for the positivity.  For a separable
$\gamma$ we have the useful condition proved in Theorem \ref{24}:
$\gamma \ge \gamma_A \oplus \gamma_B $ for some covariance matrices
$\gamma_A$ and $\gamma_B$.  Thus
\begin{equation*}
    tr[Z \gamma] \ge tr[Z_A \gamma_A ] + tr[Z_B \gamma_B] \ge 0
\end{equation*}
and with Eq. \eqref{27} and $\rho_{A}, \rho_{B}$
the Gaussian states with covariance matrix $\gamma_{A}, \gamma_{B}$
respectively and vanishing displacement, follows
\begin{eqnarray*}
    tr[ Z \gamma] & = & tr[\hat Z_{A}[\hat R] \rho_{A}] 
    + tr[\hat Z_{B}[\hat R] \rho_{B}]. 
\end{eqnarray*}

Additionally it is $Z_{A}, Z_{B} \ge 0$ and hence $tr[\hat Z_{A}[\hat
R] \rho_{A}] \ge 2 \, str[Z_{A}] $ and $tr[\hat Z_{B}[\hat R] \rho_{B}] \ge 2 \,
str[Z_{B}] $ from above, so we get
\begin{eqnarray*}
    tr[ Z \gamma]  &\ge & 2 \,   str[Z_{A}] + 2 \, str[Z_{B}] 
\end{eqnarray*}
reached again when both subsystems are in ground state.  
By the precondition we also have $tr[Z \gamma] \ge
1$ for all $\gamma \in \Gamma_{A|B}(\mathbbm{R}^{2N})$ so that the
condition on the symplectic trace of the entanglement witness is
\begin{equation*}
    2 str[Z_{A}] + 2 str[Z_{B}] \ge 1. 
\end{equation*}

$2. \Leftarrow$ : 
We use the same idea as for $1.  \Leftarrow $. For all $\gamma \in \Gamma_{A|B}
(\mathbbm{R}^{2N})$ and $Z^{WNF}_{A} $ and $Z^{WNF}_{ B}$ the
symplectic diagonalized submatrices of $Z$ it is :
\begin{eqnarray*}
tr[Z \, \gamma] &\ge& tr[Z_A \gamma_A ] + tr[Z_B \gamma_B] 
= tr[Z^{WNF}_A \gamma'_A ] + tr[Z^{WNF}_B \gamma'_B] \\ 
&=& tr[ Z^A_0 \gamma'^A_0] + \sum_{i\in A}  tr[z^A_i \, \mathbbm{1}\, \gamma'^A_i]
 \\
&& \qquad \qquad \qquad + tr[ Z^B_0 \gamma'^B_0] + \sum_{i\in B} tr[z^B_i \, \mathbbm{1}\,
\gamma'^B_i] \\
&\ge & 2 \, \sum_{i\in A} z^A_i + 2 \, \sum_{i\in B} z^B_i 
= 2 \, str[Z_A] + 2 \, str[Z_B] \, \ge \, 1, 
\end{eqnarray*} 
finally completing the proof.
\end{proof}\vspace{2ex}

We have now a powerful tool to decide whether a given matrix is an
entanglement witness for the given split $A|B$ or not.  And with them
we can restrict the sets a given covariance matrix is an element of. 
The picture shows how two entanglement witnesses $Z_{1}$ and $Z_{2}$
divide the set of covariance matrices in $\mathbbm{R}^{2N}$ into four
subsets.  \begin{figure}[h]
\hspace*{1.0cm} {\includegraphics[width=0.5\textwidth]{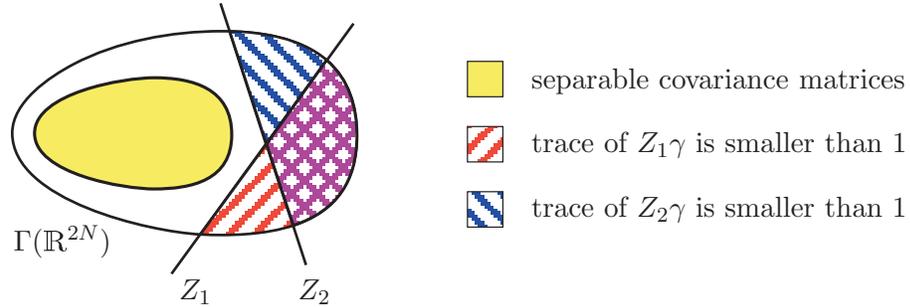}}

\vspace{-3.0cm}\hspace{8.0cm}
\begin{minipage}[t]{6.5cm}
separable covariance matrices
\\ \vspace{-0.1cm}

trace of $Z_{1} \gamma $ is smaller than $1$
\\\vspace{-0.1cm}
 
trace of $Z_{2} \gamma $ is smaller than $1$
\end{minipage}

\hspace{1.2cm}$\Gamma(\mathbbm{R}^{2N}) $\\

\vspace{-0.3cm}\hspace{3.4cm}$Z_{1} \qquad\quad Z_{2}$ 

\vspace{-0.3cm}\hspace*{1cm}{\caption{\small Entanglement witnesses
dividing $\Gamma(\mathbbm{R}^{2N})$}}

\vspace{2ex}
\end{figure} If the trace of the given CM $\gamma $ together
with the entanglement witness $Z_{1}$ ($Z_{2}$) was smaller than one,
then  $\gamma$ belongs to the red-lined (blue-lined) set
respectively.  If both measurements gave  values smaller than one, it
is in the squared set.  In those cases all states having the
covariance matrix $\gamma$ are entangled.  If, on the other hand, both
entanglement witnesses gave values greater than one, we can not say
whether the given matrix is entangled or not. But from the geometric
representation we guess that \emph{ optimal entanglement witnesses }
going exactly through the edge between separable with entangled CMs,
exist and indeed they do.  We will not introduce them here but only
give references \cite{42}.

 \section{Estimation of the Degree of  Entanglement}

We now come to the main results of this thesis. We have so far learned a
lot about Gaussian states, their covariance matrices and their
entanglement properties. When producing a Gaussian two-mode state 
in the lab usually one wants to make sure 
that the outcoming state behind the particular setup is really the
desired state.  Especially it is often necessary to check whether the
state has the expected entanglement properties used for quantum
cryptography or other purposes. For that reason one is interested in a simple measurement scheme 
telling the experimenter immediately what  degree of entanglement the produced state has.
In the third chapter  we learned that for Gaussian states entanglement, 
quantified in the logarithmic negativity, can easily be calculated using 
only  the covariance matrices of the states. So it should be possible 
to measure  some properties of a CM  and give an estimate of 
the degree of entanglement for the  Gaussian states having 
this particular covariance matrix.
 In the previous section we introduced the concept of entanglement
witnesses and saw that they are very practical and provide a natural way to distinguish 
separable from entangled states. But one could do even better. It is not only possible to 
decide whether a state is entangled or not but with practicable measurements one can estimate 
a lower bound of entanglement of a given unknown Gaussian state. 
We show how this can be done and start off with the definition of $p -$separability.
\begin{defi}
For any $\gamma \in \Gamma(\mathbbm{R}^{2N}) $ covariance matrix of a
bipartite system $A|B$
and any $p \in (0,1]$ we define \setlength{\mathindent}{2cm}
\begin{equation*}
\gamma \mbox{ is } p - \mbox{separable}\index{$p-$separable} \iff
{\gamma \over p } \mbox{ is separable}
\end{equation*}
with respect to the split $ A|B$ respectively. The sets of $p-$separable covariance matrices 
are again closed, convex subsets of $\Gamma(\mathbbm{R}^{2N})$ and 
will be denoted by $\Gamma^p_{A|B}(\mathbbm{R}^{2N})$.
\end{defi}
The definition is sensible since $\, {\gamma \over p} $ with Lemma
\ref{absorb} and $\alpha = {1 \over p} \ge1$ \,still has covariance
matrix properties.
Note, that by definition \emph{separability }is equal to
$1-$\emph{separability} and for $p_1 \ge p_2 $ follows:
$\gamma \mbox{ is } p_1 - \mbox{separable} \Rightarrow \gamma \mbox{
is } p_2 - \mbox{separable}$ and the sets $\Gamma^p_{A|B}(\mathbbm{R}^{2N})$ 
include each  other like the shells of an  onion 
$\Gamma^1_{A|B}(\mathbbm{R}^{2N}) \subset \Gamma^{0.5}_{A|B}(\mathbbm{R}^{2N}) 
\subset \Gamma^{0.2}_{A|B}(\mathbbm{R}^{2N}) \subset \Gamma(\mathbbm{R}^{2N}),$ 
as shown in the figure.

 \vspace{-0ex}
\begin{figure}[h] \hspace*{2.3cm}
{\includegraphics[width=0.5\textwidth]{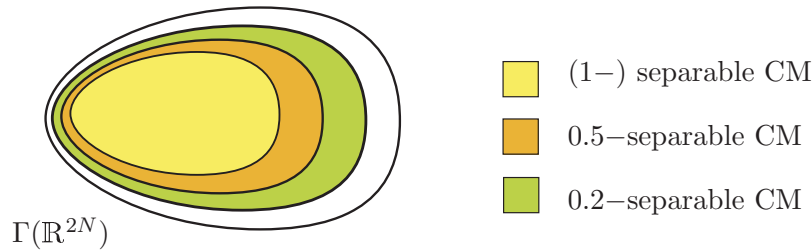}}

\vspace{-2.5cm}\hspace{9.3cm}
\begin{minipage}[t]{6.5cm}
$(1-)$ separable CM 
\\\vspace{-0.1cm}

$0.5 -$separable CM 
\\\vspace{-0.1cm}

$0.2 -$separable CM
\end{minipage}
\vspace{-0.2cm}
\hspace{2cm}$\Gamma(\mathbbm{R}^{2N})$

\vspace*{- 0.3 cm}\hspace*{0cm}\hfill\caption{\small
$p-$Separability}\label{psep} \end{figure}

\newpage
Using the definition of $p-$separability  Theorem \ref{24} can be reformulated to
\begin{theorem}[$p-$Separability of CMs]\label{58}
  Let $\gamma $ be a,  with respect to the two  parties $A$ and $B$,
  $p-$separable covariance matrix, i.e. $\gamma \in  \Gamma^p_{A|B}(\mathbbm{R}^{2N})$.  
  Then there exist proper  covariance matrices 
$\gamma^{A} \in \Gamma(\mathbbm{R}^{2N_A})$ and $ \gamma^{B} \in \Gamma(\mathbbm{R}^{2N_B})$
  corresponding to the  parties, such that
\setlength{\mathindent}{2cm}
  \begin{equation*}
    \gamma \ge p \,\, (\gamma^{A} \oplus  \gamma^{A}).
  \end{equation*}
\end{theorem}
And finally, $p -$separability of covariance matrices can easily be
expressed with entanglement witnesses as the following lemma shows:
\begin{lemma} \label{psep1}
  For $\gamma $ CM of a bipartite system, $p \in (0,1]$  and $Z$ being an entanglement
    witnesses, we find the relation:\setlength{\mathindent}{2cm}
    \begin{eqnarray*}
       \gamma \mbox{ is } p -\mbox{separable } (A|B) &\iff& \, \forall Z \in
      \mathcal{Z}_{A|B}(\mathbbm{R}^{2N}) : \, tr[Z \gamma] \ge p \\
      \negthinspace  \negthinspace  \negthinspace  \negthinspace  \negthinspace  \negthinspace
      \negthinspace  \negthinspace  \negthinspace  \negthinspace  \negthinspace  \negthinspace 
      \negthinspace  \negthinspace  \negthinspace  \negthinspace  \negthinspace  \negthinspace 
      \negthinspace  \negthinspace  \negthinspace  \negthinspace  \negthinspace  \negthinspace
      \mbox{ or } \quad       \gamma \mbox{ is not } p -\mbox{separable } (A|B)   
      &\iff& \, \exists Z \in  \mathcal{Z}_{A|B}(\mathbbm{R}^{2N}) : \, tr[Z \gamma] < p.
    \end{eqnarray*} 
\end{lemma}
\begin{proof}
    With the definition of $p-$separability, Theorem \ref{24}  and Lemma \ref{gammazab} the
    stated relations follow immediately.
\end{proof}\vspace{2.0ex}

For $p < 1$, $p-$separability is  weaker than the usual  separability,
but when blown up with a factor ${1 \over p}$ they would reach the set of separable 
covariance matrices $\Gamma_{A|B}(\mathbbm{R}^{2N})$. In this sense, 
the smaller the number  $p$ is, the more entangled can the covariance matrix be.
We observe, that  when measuring an entanglement witness $Z$, the outcome 
$p =tr [ Z \gamma]$ does not only distinguish separable 
($p \ge 1$ for all witnesses) from entangled ($ p < 1$ for one particular 
witness) covariance matrices,  furthermore it orders the entangled CMs in
sets of $p-$separable CMs.\\

 An experimenter who produces Gaussian states 
with a certain degree of entanglement is naturally curious if he succeeded 
in doing so.  When he  measures an entanglement witness $Z_0$ on a 
given unknown Gaussian state $\rho$ having a vanishing displacement 
and covariance matrix $\gamma$ the outcome is given by 
\setlength{\mathindent}{2cm}
\begin{equation*}
  m = \langle \hat Z_0 [\hat R\,] \rangle_{\rho} = tr [ Z_0 \gamma] 
< p: = m + \epsilon ,\quad  \forall \, \epsilon > 0.
\end{equation*}
The measurement outcome tells him now immediately that the state is \emph{ not } 
$p-$separable  for all  $\epsilon  > 0$.
But maybe this is not the best estimate he could do. It is possible 
that with a different entanglement witness a smaller outcome $m$ 
could be measured and the state would be even more entangled.
Since he produced the Gaussian state $\rho$, it is not completely unknown to him;
he actually has a strong conjecture what the state will
be and can use his knowledge to find an \emph{ mini\-mal entanglement
witness} for the covariance matrix associated to the state $\rho$. 
In contrast to optimal entanglement witnesses it
is not the witness going exactly through the edge between entangled
and separable states, but it is the one that gives the best estimate
of the degree of entanglement the supposed state has. Obviously the 
minimal entanglement witnesses depend on the covariance matrix 
for whom they are minimal. We will understand the minimal entanglement witnesses
$Z_{min}(\gamma)$ to a given bipartite  CM $\gamma$  as being those witnesses 
$Z_{min}(\gamma) \in \mathcal{Z}_{A|B}(\mathbbm{R}^{2N}), $ which give  
\begin{equation*}
 tr [Z_{min}(\gamma)\,\gamma ] =  p_{min}(\gamma),
\end{equation*}
  while all other 
EWs $Z \in \mathcal{Z}_{A|B}(\mathbbm{R}^{2N})$ give $tr [Z \gamma ] > \,p_{min}$.  
The number $p_{min}(\gamma)$ is  then the  smallest value of the outcomes $p$, 
one could measure for the CM $\gamma$.

\begin{figure}[h]

\hspace*{3.3cm}{\includegraphics[width=0.5\textwidth]{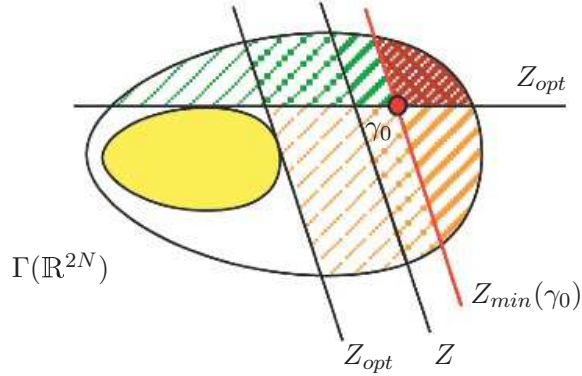}}

\vspace{-3.7cm}\hspace{9.2cm}\begin{minipage}[h]{4.0cm}$ Z_{opt}$\\

\vspace{-0.3cm}\hspace{-2.0cm}$\gamma_{0}$ \end{minipage}\\

\vspace{1.6cm}

\hspace{8.6cm}$Z_{min}(\gamma_{0})$\\

\vspace{-1.3cm}\hspace{2.5cm}$\Gamma(\mathbbm{R}^{2N})$\\

\vspace{0.2cm}

\hspace{6.9cm}$Z_{opt}$ \hspace{0.4cm}$Z$

\vspace*{- 0.3 cm}\hspace*{-2cm}\caption{\small Optimal and minimal
entanglement witnesses}\label{schnitt} \end{figure}\vspace{-2ex}

\vspace{2ex}
In Fig. \ref{schnitt} the coloured areas are those
where entanglement witnesses gave values smaller than one, so the CMs
in these areas must be entangled.  The minimal  EW $Z_{min}(\gamma_{0})$ gave
$tr[Z_{min}(\gamma_{0}) \, \gamma_{0}] = p_{min}(\gamma_{0})$ for the covariance
matrix $\gamma_{0}$, symbolised by the red dot.\vspace{2ex}

We see  that it should be  possible to give lower bounds for the degree 
of  entanglement of a Gaussian state  using entanglement witnesses.
To  estimate the degree of entanglement quantified in the logarithmic
negativity,  we will need that the PPT-criterion is necessary and
sufficient, which is the case in $1_{A} \times N_{B}$ mode systems. 
We learned in Lemma \ref{25} that the logarithmic negativity of a
Gaussian state can be calculated from its covariance matrix:
\begin{equation}\label{logneg1}
E_{\mathcal{N}} (\gamma) = - \sum_i \min (\ln \gamma_i^{T_A}, 0).
\end{equation}
In the following lemma  all the things we learned are merged together 
and we find how the outcomes of the measurements of entanglement 
witnesses bind the logarithmic negativity of the measured state.

\begin{lemma}\label{pse} 
    In case the PPT-criterion is necessary and sufficient:\\
    If an experimental
    setup measuring the trace of an entanglement witness $Z$ with a 
    covariance matrix $\gamma$ belonging to an unknown Gaussian $ N-$ mode state
    gave the outcome $m \in (0,1)$, the logarithmic negativity of
    the state is bounded from below by
    \begin{eqnarray*}
	E_{\mathcal{N}}(\gamma) &\ge & \ln {1 \over m}.
    \end{eqnarray*}
     In the two-mode case, where only one symplectic eigenvalue of
     $\gamma^{T_{A}}$ can be smaller than one, the minimal witness
     $Z_{min}$ giving the smallest possible value $m_{min}$ gives exactly
     the logarithmic negativity the state possesses.
 \end{lemma}
 \begin{proof} 
     With the symplectic eigenvalues of the partially transposed covariance
     matrix $\gamma^{T_{A}}$ defined in Eq. \eqref{gTA} and Lemma \ref{psep1}
     we see that forall $p \in (0,1]$ \setlength{\mathindent}{0.0cm}
     \begin{eqnarray*} 
	 \forall i: \gamma_i^{T_A} \ge 1 \! &\iff&\! \gamma \mbox{ is separable
	 } \quad \mbox{(PPT!),} \\
	 \forall i: \gamma_i^{T_A} \ge p\! &\iff&\! \gamma \mbox{ is } p-
	 \mbox{separable } \! \iff \forall Z \in
	 \mathcal{Z}_{A|B}(\mathbbm{R}^{2N}) : tr[Z \gamma] \ge p,\\
	 \exists i: \gamma_i^{T_A} < p\! &\iff&\! \gamma \mbox{ is not } p-
	 \mbox{separable} \!  \iff
	 \exists Z \in \mathcal{Z}_{A|B}(\mathbbm{R}^{2N}) : tr[Z \gamma] < p.
     \end{eqnarray*} \setlength{\mathindent}{2cm}\vspace{-1ex}

     From the measurement we     know that there exists an entanglement
     witness giving the result $m$. 
     We hence know that there exists at least one symplectic eigenvalue
     $\gamma^{T_{A}}_{1}$ of $\gamma^{T_{A}}$ which is smaller than $p = m
     + \epsilon$ for all $\epsilon > 0$.  Thus we can give a nontrivial lower bound
     for the logarithmic negativity of the state.  
     \begin{eqnarray*}
	 E_{\mathcal{N}}(\gamma) 
	 &=& - \sum_{i=2}^{N} \min (\ln \gamma_i^{T_A}, 0) 
	 - \min(\ln \gamma_{1}^{T_{A}}, 0)\\
	 &>& - \ln p = \ln {1 \over m + \epsilon} 
     \end{eqnarray*}
    and since this inequality is true for $p = m +\epsilon, \;   \forall
    \epsilon > 0$, the proof of the first statement is complete.  For the
    second part we assume a symplectic eigenvalue $\gamma_{1}^{T_{A}}$
    smaller than one and we get
         \begin{eqnarray*}
	 E_{\mathcal{N}}(\gamma) 
	 &=& -  \min (\ln \gamma_2^{T_A}, 0) - \min(\ln \gamma_{1}^{T_{A}},
	 0)\\
	 &=& 0 +\ln {1 \over \gamma_{1}^{T_{A}}}.
     \end{eqnarray*}
     Since the outcome of our measurement gave the
     value $m_{min}$
     for the given covariance matrix, we conclude that the symplectic
     eigenvalue $\gamma_{1}^{T_{A}}$ is smaller than or equal to $m_{min}$. 
     On the other hand by definition of the minimal measurement outcome
     $m_{min}$ it is for all $Z \in Z_{A|B}(\mathbbm{R}^{2N}) : \, tr[Z
     \gamma] \ge m_{min}$ and hence for all symplectic eigenvalues
     $\gamma_{i}^{T_{A}} \ge m_{min}$.  Hence $\gamma_{1}^{T_{A}}$ is equal
     to $m_{min}$ and the logarithmic negativity is given as stated.
 \end{proof}\vspace{2ex}

The lemma provides an easy method to estimate a lower bound of
entanglement by measuring different entanglement witnesses.
The estimated  logarithmic negativity becomes larger 
the smaller the minimal measurement outcome was, respecting the ordering 
induced by the sets of  $p-$separable covariance matrices.  
Entanglement witnesses also detect bound entangled states, being
those entangled states having a positive partial transpose. If existent, they
destroy the sufficiency of the PPT-criterion and measured in terms of the
logarithmic negativity they have a vanishing degree of entanglement. 
Unless bound entangled states exist 
the logarithmic negativity  is strictly positive for entangled 
covariance matrices. A possible goal for future research could be a
computable entanglement measure also  respecting the $p-$ordering 
but giving nonzero values for bound entangled states.\\

Summarising our achievements, we see  that an experimenter, wanting 
to know the degree of entanglement his produced Gaussian state possesses, 
simply  has to measure some entanglement witnesses and the results 
instantly tell him how much the state is minimally entangled. With little 
additional  knowledge about the state it is furthermore  possible 
to optimise the measurements, trying only minimal entanglement witnesses.
In the following example we see  that sometimes  even the knowledge of 
the structure of the covariance matrix is sufficient to determine minimal 
entanglement witnesses  for it. \vspace{2ex}

\begin{exa}\label{exa20}
    Let $\gamma $ be of the special form
    \begin{equation*}
	\gamma = \left(
	\begin{array}{cccc}
	    a&0&b&0\\0&a&0&-b\\b&0&a&0\\0&-b&0&a
	\end{array}\right) \quad \mbox{  with } a\ge 1\mbox{ and } 0 \le b\le a.
    \end{equation*} 
     Then the symplectic eigenvalues of the $\gamma^{T_{A}}$ are
     $\gamma^{T_{A}}_{1} = a - b$ and $\gamma^{T_{A}}_{2} = a + b$.  The
     logarithmic negativity of $ \gamma$ is then given by
     $E_{\mathcal{N}}(\gamma) = \ln {1 \over a -b }$ and the minimal
     measurement outcome for an entanglement witness is $m_{min} = a - b$. 
     An entanglement witness which allows to measure the smallest value
     $m_{min}$ is given by:
     \begin{equation*}
	Z_{min}=\frac{1}{4} \left(
	\begin{array}{cccc}
	    1&0&-1&0\\ 0&1&0&1\\-1&0&1&0\\ 0&1&0&1
	\end{array}
	\right) \quad \mbox{ with } tr[Z_{min} \gamma] = a - b.
    \end{equation*}
    A small check that $Z_{min}$ is really an entanglement witness in
    $\mathcal{Z}_{A|B}(\mathbbm{R}^{2N})$: \, $ Z_{min} $ is positive since its
    eigenvalues are with Eq. \eqref{99} given by  $\lambda_{1,3} = {a + b \over 4} \ge 0$ and
    $\lambda_{2,4} = { a -b \over 4} \ge 0.$ 
    The symplectic eigenvalue of $Z_{A} = Z_{B} = {1 \over 4} \left(
    \begin{array}{cc}
	1&0\\ 0&1
    \end{array}\right)$ is $z_{1} = {1 \over 4}$ hence the condition
    on the symplectic trace of the submatrices is fulfilled since  
    $str[Z_{A}] + str[Z_{B}] = {1 \over 2} $.
    The Hamiltonian to be measured  is given by
    \begin{eqnarray*}
	\hat Z[\hat R] = 2 \sum_{k,l=1}^{4} Z_{kl} \hat R_{k} \, \hat R_{l} =
	{(\hat x_{1} - \hat x_{2})^{2} \over 2} + {(\hat p_{1} + \hat
	p_{2})^{2} \over 2}.
    \end{eqnarray*} 
    and is just a beam splitter mixing the two modes followed by a  measurement of  the position and momentum.
\end{exa}
\vspace{2ex}

In the example we were
sure to have a CM of the given form with entries $a$ and $b$ unknown
to us. We only used the knowledge about the structure of the CM and 
found a possible minimal entanglement witness.  When measuring the 
proposed EW we got the exact logarithmic negativity independent of 
the actual values in $\gamma$. Similar optimisations could  be done 
for every particular structure a CM can have and the better the 
knowledge of the CM the more exact one can determine the appropriate
minimal entanglement witness.

\section{Duan Criterion} \label{duan}


The Duan-criterion proposed in 2000 by
L.M. Duan et. al., Ref.  \cite{D1}, is a special case of the theory of
entanglement witnesses.
\begin{lemma}[Duan-criterion]\hspace{7cm}
   
    The criterion states that given a
separable two-mode state $\rho = \sum_{i} \, p_{i} \, \rho_{i}^{A}
\otimes \rho_{i}^{B} $ every pair of operators
    \begin{eqnarray}
      \begin{split}
	\hat u &= |a| \hat x_{1} + {1 \over a} \hat x_{2}, \\
	\hat v &= |a| \hat p_{1} - {1 \over a} \hat p_{2}
      \end{split} \quad \quad a \in \mathbbm{R} , a \not = 0,
    \end{eqnarray}
    has to fulfil the relation
    \begin{eqnarray}\label{duan0}
	\Delta^{2} \hat u + \Delta^{2} \hat v \ge a^{2} + {1 \over a^{2}},
    \end{eqnarray}
    where $\Delta^{2} \hat u  = \langle \hat u^{2} \rangle - \langle \hat u \rangle^{2}$ 
    and similarly $ \Delta^{2} \hat v$ are evaluated in the state $\rho$.
    
    A necessary condition on a state to be entangled is thus that there 
    exists a value  $a$ such, that the above relation is violated.  For Gaussian
    states, Duan et. al. showed that the condition  is also  sufficient.
\end{lemma}
\vspace{2ex}

We translate Duan's proposal in the language of entanglement
witnesses.  Without loss of generality we assume that the states to be measured 
have  vanishing displacements. Then the measured quantity  following 
Duan's proposal is given by
\setlength{\mathindent}{0cm}
\begin{eqnarray}\label{duan1}
  \Delta^{2} \hat u + \Delta^{2} \hat v &=& 
  \langle \hat u^{2} \rangle +  \langle \hat v^{2} \rangle\\
  &=&  a^{2}  \langle\hat x_{1}^{2} \rangle   
  + {1 \over a^{2}}  \langle\hat x_{2}^{2} \rangle 
+  a^{2} \langle \hat p_{1}^{2} \rangle 
  + {1 \over a^{2}} \langle \hat p_{2}^{2} \rangle
  + 2  \, {|a| \over a} \left(   \langle\hat x_{1 } \hat x_{2} \rangle
  -   \langle\hat p_{1} \hat p_{2} \rangle\right). \nonumber
\end{eqnarray}
In the language of covariance matrices we find
\setlength{\mathindent}{0cm}
\begin{eqnarray}
  \Delta^{2} \hat u + \Delta^{2} \hat v &=& 
  {a^{2} \over 2} \, \left(\gamma_{11}  +\gamma_{22} \right)
  + {1 \over  2 a^{2}}   \, \left( \gamma_{33} +  \gamma_{44} \right)
  + {|a| \over a} \,\left(   \gamma_{13}-  \gamma_{24}\right).
\end{eqnarray}
We put  Eq. \eqref{duan1} in an appropriate form to read off the 
corresponding entanglement witnesses. 
\setlength{\mathindent}{0cm}
\begin{eqnarray}
1 \, \le  \,  { \Delta^{2} \hat u + \Delta^{2} \hat v \over a^{2} + {1 \over a^{2}}} 
&=&  {1 \over 2 (a^{2} + {1 \over a^{2} })} \, 
\left[a^{2}  \,  \left(\gamma_{11}  +\gamma_{22} \right)
+ {1 \over   a^{2}} \,   \left( \gamma_{33} +  \gamma_{44} \right) \right] \nonumber\\
&& + {1 \over 2 (a^{2} + {1 \over a^{2} })} \, 2 \, {|a| \over a} \, 
\left(   \gamma_{13}-  \gamma_{24}\right).
\end{eqnarray}
One  family of  entanglement witnesses  $Z_{a}$ which give 
$tr[\hat Z_{a} \, \rho ] = { \Delta^{2} \hat u 
+ \Delta^{2} \hat v \over a^{2} + {1 \over a^{2}}}$ when measured in an about 
zero centered state $\rho$ has the matrix representation
\setlength{\mathindent}{2cm}
\begin{eqnarray*}
    Z_{a}= {1 \over 2 (a^{2} + {1 \over a^{2} } ) } \,\,   \left(
    \begin{array}{cccc}
	a^{2} & 0& {|a| \over a}& 0\\	
	0&  a^{2}& 0& - { |a| \over a}\\ 
	{|a| \over a}&0&{1 \over a^{2}}& 0\\ 
	0&- {|a| \over a} &0&{1 \over a^{2}}
    \end{array} \right).
\end{eqnarray*}
By construction these witnesses, with arbitrary $a \not = 0$,  give for all separable 
states  values greater or equal to one, see  Eq. \eqref{duan0}.
For all $a \not =0 \,$ $Z_{a}$ is then indeed an entanglement witness since it is
a positive, symmetric matrix fulfilling the condition on the symplectic trace  of its submatrices
\begin{equation*}
  str[Z_{a \, A}] + str[Z_{a \, B}] = {1 \over 2 (a^{2} + {1 \over a^{2} })} \,\, 
  (a^{2} + {1 \over a^{2}}) =  {1 \over 2}
\end{equation*}
and  give values greater or equal to one for all separable covariance matrices.
The sufficiency is a bit surprising since the family of these entanglement
witnesses is not the whole set $\mathcal{Z}_{A|B}(\mathbbm{R}^{2N})$,
for a proof see the original paper \cite{D1}.\\

We conclude that the theory of entanglement witnesses is more general
than the Duan-criterion, since it can be used for  multi-mode states as well. 
We also note that the set of necessary EWs could be smaller than
defined in Def. \ref{entwit} and characterised by Theorem \ref{47}, as the
Duan-criterion shows.  Further investigations should analyse which
entanglement witnesses are redundant.  For example the set of EWs
could be equipped with an equivalence relation  collecting
entanglement witnesses, which can be transformed into each other by
local symplectic transformations in  equivalence classes.  This
would not change the theory of EWs since the entanglement properties of
the states are preserved under LOCC.

  \chapter{Summary and Outlook}

In the thesis we investigated entanglement properties of Gaussian
states and how well one can estimate them for a given state with real
experimentally available measurements.
In the second chapter we started to dive into the language of continuous
variable states.  We learned what symplectic transformations are and
how they are related to unitary transformations.  We introduced the
quantum characteristic function, in analogy to the classical Fourier
transformed of a probability distribution, describing a quantum statistical
system.  Finally we formulated the definition of entanglement of
bipartite states and cited the famous PPT-criterion allowing to judge
if a state is entangled or not.  In the third chapter we defined
Gaussian states and found that well known states like the coherent,
the squeezed and the thermal states of a harmonic oscillator  belong
to the family of Gaussian states.  We recognised that for Gaussian
states the covariance matrix plays an extraordinary role, describing
all important properties like the Heisenberg uncertainty relation,
purity, squeezing and especially entanglement of the state.  We
investigated two normal forms of covariance matrices and formulated
conditions on the CM's characteristic numbers connected to them.
We went on to exploit general operations one could implement on
Gaussian states, without destroying this important property.  We saw
that in fact many experimentally available operations are of
Gaussian nature,  and  learned what homodyne measurements do
and how a projection of a two mode Gaussian state on coherent states
of its second mode changes the state's covariance matrix.  

In the fifth chapter we gave a first try on estimating the
entanglement properties of an unknown Gaussian two-mode state.  Instead of
measuring all entries of the state's covariance matrix we proposed a
scheme allowing to calculate the exact symplectic eigenvalues of the CM
and its partial transposed with nine kinds of measurements.
  With them we can determine the degree of
entanglement the Gaussian state possesses, quantified in the
logarithmic negativity.  Many possible ways lead from this ansatz; one
could ask if it is possible to determine or well estimate the entanglement
properties of a given Gaussian two-mode state with even less than nine
measurements.  Other properties like purity or squeezing could be
investigated with similar strategies and probably less measurements.  
In the next step one should try to find schemes for entanglement estimation
for more mode states and non-Gaussians.

Finally, we introduced the  fantastic concept of entanglement witnesses
allowing us to give a lower bound for the logarithmic negativity of a
Gaussian state by measuring quadratic Hamiltonians.  Entanglement
witnesses show a lot of nice properties, e.g., they detect bound states,
being those entangled states having a PPT.  In comparison to other
measurement requirements they are easily available and with a  
minimal entanglement witness for a given CM, it is actually possible to give
the exact logarithmic negativity.  
Many questions concerning EWs remain open and  should be answered in the near future.  
We give now some proposals for further research issues:
\begin{itemize}
    \item[]{ When discussing the Duan-criterion in Section \ref{duan} we
      saw that  the set of entanglement witnesses is bigger than necessary and
      should be reduced to a minimal set.  A first step in this direction
      would be a equivalence relation grouping all EWs equivalent up to
      local symplectic transformations in equivalence classes.  Further
      reductions can probably be found.} 
    \item[]{Bound entangled states  could maybe be characterised by the 
      entanglement witnesses detecting them. Is it then  possible to find
      a nontrivial lower and upper bound for the minimal $p_{min}$ for all
      bound entangled states?} 
    \item[]{In our discussion we only used the
      logarithmic negativity.  One could check if the entanglement witnesses
      allow statements about the degree of entanglement of Gaussian states
      measured in other entanglement measures.  Additionally the
      $p-$separability seems to imply an ordering concerning the degree of
      entanglement in the set of covariance matrices.  It should be
      investigated if the minimal $p_{min}(\gamma)$ itself is a proper
      entanglement measure for the CM $\gamma$ or if known entanglement measures respect the
      $p-$ordering as the logarithmic negativity does.} 
    \item[]{We gave a
      proof that for two-mode Gaussian states the logarithmic negativity is
      exactly given by the logarithm of the inverse minimal $p_{min}(\gamma)$ one
      could measure for the CM $\gamma$ belonging to the state.  Is that
      also possible for more mode states?  And are there other strategies to
      determine the minimal entanglement witness for a particular class of
      covariance matrices like in Example 9?} 
    \item[]{The entanglement
      witnesses were formulated only on the covariance matrices.  Hence they
      decide about the entanglement properties of Gaussian states and give a
      sufficient condition for non-Gaussian states having an entangled CM. Is
      it possible to generalise the concept of entanglement witnesses to
      higher moments so that it applies to other classes of continuous
      variable states?  } 
    \item[]{And finally, for systems split in more
      than two parties what definition of entanglement shall one choose and
      how can entanglement witnesses used for  multi-partite
      entanglement?  }
\end{itemize}

 \begin{appendix}

\chapter{Definitions and Proofs}
\begin{defi}\label{Poisson}
The Poisson bracket is defined by
\begin{equation*}
\{ F, G \} = \sum_{i=1}^{N} 
{\partial F \over \partial x_i} {\partial G \over \partial p_i} -
 {\partial G \over \partial x_i} {\partial F \over \partial p_i}.
\end{equation*}
The commutator of two $M \times M$ matrices or operators $A$ and $B$ is defined by 
\begin{equation*}
    [A, B] = AB -BA.
\end{equation*}
The anti-commutator of two $M \times M$ matrices or operators $A$ and $B$ is defined by 
\begin{equation*}
    \{A, B\}_{+} = AB + BA.
\end{equation*}
\end{defi}

\begin{defi}\label{det=1}
A \emph{ symplectic vector space $(V, \omega )$} is a finite dimensional real vector space $ V$ 
equipped with a distinguished bilinear form $ \omega$ which is antisymmetric and 
nondegenerate, i.e. $ \omega( u, v) = - \omega(v, u) \quad u, v \in V $ and for every 
$u \not= 0 \in V $ there is a $ v \in V$ satisfying $ \omega(u, v) \not= 0$. 
\end{defi}

For example the standard space  $ ( \mathbbm{R}^{2N} , \omega_0 ) $ 
is a symplectic vector space with  
$\omega_0 (u, v) = v^{T} \sigma u $ for all $ u, v \in \mathbbm{R}^{2N}$
where $ \sigma = \bigoplus_{i= 1}^N
\left(\begin{array}{rr}0&1\\-1&0\end{array}\right)$.

\begin{defi}
A linear map $ S : V \rightarrow V $ on a symplectic vector space $ (V, \omega) $ 
is called \emph{ symplectic}  or \emph{ canonical} if 
\begin{equation*}
S^{*}\omega = \omega
\end{equation*}
where $S^* \omega $ is the pullback 2-form given by $ S^* \omega(u, v) = \omega (Su, Sv)$
\end{defi}

In the standard space   $ ( \mathbbm{R}^{2N} , \omega_0 ) $ a matrix $S $ is symplectic iff 
$\omega_0 (u, v) = v^{T} \sigma u = \omega_0(Su, Sv) = (S v)^{T} \sigma S u $
for all $ u, v \in \mathbbm{R}^{2N}$, that is $S^T \sigma S = \sigma
$.

It follows immediately that $ \det^2 S = 1$. To fix the sign 
we observe that the exterior power 
$\Omega = \omega_0 \wedge \omega_0 \wedge ... \wedge \omega_0 $ with 
$ N$ factors is a volume form since $\omega_0 $ is nondegenerate. 
We know that there is only  one  alternating $N-$ form  on $V $ which gives  
the volume spanned by the vectors $u_i$ with $ i= 1, ... N $ it is acting on. 
This is the determinant itself or  the determinant multiplied with an arbitrary nonzero number.\\

If $ S $ is a matrix in  $  \mathbbm{R}^{2N} $ then 
\begin{eqnarray*}
\Omega (u_1, ..., u_{2N}) & = & S^{*}\Omega (u_1, ..., u_{2N}) 
= \Omega (S u_1, ...,S u_{2N})\\
&=& c \, \det [S u_1, ...,S u_{2N}] \nonumber
 = c \, \det S \, \det [ u_1, ..., u_{2N}] \\ \nonumber
&=& \det S \,  \Omega ( u_1, ..., u_{2N}) 
\end{eqnarray*}
 
All symplectic matrices therefore have determinant plus one.  The proof was
taken from Ref.  \cite{HZ}.

\begin{lemma}\label{traceweyl} The trace of an $N-$mode Weyl operator $\hat
W_{\xi}$, $\xi \in \mathbbm{R}^{2N}$ is given by $tr [ \hat W_{\xi}] =
(2\pi)^{N} \delta^{2N}(\xi)$, where $\delta^{2N}(\xi) = \delta(\xi_{1}
) \cdot \ldots \cdot \delta(\xi_{2N} )$ denotes the  $2N-$dimensional delta
distribution.
\end{lemma}\setlength{\mathindent}{0cm}
\begin{proof}
 With the properties of the coherent states (see Section \ref{csts}) we
 can calculate the trace of every operator in this overcomplete  basis.
 \begin{eqnarray*}
	tr [\hat A] &=& \sum_{n=0}^{\infty} \; \langle n | \hat
	A | n \rangle 
	= {1 \over \pi} \sum_{n=0}^{\infty} \; 
	\int  \dif \Re(\alpha) \dif \Im(\alpha) \; \langle n | \alpha \rangle
	\langle \alpha | \hat A | n \rangle \\
    	&=& {1 \over \pi} \; \int  \dif \Re(\alpha) \dif \Im(\alpha) \;
	\sum_{n=0}^{\infty} 
	\langle \alpha | \hat A | n \rangle \langle n | \alpha \rangle \\
	&=& {1 \over \pi} \; \int  \dif \Re(\alpha) \dif \Im(\alpha) \;
	\langle \alpha | \hat A | \alpha \rangle
\end{eqnarray*}
With the tensor product structure of the Weyl operators $ \hat W_{\xi}
= \bigotimes_{j = 1}^{N} \hat W_{\eta^{j}} $, $\xi = (\eta^{1},
\ldots, \eta^{N})^{T} \in \mathbbm{R}^{2N}, \eta^{j} \in \mathbbm{R}^{2}$, it
is sufficient to calculate only the one-mode case.  Using the identity
for the one mode $\hat W_{\eta} = \hat D(-\beta)$ for $\beta =
{\eta_{1} + i \eta_{2} \over \sqrt{2}}$ and furthermore $\hat
D(\beta)\hat D(\alpha) = e^{- i\Im(\beta \alpha^*)} \, \hat D(\beta +
\alpha)$, we start to calculate the trace of the Weyl operators.
\begin{eqnarray*}
    tr [ \hat W_{-\eta} ] &=& {1\over \pi} \int \dif \Re(\alpha) \dif
    \Im(\alpha) \; \langle \alpha | \hat D(\beta) | \alpha \rangle \\
    &=& {1 \over \pi} \int \dif \Re(\alpha) \dif \Im(\alpha)  \; \langle 0| \hat
    D(-\alpha) \hat D(\beta) \hat D(\alpha) | 0\rangle \\
    &=& {1 \over \pi} \int \dif \Re(\alpha) \dif \Im(\alpha) \; e^{- 2 i
    \Im(\beta \alpha^{*}) } \langle 0| \hat D(\beta) | 0\rangle \\
      &=& {1 \over \pi} e^{ -\Re(\beta)^{2} - \Im(\beta)^{2} } \; 
      \int \dif \Im(\alpha)  \; e^{2 i\Re(\beta) \Im(\alpha) }
      \int \dif \Re(\alpha)  \; e^{-2 i \Im(\beta) \Re(\alpha)} \\
       &=& {1 \over \pi} e^{ -{\eta_{1}^{2} + \eta_{2}^{2} \over 2} } \; \int
       \dif \Im(\alpha) \; e^{\sqrt{2} i \eta_{1} \Im(\alpha) } \int \dif
       \Re(\alpha) \; e^{- \sqrt{2} i \eta_{2}\Re(\alpha)} \\
      &=& {1\over \pi} e^{ -{\eta_{1}^{2} + \eta_{2}^{2} \over 2} }  \; 
      2 \pi \; \delta(\sqrt{2} \eta_{1} ) \; 2 \pi \; \delta(\sqrt{2}
      \eta_{2}) \\
      &=& 2\pi \; \delta( \eta_{1}) \; \delta( \eta_{2})
\end{eqnarray*}
Immediately we see that for the $N-$mode Weyl operators 
$ tr [\hat W_{\xi}] = tr [\bigotimes_{j = 1}^{N} \hat W_{\eta^{j}} ] 
= \prod_{j=1}^{N} tr [ \hat W_{\eta^{j}}] = (2\pi)^{N} \;
\prod_{j=1}^{2N} \delta( \xi_{j})$,  as  stated.  
\end{proof}

\begin{lemma}[Baker-Campbell-Hausdorff]\label{BCH}\hspace{15cm}

  For $M \times M$ matrices or operators $A$ and $B$
    \begin{equation}
	e^{ A} \,  B \, e^{-  A } 
	= B + [A,B] + {[A, [A,B]]\over 2! } + {[A, [A, [A,B]]]\over 3! } \dots 
    \end{equation}
    For $ C= [A, B] $ and $ [A, C] = 0 = [B, C]$:
    \begin{equation}
  e^{A + B }= e^{A} \cdot e^{B} \cdot e^{- {C \over 2}}. 
    \end{equation}
\end{lemma}

\begin{lemma}\label{posmat}
For every two  positive $N \times N$ matrices $A $ and  $B$
: $ tr[A B] \ge 0$
\end{lemma}
\begin{proof}
 For every positive matrix $A$ the square root is defined by
$\sqrt A := O \sqrt D O^T$ were $D$ is the diagonal matrix of the (positive) 
eigenvalues of $A$ and $O$ the orthogonal matrix which diagonalizes
$A$.  Since the trace is invariant under cyclic permutations it is $
tr [A B] = tr[\sqrt A \sqrt A B] = tr[\sqrt A B \sqrt A] \ge 0$
\end{proof}

\begin{lemma} \label{8}
    The spectrum of a matrix $C= A B$ with $A$ and 
    $B$ square matrices and one of them invertible  
    is the same as the the spectrum of the matrix $D = B A$. 
 \end{lemma}
    \begin{proof}
        The eigenvalues of the product $A B$ are determined by the equations
        \begin{equation*}
	0= (A B - \lambda_{i} \mathbbm{1}) x_{i}
	=(A - \lambda_{i} B^{-1}) B x_{i}
	= B^{-1} (B A - \lambda_{i}) B x_{i}.
        \end{equation*}
        Since $B^{-1} $ is nonsingular the same eigenvalue equations 
        hold for the product $B A$.
    \end{proof}

\begin{lemma}
The symplectic eigenvalues of a positive matrix are invariant under
symplectic transformations $S \in Sp(2N, \mathbbm{R})$.
\end{lemma}
\begin{proof}
    The symplectic eigenvalues of a positive matrix $ A $  are given by  the (usual) 
eigenvalues of $ - i \sigma A$, determined by  $ [-i \sigma A - \lambda_i \mathbbm{1} ] x_i = 0$,
 with all positive $\lambda_i$ symplectic eigenvalues of $A$.  Let now $S $ denote a symplectic
transformation so that $A = S A' S^T$.  It follows that $A'$ has the
same eigenvalues as $A$ since $ [-i \sigma S A' S^T - \lambda_i
\mathbbm{1} ] x_i = S^{-T} [-i \sigma A - \lambda_i \mathbbm{1} ] S^T
x_i = 0 $.
\end{proof}

\end{appendix}

\cleardoublepage
\addtocontents{toc}{\contentsline{chapter}{Bibliography}{\thepage}}

\printindex

 \backmatter
 
 \fancyhead[RO]{\thepage} 
 \fancyhead[RE]{\bf INDEX}
 \fancyhead[LO]{\bf INDEX}
 \fancyhead[LE]{\thepage}
 \fancyfoot[CO,CE]{}

\chapter{Index}
\begin{itemize}
    \item[]\hspace{-0.9cm} $|\alpha \rangle $, coherent
    state vectors with complex $\alpha$ \dotfill 20 
     
    \item[]\hspace{-0.9cm} $A^{WNF} $, Williamson normal form of a positive, even
    dimensional matrix $A$ \dotfill 24
   
    \item[]\hspace{-0.9cm} $\mathcal{B}(\mathcal{H})$, bounded linear operators on
    $\mathcal{H}$ \dotfill 10

    \item[]\hspace{-0.9cm} CCR, canonical commutation relations \dotfill 6

    \item[]\hspace{-0.9cm} CM, covariance matrix \dotfill 10
    
    \item[]\hspace{-0.9cm} CVS, continuous variable states \dotfill 5

    \item[]\hspace{-0.9cm} $\hat D(\alpha)$, displacement operators, elements of $\mathcal{B}(\mathcal{H})$
      \dotfill 13 

    \item[]\hspace{-0.9cm} $\vec d$, displacement vector,
    element of $\mathbbm{R}^{2N}$, also $\vec D = \sigma \vec d $ \dotfill 10

    \item[]\hspace{-0.9cm} EW, entanglement witness, elements of  $\mathcal{Z}_{A|B}(\mathbbm{R}^{2N})$ \dotfill 55, 59

    \item[]\hspace{-0.9cm} $\Gamma_{A|B}(\mathbbm{R}^{2N})$, closed convex set of $2N
    \times 2N$ covariance matrices,\\ separable with respect to the split
    $A|B$ \dotfill 58   

    \item[]\hspace{-0.9cm} $\Gamma^p_{A|B}(\mathbbm{R}^{2N})$, closed convex set of $2N
    \times 2N$ covariance matrices,\\ $p-$separable with respect to the split
    $A|B$ \dotfill 64

    \item[]\hspace{-0.9cm} $\Gamma(\mathbbm{R}^{2N})$, closed convex set of $2N \times 2N$
    covariance matrices \dotfill 58
   
    \item[]\hspace{-0.9cm} $\gamma$, covariance matrix, element of
    $\Gamma(\mathbbm{R}^{2N})$, also $\Gamma = \sigma \gamma \sigma^{T}$ \dotfill 10
   
    \item[]\hspace{-0.9cm} $\Gamma^{SNF}$, Simon normal form of a two-mode covariance
    matrix $\Gamma$ \dotfill 27
    
    \item[]\hspace{-0.9cm} $\Gamma^{T_{A}}$, partially transposed covariance matrix \dotfill 30

    \item[]\hspace{-0.9cm} $\mathcal{H}$, Hilbert space, in our case the
    Hilbert space is the $\mathcal{L}^{2}(\mathbbm{R}^{N}, \mathbbm{C})$,\\
    the space of square integrable complex functions over $\mathbbm{R}^{N}$ \dotfill 10

    \item[]\hspace{-0.9cm} $K(N)$, passive symplectic transformations \dotfill 7
    
    \item[]\hspace{-0.9cm} LOCC, local operations and classical communication \dotfill 16
    
    \item[]\hspace{-0.9cm} $\Pi(N)$, active symplectic transformations \dotfill 8 

    \item[]\hspace{-0.9cm} $p-$separable \dotfill 64

    \item[]\hspace{-0.9cm} QIT, quantum information theory \dotfill 5

    \item[]\hspace{-0.9cm} $\hat R =(\hat R_1, ~...~,\hat R_{2N})^T =(\hat x_1,\hat
    p_1,~...~,\hat x_N,\hat p_N)^T $ \dotfill 6

    \item[]\hspace{-0.9cm} $\rho^{T_{A}}$, partial transpose of a state,
     \dotfill 17
    
    \item[]\hspace{-0.9cm} $\rho_{t}$, thermal state of a harmonic
    oscillator \dotfill 23 
    
    \item[]\hspace{-0.9cm} $ S_{A|B}(\mathcal{H})$, set of separable states
    $\rho$ on $\mathcal{H}_{A} \otimes \mathcal{H}_{B}$, subset of  $S(\mathcal{H})$ \dotfill 16
    
    \item[]\hspace{-0.9cm} $S(\mathcal{H})$, closed convex set of states on a Hilbert space
    $\mathcal{H}$,\\ subset of $\mathcal{B}(\mathcal{H})$ \dotfill 10

    \item[]\hspace{-0.9cm} $\sigma =\bigoplus_{i= 1}^N
    \left(\begin{array}{rr}0&1\\-1&0\end{array}\right)$, symplectic
    matrix \dotfill 5 
    
    \item[]\hspace{-0.9cm} Simon invariants \dotfill 27

    \item[]\hspace{-0.9cm} $Sp(2N, \mathbbm{R})$, set of symplectic  transformations \dotfill 6

    \item[]\hspace{-0.9cm} states or density matrices $\rho$ in $S(\mathcal{H})$ \dotfill 10 

    \item[]\hspace{-0.9cm} $str[A]$, symplectic trace of a positive, symmetric, \\ 
    even dimensional matrix $A$ \dotfill 61

    \item[]\hspace{-0.9cm} symplectic eigenvalues \dotfill 24

    \item[]\hspace{-0.9cm} $\hat W_{\xi} $, Weyl operators, elements of $\mathcal{B}(\mathcal{H})$ \dotfill 12

    \item[]\hspace{-0.9cm} $|\zeta\rangle$, squeezed vacuum state vector, with complex $\zeta$ \dotfill 21

    \item[]\hspace{-0.9cm} $\mathcal{Z}_{A|B}(\mathbbm{R}^{2N})$, set of real symmetric matrices,\\
    with $Z \in \mathcal{Z}_{A|B}(\mathbbm{R}^{2N}) \, \iff Z \ge 0 $ and
    $str[Z_{A}] + str[Z_{B}] \ge{1 \over 2} $ \dotfill 59, 61
    
    \item[]\hspace{-0.9cm} $\mathcal{Z}(\mathbbm{R}^{2N})$, set of real symmetric matrices,\\
    with $Z \in \mathcal{Z}(\mathbbm{R}^{2N}) \, \iff Z \ge 0 $ and $str[Z] \ge{1 \over 2} $ \dotfill 59, 61 
    
    \item[]\hspace{-0.9cm} $\hat Z[\hat R]$, denotes the  Hamiltonian defined by a matrix $Z
    \in \mathcal{Z}_{A|B}(\mathbbm{R}^{2N}) $\dotfill 59
    
\end{itemize}

\cleardoublepage

 \fancyhead[LO]{\bf } \fancyhead[RO]{\thepage}
 \fancyhead[RE]{\bf} \fancyhead[LE]{\thepage} \fancyfoot[CO,CE]{}
 \vspace{3cm}\bfseries \huge Acknowledgements \normalsize \rm \\\vspace{3ex}

In the first place I am very grateful to Jens Eisert giving me the
opportunity to explore a challenging subject and helped me all along my way
with great discussions in a friendly and open atmosphere.  I want to
thank Martin Wilkens, who invited me to join his group at an
early stage so that I became  well-integrated and made many friends.  He
always lent me  an ear  when I found myself in trouble.  The Quantum
Optics Group gave me a real nice experience (thanks for the coffee) and I am
especially grateful to Timo Felbinger for all his assistance in keeping my
computers running.  He was the one who answered hundreds of those
little but not so easy questions, always enlightening me. \\

I thank my parents supporting me whenever I needed them and especially my sister
Clara who gave me a smile and big hugs when I was down. 
Thanks also to Katrin Junghans who helped me to draw  the nice  pictures.
Finally I want to thank Andreas Keil who helped me in every possible way,
including our late night  discussions, keeping me alive when I forgot to
care about food and sorting out mistakes and misprints in this
thesis.\\

I want to thank the Studienstiftung des deutschen Volkes for financial
support and the Wilhelm und Else Heraeus Stiftung for giving me the
opportunity to join two well-organised and instructive summer schools
on QIT.\\

Potsdam, 30.09.2003 \hfill Janet Anders\\

\cleardoublepage

 \fancyhead[LO]{\bf } \fancyhead[RO]{\thepage} \fancyhead[RE]{\bf}
 \fancyhead[LE]{\thepage} \fancyfoot[CO,CE]{}
 
\end{document}